\documentclass[fleqn,usenatbib]{mnras}

\usepackage{newtxtext,newtxmath}

\usepackage[T1]{fontenc}
\usepackage{xcolor}
\usepackage{amsmath}
\usepackage{comment}
\usepackage{fix-cm}

\DeclareRobustCommand{\VAN}[3]{#2}
\let\VANthebibliography\thebibliography
\def\thebibliography{\DeclareRobustCommand{\VAN}[3]{##3}\VANthebibliography}

\newcommand{\msun}{\,\rm M_\odot}

\newcommand{\I}{$\scriptstyle\mathrm{I}$}

\usepackage{graphicx}	
\usepackage{amsmath}	
\usepackage{pifont}
\newcommand{\cmark}{\ding{51}}%
\newcommand{\xmark}{\ding{55}}%
\usepackage{hyperref}

\title[BH accretion in dwarfs]{The MandelZoom project I: modelling black hole accretion through an $\alpha$-disc in dwarf galaxies with a resolved interstellar medium
}

\author[E. -J. Shin et al.]{
Eun-jin Shin,$^{1,2}$\thanks{E-mail: ejs237@cam.ac.uk}
Debora Sijacki,$^{1,2}$
Matthew C. Smith,$^{3}$
Martin A. Bourne,$^{1,2,4}$ and
Sophie Koudmani $^{4,5}$
\\
$^{1}$Institute of Astronomy, University of Cambridge, Madingley Road, Cambridge CB3 0HA, UK\\
$^{2}$Kavli Institute for Cosmology, University of Cambridge, Madingley Road, Cambridge CB3 0HA, UK\\
$^{3}$Max-Planck-Institut f{\"u}r Astrophysik, Karl-Schwarzschild-Str. 1, D-85748, Garching, Germany\\
$^{4}$Centre for Astrophysics Research, Department of Physics, Astronomy and Mathematics, University of Hertfordshire, College Lane, Hatfield, AL10 9AB, UK\\
$^{5}$St Catharine's College, University of Cambridge, Trumpington Street, Cambridge CB2 1RL, UK\\
}

\date{MNRAS, submitted}

\pubyear{2025}

\begin{document}
\label{firstpage}
\pagerange{\pageref{firstpage}--\pageref{lastpage}}
\maketitle

\begin{abstract}
While mounting observational evidence suggests that intermediate mass black holes (IMBHs) may be important in shaping the properties of dwarf galaxies both at high redshifts and in the local Universe, our theoretical understanding of how these IMBHs grow is largely incomplete. To address this, we perform high-resolution simulations of an isolated dwarf galaxy with a virial mass of $10^{10}\msun$ harbouring a $10^4\msun$ IMBH at its centre at a peak spatial resolution of $\lesssim 0.01$~pc. Within the fully multi-phase interstellar medium (ISM), we incorporate explicit sampling of stars from the initial mass function, photo-ionization, photoelectric heating, individual supernovae (SNe), as well as a Shakura-Sunyaev accretion disc model to track the evolution of BH mass and spin. We find that a nuclear star cluster (NSC) effectively captures the ISM gas and promotes formation of a circumnuclear disc (CND) on scales of $\lesssim 7$~pc. Simultaneously, gravitational torques from the NSC reduce CND angular momentum on (sub-)parsec scales, circularizing the gas onto the $\alpha$-accretion disc and promoting sustained IMBH growth at $\sim 0.01$ of the Eddington rate. While in the innermost regions ($\lesssim 0.5$~pc), star formation is highly suppressed, the CND is susceptible to fragmentation, leading to the formation of massive, young stars. Interestingly, despite an in-situ SN rate of $0.3$~Myr$^{-1}$, the dense CND persists, sustaining BH accretion and leading to its net spin-up. Our study demonstrates the complexity of IMBH accretion within a multi-phase ISM, and paves the way for next-generation studies where IMBH growth in a fully cosmological context can be captured. 
\end{abstract}




\section{Introduction}
Intermediate-Mass Black Holes (IMBHs), defined as BHs with masses ranging from $100$ to $10^6 \,\msun$, are a critical puzzle piece in our understanding of the assembly of supermassive black holes (SMBHs; $10^6-10^{11}~\msun$). As possible analogues of massive galaxies hosting SMBHs \citep[e.g.,][]{Magorrian+1998, Ferrarese+Merritt2000,Haring+Rix2004, Kormendy+Ho2013, Reines+Volonteri2015}, lower-mass galaxies, and in particular dwarfs, are considered promising sites for IMBH formation and growth. This expectation is supported by the well-established scaling relations between BH masses and host galaxy properties, such as the stellar velocity dispersion, bulge and stellar mass \citep[e.g.,][]{Greene+Ho2007, Greene+2020, Reines2022, Gultekin+2022}. However, detecting such low-mass BHs remains challenging, as their limited dynamical influence and intrinsically low luminosities make them elusive observational targets \citep{Greene+2020}. Nevertheless, the number of IMBH candidates identified in both local and high-redshift galaxies is steadily increasing, reflecting growing observational efforts and interest in this population \citep[e.g.,][]{Farrell+2009, Davis+2020, Kim+2020, Pechetti+2022, 
Pacucci+2023, Bykov+2024, Davis+2024, Haberle+2024, Maiolino+2024, Sacchi+2024, Sanchez+2024}. Fundamental questions, however, remain: how do BH seeds form and evolve within their host galaxies? How is multiphase gas funnelled into the vicinity of BHs? How is this inflow ultimately circularized to a BH accretion disc?

Nuclear Star Clusters (NSCs), frequently found at the centres of dwarf galaxies, are particularly favourable for the formation and growth of IMBHs. With masses ranging from $10^4\, \msun$ to $10^8\, \msun$ and effective radii of a few parsecs or less, NSCs are among the densest stellar systems in the Universe \citep{Neumayer+2020}.  Despite their prevalence, the origin of NSCs remains an open question. Recent studies suggest they form through multiple channels: from the remnants of globular clusters (GCs) that spiral inward due to dynamical friction \citep[e.g.,][]{Tremaine+1975, Capuzzo-Dolcetta1993, Capuzzo-Dolcetta+Miocchi2008, Agarwal+Milosavljevic2011, Portaluri+2013, Gnedin+2014, lyu+2025, Poulain+2025}, in-situ star formation at the galactic centre \citep[e.g.,][]{Loose+1982, Milosavljevic2004, Walcher+2006, Schinnerer+2007, Antonini+2015, Paudel+Yoon2020, Fahrion+2021, Fahrion+2022, Partmann+2025}, or via a combination of both of these processes. The presence of NSCs is observed in galaxies spanning a wide range of masses, morphologies, and gas contents, with their nucleation appearing to depend on a complex interplay of these factors \citep[e.g.,][]{Georgiev+2009b, den_Brok+2014, Georgiev+2016}. In the Virgo Cluster, approximately $30\%$ of galaxies with stellar masses greater than $10^7 \,\msun$ host NSCs at their centres and this fraction rises to $80\%$ for galaxies with stellar masses above $10^9\,\msun$ \citep{Sanchez-Janssen+2019, Neumayer+2020, Carlsten+2022, Hoyer+2021}. Meanwhile, recent observations from the {\it James Webb Space Telescope} ({\it JWST}) have revealed extremely dense (reaching $10^{5}\,\msun~{\rm pc}^{-3}$) and clustered star-forming environments, such as the Cosmic Grapes galaxy \citep{Fujimoto+2024, Topping+2024} and the Cosmic Gems \citep{Adamo+2024}. These findings suggest that such high-density environments may not be rare, but perhaps the norm, in high-redshift galaxies.

The dense environment of an NSC provides favourable conditions for the emergence of IMBH seeds. Studies have shown that dense, metal-poor star clusters, if sufficiently compact, can trigger repeated stellar collisions, leading to the formation of very massive stars (VMSs) with masses up to $\sim 10^3\,\msun$, which collapse directly into BHs without losing mass in the process \citep{Spitzer+Stone1967, Spitzer+Hart1971, Peebles1972, Begelman+Rees1978, Mezcua2017, Stone+2017}. These formation process has been further explored using Monte Carlo method \citep[e.g.,][]{Freitag+2006, Gonzalez-Prieto+2024} and direct $N$-body simulations \citep{Portegies_Zwart+McMillan2002, Katz+2015, Arca_Sedda+2023, Rantala+2024,Fujii+2024}. Being the most massive members of star clusters, IMBH seeds experience dynamical friction, causing them to sink towards the NSC centre, where they may encounter more material or undergo hierarchical BH mergers that facilitate their growth \citep[e.g.,][]{Fujii+2024, Rantala+2024}. The NSC not only expands the region of gravitational influence but also increases the escape velocity, thereby mitigating the ejection of BHs through gravitational-wave-induced recoil kicks (which can reach up to $\sim 5000~{\rm km}~{\rm s}^{-1}$) or three-body interactions, both of which are major obstacles to hierarchical BH growth \citep{Neumayer+2020}. Furthermore, tidal disruption events (TDEs) within NSCs \citep{Stone+2017, Lee+2023, Rizzuto+2023, Chang+2025, Rantala+2024}, which provide additional accretion from disrupted stellar material and gas inflow facilitated by the NSC’s gravitational potential \citep{Fujii+2024, Partmann+2025} may promote BH growth.

Once BHs enter the IMBH regime, gas accretion may become an important channel for their further growth. Decades of research have sought to understand how gas loses its angular momentum and is transported from galactic scales to the central BH. Mechanisms such as galaxy mergers \citep[e.g.][]{Hernquist1989, Barnes+Hernquist1991} and secular processes, including bars and spiral waves, can induce gas angular momentum transport and gas inflows to sub-kiloparsec scales near the galactic centre \citep[e.g.][]{Laine+2002, Hopkins+Quataert2010, Emsellem+2015}. At scales of $ \lesssim 100$~pc, a cascade of non-axisymmetric gravitational instabilities operating across a range of spatial scales, such as ``bars-within-bars'' mechanism \citep[e.g.,][]{Shlosman+1989, Escala2007, Levine+2010, Hopkins+Quataert2011, Mayer+2015} may lead to further angular momentum transport. As gas continues to flow inward, it circularizes in the nuclear region within a few to tens of parsecs, forming a circumnuclear disc (CND). 

Within the CND, a multiphase mixture of gas and dust, spanning a range of temperatures and ionization states, arises from shocks, fragmentation, star formation, and feedback from both the active galactic nucleus and surrounding stars \citep[e.g.,][]{Maciejewski2004, Lupi+2016, Schartmann+2018, Hopkins+2024a}. On even smaller scales, within the central few parsecs, structures such as circumnuclear rings, mini-spirals, and mini-bars further accelerate angular momentum transport, guiding the gas toward the accretion disc \citep{Maciejewski2004, Mayer+2015, Trani+2018, Emsellem+2015}. At sub-parsec scales ($0.1-0.01$~pc), the gas finally circularizes into the accretion disc, where it can be funnelled onto the BH.

A classic approach to modelling BH accretion in galaxy formation simulations is the Bondi-Hoyle-Lyttleton accretion prescription, which estimates the accretion rate under the assumption of simple radial inflow. Thus, it neglects all of the above-mentioned complexities of gas transport and angular momentum transfer. On the other hand, the only accretion disc model with a global analytical solution, the Shakura-Sunyaev, or as often called, the $\alpha$-accretion disc model \citep{Shakura+Sunyaev1973}, accounts for effective viscosity the angular momentum of accreting gas through a geometrically thin and optically thick disc towards the BH, allowing for a more accurate prediction of BH mass growth. Furthermore, gas accretion not only affects the BH mass but also its spin. \cite{Bardeen+Petterson1975} studied the evolution of a spinning BH within the $\alpha$-disc framework, demonstrating how Lense-Thirring precession can warp and align the accretion disc. Subsequent studies \citep[e.g.,][]{Papaloizou+Pringle1983, King+2005, Lodato+Pringle2006} have explored how viscosity and internal torques influence this alignment process. BH spin is not a passive parameter merely influenced by accretion; it plays a crucial role in shaping radiative efficiency and launching relativistic jets \citep[e.g.,][]{King+Pringle2006, Sijacki+2009, Tchekhovskoy+2011, Talbot+2021, Talbot+2022}. Additionally, in binary mergers, BH spin at coalescence influences recoil velocities and post-merger evolution \citep[see e.g.,][]{Campanelli+2007, Schnittman2007, McKinney+2012, Gerosa+15, Bourne+2024}. 

In recent years, there has been an increased effort to improve the modelling of BH spin evolution in simulations. This includes advancements in idealized hydrodynamical simulations \citep[e.g.,][]{Fiacconi+2018, Talbot+2022, Koudmani+2024, Bourne+2024}, cosmological simulations \citep[e.g.,][]{Sijacki+2009, Dotti+2013, Dubois+2014, Bustamante+Springel2019, Dubois+2021, Dong-Paez+2023, Husko+2024, Rennehan+2024, Sala+2024, Beckmann+2023, Peirani+2024}, as well as in general relativistic magneto-hydrodynamical (GRMHD) simulations \citep[e.g.,][]{Cui+2023, Fedrigo+2024}. However, a key challenge remains in accurately modelling BH mass and spin evolution based on gas flows from the well-resolved multiphase ISM subject to realistic stellar feedback. This process requires capturing a wide range of dynamical scales, from tens of kpc down to the accretion disc scale ($\lesssim 0.1$~pc).

In this work, we aim to understand IMBH growth by more precisely modelling the physical processes from a well-resolved ISM in a galaxy down to the accretion disc at $\sim 0.01$~pc. To achieve this, we introduce the {\it MandelZoom} project -- a new multi-scale simulation framework designed to capture the detailed co-evolution of the central BH and its host galaxy across a wide dynamic range. Using the moving-mesh code {\sc Arepo}, we simulate a realistic multiphase ISM by tracking the evolution of individual stars and incorporating early stellar feedback, including photo-ionization, photoelectric heating from young massive stars, and supernova (SN) explosions, following \cite{Smith+2021}. Within this resolved ISM, we show how CNDs form and investigate the impact of NSCs on the circumnuclear region. Using the super-Lagrangian refinement scheme \citep{Curtis+Sijacki2015}, our simulation reaches a resolution of $\lesssim 0.01$~pc, sufficient to reliably capture the self-gravity radius of the $\alpha$-accretion disc, which allows us to track the mass and angular momentum flux from the ISM onto the $\alpha$-disc, and hence determine BH mass and spin evolution self-consistently. We furthermore trace the evolution of the CND, which undergoes gravitational instability, warping, in-situ star formation, and the subsequent heating and disruption via stellar feedback.

This work is organized as follows. Section~\ref{sec2:methods} describes the simulation setup (Section~\ref{sec:IC}$-$\ref{sec:SFFB}), the super-Lagrangian refinement scheme (Section~\ref{sec:refinement}), and the $\alpha$-accretion disc model (Section~\ref{sec:bh-model-method}). In Section~\ref{sec3:results}, we discuss how the ISM evolves in our simulations and analyse how NSC torques impact the evolution of the CND (Section~\ref{sec:overview}$-$\ref{sec:NSCgas-properties}). We then investigate star formation and stellar feedback in the CND in Section~\ref{sec:SF-in-CND}. Finally, we explore how the BH mass and spin evolve within the $\alpha$-accretion disc framework in Section~\ref{sec:bh-evolution}. We discuss our results in Section~\ref{sec4:Discussion}, highlighting the limitations of our model and potential future improvements before summarizing our findings in Section~\ref{sec5:conclusion}.

\begin{table*}
\centering\label{tab:runs}
\caption{List of simulations performed in this study. The parameters are: the presence of CGM, whether the simulation includes super-Lagrangian refinement scheme, whether the simulation includes an NSC, BH accretion prescription adopted, newly-created stellar mass in the galactic region, $M_{\rm new\,\star, total}$, newly-created stellar mass in the nuclear region ($r< 10$~pc), $M_{\rm new\,\star, NSC}$, and the BH mass, $M_{\bullet}$, at $t=200$~Myr.}
\begin{tabular}{cccccccc} %
\hline
Setup name & CGM & Super-Lagrangian & NSC & BH accretion & $M_{\rm new\,\star, total}$ & $M_{\rm new\,\star, NSC}$& $M_{\bullet}$\\
&  &  refinement & ($r_{\rm eff}$ [pc]) &  &[10$^5\msun$]  &[$10^3\msun$]  &[$10^4\msun$]\\
\hline\hline
{\tt noCGM}& \xmark &\xmark&\xmark&\xmark&4.86&-&-\\
{\tt CGM}& \cmark &\xmark&\xmark&\xmark&4.48&-&-\\
{\tt noNSC}& \cmark &\cmark&\xmark &$\alpha$-disc&5.38&-&1.005\\
\hline
{\tt NSC-3pc}& \cmark &\cmark&3 &$\alpha$-disc&{5.43}&{6.61}&1.107\\
{\tt NSC-5pc}& \cmark &\cmark&5 &$\alpha$-disc&
{5.35}&{2.88}&1.091\\
{\tt NSC-5pc-noSL}& \cmark &\xmark&5 &$\alpha$-disc&5.32&4.56&1.144\\
{\tt NSC-5pc-Bondi}& \cmark &\cmark&5 &Bondi&5.37&3.75&4.572\\
{\tt NSC-9pc}& \cmark &\cmark&9&$\alpha$-disc&5.43&7.44&1.116\\
\hline
\end{tabular}
\end{table*}

\section{Numerical Methods}
\label{sec2:methods}
\subsection{Initial conditions}
\label{sec:IC}
In this study we adopt a Wolf–Lundmark–Melotte (WLM)-like dwarf system from \cite{Smith+2021}. We generate initial conditions (ICs) for a dwarf galaxy with a total mass of $10^{10} \msun$ using {\sc MakeNewDisk} \citep{Springel+2005}. The IC includes a dark matter halo following the \cite{Hernquist1990} profile, with a concentration parameter $c = 15$, a spin parameter $\lambda = 0.035$, and a virial radius $r_{\rm vir}=41$~kpc. The ICs contain an exponential disc with a scale radius of $1.1$~kpc. 
The gas and stellar discs have total masses of $6.83\times 10^7 \msun$ and $ 9.75\times 10^6 \msun$, respectively.
The vertical distribution of the stellar disc follows a Gaussian profile with a scale-height of $0.7$~kpc, while that of the gas disc is set to achieve hydrostatic equilibrium with an initial temperature of 10$^4$~K. We initially set the metallicity of the gas to $0.1\,\mathrm{Z_\odot}$. Our standard refinement scheme keeps the mass of gas cells within a factor of 2 of the target mass, $m_{\rm gas, target}=20\msun$, while the dark matter and pre-existing star particles have masses of $1640\msun$ and $20\msun$, respectively.
The gravitational softening length for dark matter is $20$~pc, while star particles and gas cells (in non-super-Lagrangian region) have adaptive softening lengths that can go down to a minimum of $1.75$~pc. To avoid artificial initial starbursts, we pre-process our initial conditions by evolving the system for 100~Myr with radiative cooling enabled, star formation disabled, and turbulence driven by a modified version of our fiducial SN feedback model. We further include the circumgalactic medium (CGM), nuclear star clusters (NSCs) with a range of masses and the central massive black hole in our ICs (see Table~\ref{tab:runs}, which summarizes our entire simulation suite).

The CGM regulates the galactic baryonic cycle by supplying new gas to the disc and exerting pressure that modulates the outflows. We use CGM properties, including density, temperature, pressure, and metallicity, derived from cosmological zoom-in simulations of dwarf galaxies by \citet{Koudmani+2022}. We initialize the CGM density profile to follow the Hernquist profile in order to achieve hydrostatic equilibrium with the dark matter halo potential\footnote{ {While, in principle, there is no well-defined equilibrium CGM for dwarf galaxies in a cosmological context, it is important to include a CGM component within isolated galaxy simulation setups, especially to be able to model more realistic propagation of galactic inflows and outflows \citep[e.g.,][]{Shin+2021}. Therefore, we have adopted CGM properties as found in the high-resolution zoom-in simulations of a similar mass dwarf galaxy \citep{Koudmani+2022}.}}. 
The CGM is centred within the simulation box, which has an extent of $200\times200\times4000$~kpc$^3$. The total mass of the CGM is set to $2.64\times10^7\msun$ within the virial radius, $r_{\rm vir} = 41$~kpc. In the outer regions, a uniform density gas with \( n_{\rm H} = 2.24\times10^{-6} \) cm\(^{-3}\) and a temperature of $3000$~K fills the entire simulation box. We initially set a spatially dependent mass resolution within the CGM, ranging from $20 \msun$ near the disc to $10^5 \msun$ at the edge of the box. Subsequently, a tracer-dependent refinement scheme is applied to the CGM, which will be discussed in a later section. In the simulation without a CGM (\texttt{noCGM}), we set a negligible uniform density in the outer region of the disc, which remains unchanged as it is neither refined nor derefined.

\begin{figure}
\includegraphics[width=0.5\textwidth]{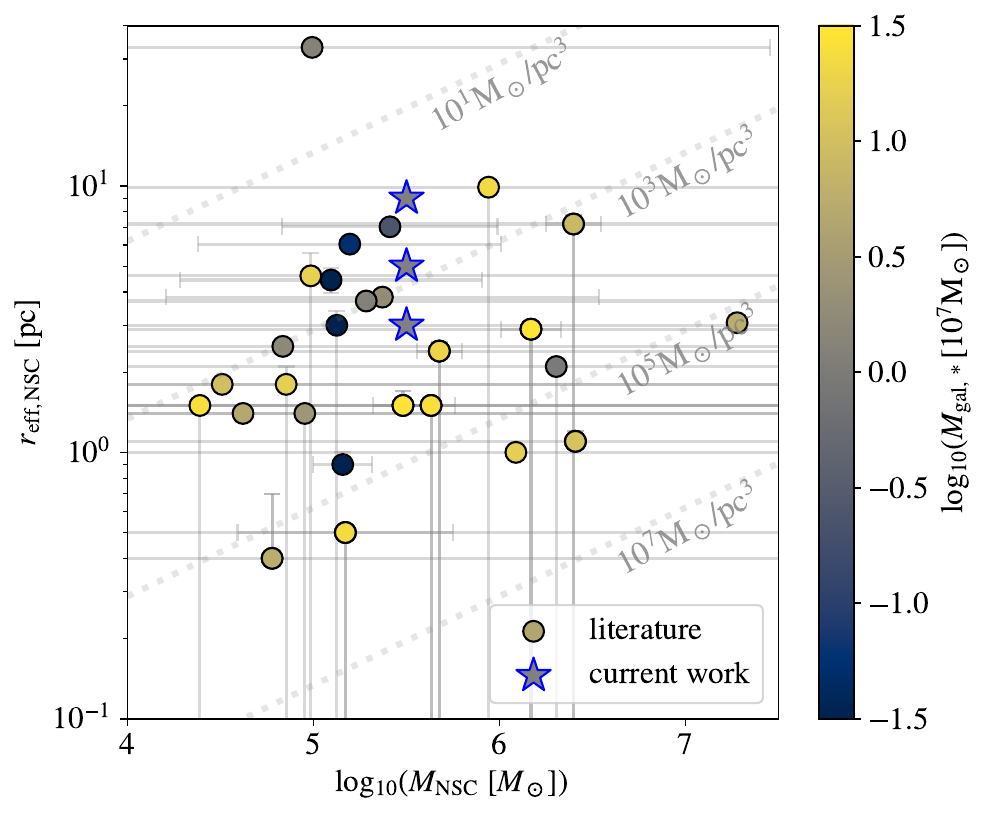}
    \caption{Effective radius versus mass of NSCs in observed dwarf galaxies with a stellar mass of $10^{5.5}-10^{8.5}\msun$ from literature \citep[{\it circle symbols};][]{Georgiev+2016} and from our current simulation work ({\it star symbols}). The colour indicates the stellar mass of host galaxies of each NSC. Our simulations span the observed range of NSC masses and effective radii, for galaxies of comparable stellar mass. For more information, see Section~\ref{sec:IC}.}
    \label{fig:NSC-parameter}
\end{figure}

Lastly, we include NSCs hosting a BH at the centre of the galactic disc to examine their impact on BH accretion in dwarf galaxies. Fig.~\ref{fig:NSC-parameter} presents the observational parameters of NSCs in dwarf galaxies selected with stellar masses ranging from $10^{5.5} \msun$ to $10^{8.5} \msun$, a range encompassing the stellar mass of our disc in the ICs ($M_{\rm gal,*}=9.75 \times 10^6 \msun$) from \cite{Georgiev+2016}. 

The NSCs hosted by dwarf galaxies of our interest have a density range of $10^1-10^6 \msun~{\rm pc}^{-3}$. In this work, we test NSCs with effective radii of $3$, $5$, and $9$~pc with the same total mass of $3.16\times10^5 \msun$, corresponding to densities of $5.85\times10^3, 1.26\times10^3$, and  $2.16\times10^2~\msun~{\rm pc}^{-3}$ at their effective radius, respectively. The NSCs are modelled using the \cite{Hernquist1990} profile and consist of collisionless stellar particles with masses of $6 \msun$, representing ``old'' stars that interact only gravitationally and do not undergo any stellar evolution or feedback\footnote{We note that for the same total mass, the Hernquist profile has a lower density at $r<30$~pc compared to \cite{King+1966} profile, which is typically used to fit observed star clusters. However, this does not affect the conclusion of our study, as having an NSC with a high density would even further strengthen our findings (see Section~\ref{sec:bh-evolution}).}. We position the NSCs, each hosting a BH with a mass of $10^4 \msun$ at its centre {, consistent with the observational relation between the NSC mass and BH mass \citep[see e.g.][for a recent compilation]{Greene+2020}}, at the location of the minimum gravitational potential within the galactic disc. The gravitational softening length for NSC particles and BHs is $0.175$~pc.

\subsection{Gravity and gas physics}
The simulations in this paper use the {\sc Arepo} gravito-hydrodynamical simulation code \citep{Springel+2010, Pakmor+2016}. Gravity is calculated with a tree-based algorithm, while hydrodynamics is computed using a finite-volume, second-order Riemann solver on an unstructured moving mesh based on Voronoi tessellation with discrete mesh-generating points. The standard mesh refinement scheme follows a quasi-Lagrangian method, keeping the mean cell mass at a target mass, $20~\msun$.

Cooling and heating rates are computed using the {\sc Grackle}\footnote{\url{https://grackle.readthedocs.io/}} library \citep{Smith+2017}, which handles a non-equilibrium primordial chemistry network of six species, H\I, H\I\I, He\I, He\I\I, He\I\I\I, and electrons, tracking the abundances of these species. The metal cooling rate is provided by pre-computed look-up tables generated by the photo-ionization code {\sc Cloudy} \citep{Ferland+2013}. We also adopt the UV background heating model by \citet{Haardt+Madau2012}, with self-shielding from the UV background based on the \citet{Rahmati+2013} prescription as implemented in {\sc Grackle}.

\subsection{Star formation and feedback}
\label{sec:SFFB}

We include star formation and the stellar feedback linked to individually tracked massive stars using the scheme presented in \cite{Smith+2021}. We refer the reader to that work for a comprehensive description of the methods, along with accompanying numerical tests, but summarise the salient details here.
 
For each gas cell, we calculate the local Jeans mass as:
\begin{equation}
M_{\rm J}=\frac{\pi^{5/2}c_{\rm s}^3}{6{\rm G}^{3/2}\rho^{1/2}}\,,
\label{eq:1}
\end{equation}
where $c_{\rm s}$ is the sound speed, G is the gravitational constant, and $\rho$ is the gas density of the cell. Cells that satisfy $M_{\rm J}< N_{\rm J}m_{\rm cell}$ are assigned a non-zero star formation rate (SFR). We choose the Jeans number, $N_{\rm J}=8$, which is a free parameter for resolving the Jeans mass. Star forming cells are assigned a SFR
\begin{equation}
\dot{m}_{\rm star}=\epsilon_{\rm SF}\frac{m_\mathrm{cell}}{t_{\rm ff}}\,,
\label{eq:2}
\end{equation}
where $t_{\text{ff}} = \sqrt{{3\pi}/{32G\rho}}$ is the local free-fall time and $\epsilon_{\rm SF}=0.02$ is star formation efficiency, following observed efficiencies in dense gas \citep[e.g.,][]{Krumholz+Tan2007}. In addition to other timestep limiters (e.g. CFL or gravity timestep), we ensure that the gas consumption time is resolved by limiting the timestep of a cell to a maximum of $0.1m_\mathrm{cell}/\dot{m}_{\rm star}$. Star forming cells are stochastically converted to collisionless star particles based on their assigned SFR.

At high mass resolution, treating star particles as if they represent a single stellar population (SSP) that independently fully sample the stellar initial mass function (IMF) is a poor approximation. For example, \cite{Smith2021} demonstrated that such an approach cannot capture the small-scale spatio-temporal clustering of ionizing sources (rare but bright OB stars) which has a significant impact on the efficiency of photo-ionization feedback to regulate the star formation in the ISM. Therefore, when a star particle is formed, we explicitly populate it with individual stellar masses drawn from the IMF. All stellar feedback channels are then tied directly to the evolution of those individual stars\footnote{To reduce memory requirements, once the sampling for a particle is complete, we discard the record of stars with a mass $<5~\mathrm{M_\odot}$ since these are not relevant to any feedback channel included in this work.}. At the mass resolution adopted in this work, massive stars that drive feedback ($\gtrsim 8~\mathrm{M_\odot}$) are typically represented by individual star particles. The sampling method adopted is the ``adjusted target'' scheme from \cite{Smith2021}. This method accurately preserves both the shape and normalisation of the input IMF across multiple star particles (which is non-trivial but necessary to avoid biasing the stellar feedback budget) while not requiring cell-to-particle or particle-to-particle mass transfers which increases complexity. The scheme in general allows a discrepancy between the dynamical mass of a star particle and its subgrid stellar inventory, but ensures that this is minimized as much as possible at the individual particle level and that the total dynamical and inventory masses are consistent over the population of star particles. For more details, we refer the reader to \cite{Smith2021}. 

We adopt a \cite{Kroupa2001} IMF between $0.08-100~\mathrm{M_\odot}$ and interpolate lookup tables as a function of zero age main sequence (ZAMS) mass for the stars in order to obtain quantities needed for stellar feedback. For simplicity, we do not interpolate the metallicity of the stars, instead adopting a single table corresponding to a metallicity of $0.1~Z_\odot$ (which is the metallicity of the ISM in our initial conditions). This approximation is valid because none of the relevant stellar properties vary strongly enough with metallicity relative to the evolution of metallicity over the course of our simulations. We obtain stellar lifetimes from the PARSEC grid of stellar tracks \citep{Bressan+2012} and we omit the effects of binary evolution. FUV ($6-13.6$~eV) and ionizing photon luminosities ($>13.6$~eV) originate from the OSTAR2002 grid of stellar models \citep{Lanz+2003} as compiled by \cite{Emerick+2019}, assuming the ZAMS value throughout the stellar lifetime.

We include photoelectric heating of dust grains by a spatially and temporally varying interstellar radiation field (ISRF) sourced by the individual stars in the domain and pass this as an additional heating term to {\sc Grackle}. The energy density in the FUV band at the location of each gas cell is calculated efficiently by approximating the radiation transport as being composed of a local attenuation at the source and sink with optically thin transport elsewhere. This allows the summation of fluxes to be carried out as part of the gravity tree walk at negligible cost. The dust-to-gas ratio is set as a function of metallicity using the broken power law of \cite{Remy-Ruyer+2014}. H\I\I\, regions around photo-ionizing sources are modelled using an anisotropic Str\"omgren-type approximation. The balance between recombination rate and ionizing photon luminosity is calculated within 12 independent angular pixels around the source. This mitigates the mass-biasing error inherent to other similar methods that instead compute the recombination rate within a sphere. Overlapping H\I\I\, regions from multiple sources are consistently handled. The maximum size of an H\I\I\, region is capped at 50~pc\footnote{We choose $r_{\rm ion,max}=50$~pc based on the PHANGS-MUSE H\I\I\, region observations from \cite{Santoro+2022}. Also, see Appendix C in \cite{Smith+2021} for a discussion of this choice.}. If a cell is identified as belonging to an H\I\I\, region, it is heated to $10^4$~K and is forbidden from cooling below that temperature while it remains tagged as part of a region.

When a star reaches the end of its life, it ceases to contribute radiation. If it has a ZAMS mass in the range $8-35~\mathrm{M_\odot}$, it triggers a core-collapse SN. We obtain the progenitor-mass-dependent ejecta mass and metallicity from \cite{Chieffi+2004} and adopt a constant energy of $10^{51}~\mathrm{erg}$ for all SNe, which are modelled using the approach introduced in \cite{Smith+2018}. Namely, mass, metals, energy and momentum are injected into the gas cell containing the star particle and its immediate neighbours (those that share a face with the host cell). Feedback quantities are distributed amongst the cells in a manner that ensures an isotropic injection (which is non-trivial in a Lagrangian code). We adopt a mechanical feedback scheme which compensates for the missing momentum when the Sedov-Taylor phase of the SN remnant is unresolved. In practice, at the resolution used in this work, the majority of SNe are already natively well resolved. We do not include Type Ia SNe or stellar winds in this work, nor do we include walkaway/runaway OB stars.

\subsection{Refinement scheme}
\label{sec:refinement}
\subsubsection{CGM region: disc tracer-dependent refinement}
\label{sec:refinement-cgm}

As briefly mentioned in Section~\ref{sec:IC}, {\sc Arepo}'s standard refinement scheme enforces the mass of gas cells within a factor of 2 of the target mass, which is $20\msun$ in our simulation suite. However, applying this level of resolution across the entire CGM is numerically impractical. To address this, we implement a tracer-dependent refinement strategy that enforces high resolution within and around the disc and in the outflowing material. For all gas cells within the disc region, defined by a cylindrical radius of $7$~kpc and height of $10$~kpc, we assign a passive Lagrangian tracer field, $\tilde{m}_{\rm disc}$, which is set to the cell mass, $m_{\rm cell}$ at $t=0$. Refinement or de-refinement occurs when $\tilde{m}_{\rm disc}/m_{\rm cell} > 0.1$, meaning that only gas cells with a tracer field comprising more than ten percent of the cell mass are eligible to be split or merged to match the cell mass to the target mass.
\begin{figure*}
\centering
\includegraphics[width=\textwidth]
{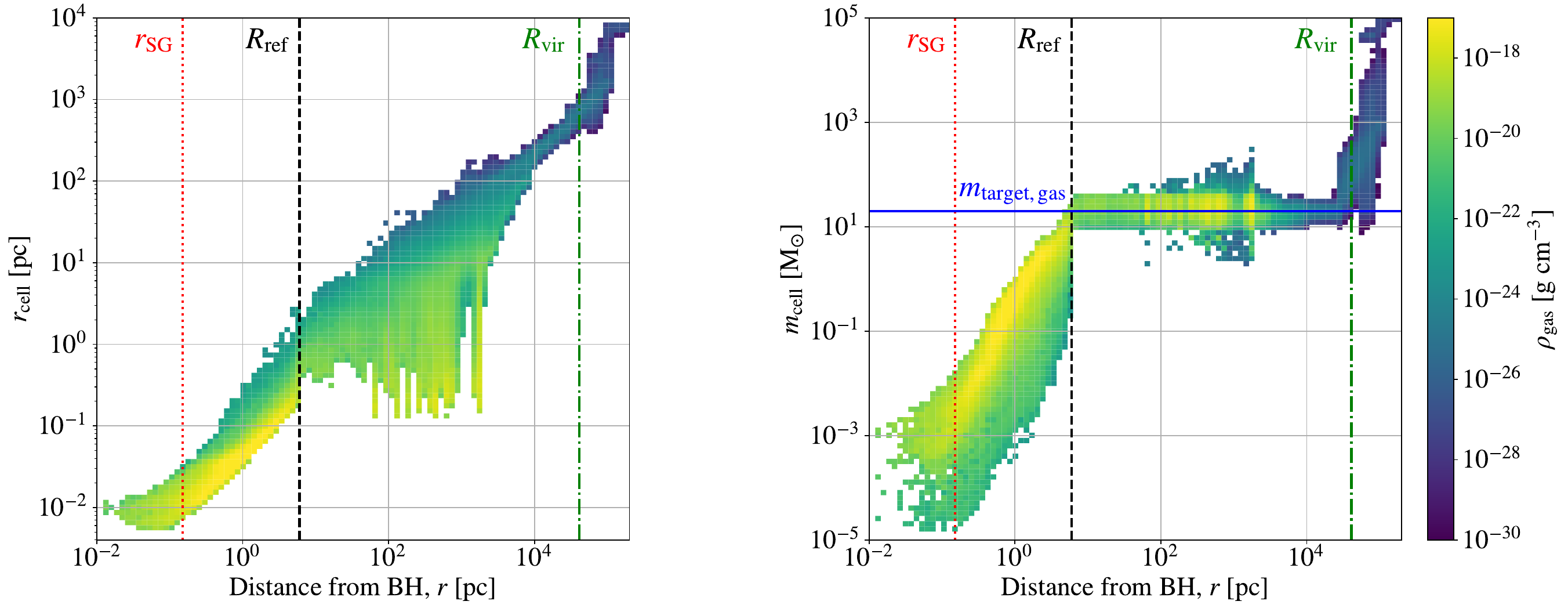}
    \caption{Distribution of spatial ({\it left}) and mass ({\it right}) resolution as a function of radial distance from the BH for the gas cells in {\tt NSC-5pc} run at $t=135$~Myr. Colours represent the cell density in each two-dimensional bin. The {\it red dashed} line indicates the typical self-gravity radius, $r_{\rm SG} \sim 0.15$~pc. The {\it black dotted} line marks the refinement radius, $R_{\rm ref} = 6$~pc. The {\it green dash-dotted} line denotes the virial radius, $r_{\rm vir} = 41$~kpc, and the {\it blue solid} line denotes the gas target mass for the refinement in the simulation. The innermost region within $r_{\rm SG}$ is very well resolved. For more information, see Section~\ref{sec:refinement}.}
    \label{fig:SL-pdf}
\end{figure*}
\begin{figure}
\includegraphics[width=0.5\textwidth]{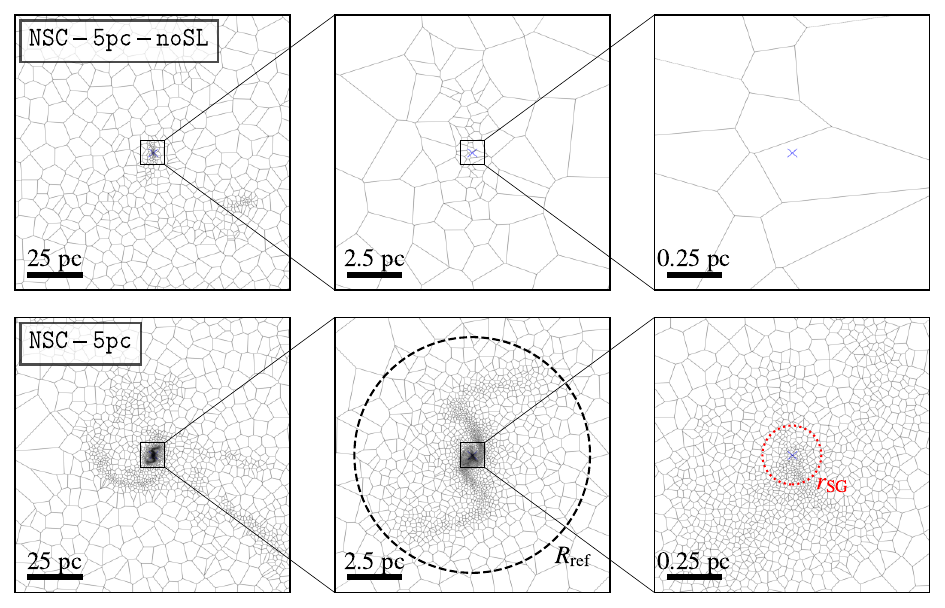}
    \caption{Slice of the Voronoi tessellation centred at the location of the BH, showing individual gas cells at $t=135$~Myr for {\tt NSC-5pc-noSL} and {\tt NSC-5pc} runs. The right panels are ten times zoomed-in images of the left panels. The {\it blue cross} mark indicates the location of the BH, and the {\it black-dashed} and {\it red-dotted} circles represent the refinement radius, $R_{\rm ref}= 6$~pc, and the self-gravity radius, $r_{\rm SG}\sim 0.15$~pc, respectively. For more information, see Section~\ref{sec:refinement-SL}.
    }
   \label{fig:SL}
\end{figure}
\subsubsection{Super-Lagrangian refinement around BHs}
\label{sec:refinement-SL}
To resolve the mass and angular momentum transport from ISM to the self-gravity radius of the accretion disc and the BH system, we use the super-Lagrangian refinement scheme proposed by \cite{Curtis+Sijacki2015}. The refinement scheme ensures that the radius of the cells in the refinement region decreases linearly with radius as they approach the BH.  {This method has been extensively tested for potential spurious numerical artefacts in \cite{Curtis+Sijacki2016, Bourne+2024, Koudmani+2024}, and shown to robustly resolve gas dynamics around the BH.}
Gas cells within a spherical region of radius $R_{\rm ref}=6~{\rm pc}$ are split or merged to keep their cell radius within the range $r_{\rm cell}$ within $R^{\rm (cell)}_{\rm max}(r)/C < r_{\rm cell} < R^{\rm (cell)}_{\rm max} (r)$, where $C=4$ is a refinement factor that controls the allowed range of cell sizes.
The maximum allowed cell radius, $R^{\rm (cell)}_{\rm max} (r)$, increases linearly from $R^{\rm (cell)}_{\rm max,0}=0.01~{\rm pc}$ at the BH location, sufficient to very well resolve the self-gravity radius of the accretion disc, $r_{\rm SG} \gtrsim 0.1$~pc, to $R^{\rm (cell)}_{\rm max,1}=0.8~{\rm pc}$ at $r=R_{\rm ref}$. With this setup, the minimum cell size near the BH becomes $R^{\rm (cell)}_{\rm max,0}/C=2.5\times10^{-3}~{\rm pc}$. To prevent the formation of extremely low-mass star particles in the refinement region, we explicitly prevent star formation in gas cells with masses less than $0.08\msun$.

Fig.~\ref{fig:SL-pdf} shows the distribution of gas cell sizes and masses as a function of the radial distance from the BH in the {\tt NSC-5pc} run at $t=135$~Myr, with the colour indicating the gas density. For $r> 10$~kpc, the size and mass resolution remain unchanged unless more than 10\% from the gas disc cell is transferred to the CGM cells. In the region $R_{\rm ref}<r< 10$~kpc, the refinement scheme maintains the mass resolution within a factor of 2 of $m_{\rm target, gas}=20\msun$, by construction. Note that the deviation in both size and mass of cells within the region $r \sim 6~{\rm pc} -1~{\rm kpc}$ is due to the formation of dense structures, such as gas clumps, and the star formation feedback in the galactic disc. At $r= r_{\rm SG}$, the cell size is less than $0.04$~pc and reaches down to $ \lesssim 0.01$~pc, while the cell mass ranges from $10^{-5}\msun < m_{\rm cell}<10^{-2}\msun$.

Fig.~\ref{fig:SL} shows Voronoi slice plots of gas cells near the BH, comparing runs with and without the super-Lagrangian refinement scheme. The presence of the NSC increases the gas density near the BH, reducing the gas cell size in both the {\tt NSC-5pc-noSL} and {\tt NSC-5pc} runs. In the two {\it left} panels, outside the refinement region ({\it black-dashed} circle), the cell size remains unaffected by the refinement scheme.  However, within the refinement region, as shown in the zoomed-in region in the {\it middle} and {\it right} panels, the cell size varies significantly depending on whether the super-Lagrangian refinement scheme is applied. Specifically, with super-Lagrangian refinement switched on, the self-gravity radius of the accretion disc, $r_{\rm SG}\sim0.15$~pc (marked by the {\it red-dotted} circle; see Fig.~\ref{fig:appendix-bh-evolution} for its evolution), is resolved by 2072 cells in this snapshot.  {The Voronoi mesh in our simulation remains regular and `round'-shaped at all spatial scales, seamlessly bridging regions with different resolutions, and exhibiting no visible artefacts at refinement boundaries.} By employing the super-Lagrangian refinement, our simulation captures hydrodynamical processes from the ISM down to the self-gravity radius of the accretion disc, allowing for a more accurate and self-consistent study of the co-evolution between the BH and the host galaxy.

\subsection{Black hole accretion model}
\label{sec:bh-model-method}
\subsubsection{$\alpha$-accretion disc and black hole spin model}
 Following \cite{Fiacconi+2018}, we assume a subgrid accretion disc that follows a steady-state \citet{Shakura+Sunyaev1973} solution and compute its evolution analytically on-the-fly. The model tracks both the mass and angular momentum evolution of the BH and its accretion disc. These two components are coupled via mass accretion and the \citet{Bardeen+Petterson1975} effect. The latter is relevant for a tilted accretion disc around a spinning BH, whereby Lense-Thirring precession causes the inner accretion disc to warp and viscous forces act to align the inner disc and the BH spin. Compared to the  \cite{Fiacconi+2018} model, we make some modifications in the treatment of mass and angular momentum inflow estimation.

In \cite{Fiacconi+2018}, the mass and angular momentum flux are estimated within a volume encompassing the 32 closest mesh-generating points, calculated with a kernel-weighted interpolation. In our simulation, which fully resolves the self-gravity radius $r_{\rm SG}$ of the accretion disc at all times, we instead directly measure the mass and angular momentum flow through the spherical surface at $r_{\rm SG}$. For mass flux estimation, we account for the gas cells satisfying $v_{\rm r} <0$ and $L < L_{\rm d}$, in a spherical shell of radius $0.5~r_{\rm SG} < r < 1.5~ r_{\rm SG}$, where $r$ is distance from the BH, $L_{\rm d}$ is specific angular momentum of accretion disc, $v_{\rm r}$ and $L$ are radial velocity and specific angular momentum of the gas cell, respectively. This approach ensures we only account for gas that is `inflowing' and can `circularize' onto the accretion disc. The mass flux, $\dot{M}_{\rm in}$ and specific angular momentum carried by the mass flux, ${L}_{\rm in}$, to the accretion disc are calculated as follows:
\begin{equation}
\dot{M}_{\rm in}=
\frac{\sum m_i v_{{\rm r}, i}}{\Delta x_{\rm shell}}\,,
\end{equation}
\begin{equation}
\vec{L}_{\rm in}=
\frac{\sum m_i v_{{\rm r}, i}\vec{L}_{i}}{\sum m_i v_{{\rm r}, i}}\,,
\end{equation}
where $m_i$, $v_{{\rm r},i}$, $L_i$ are the mass, relative radial velocity, and specific angular momentum of the $i$-th gas cell, respectively, and the summation is carried out for the inflowing and circularized gas cells within the shell with a width $\Delta x_{\rm shell} = r_{\rm SG}$. The inflowing specific angular momentum, $\vec{L}_{\rm in}$, is calculated as a weighted average of the angular momentum of the gas cells, where the weights correspond to each cell's contribution to the mass inflow. The resulting $\dot{M}_{\rm in}$ and $\vec{L}_{\rm in}$ are used to update the mass and angular momentum of the $\alpha$-disc,
\begin{equation}
 {\dot{M}_{\rm d}=\dot{M}_{\rm in}-\dot{M}_{\bullet,0},}
\end{equation}
\begin{equation}
 {\dot{\vec{J}}_{\rm d}=\dot{M}_{\rm in}{\vec{L}}_{\rm in}-\dot{\vec{J}}_{\bullet,0},}
\label{eq:6}
\end{equation}
 {where $\dot{M}_{\rm d}$, $\dot{\vec{J}}_{\rm d}$ are the mass and angular momentum inflow rate to the accretion disc, respectively\footnote{Here, the viscous torque resulting from the misalignment between the inflowing gas and the $\alpha$-disc is negligible (see Appendix~\ref{sec:appendix-torque-by-warped-disc}).}, and $\dot{M}_{\bullet,0}$, $\dot{\vec{J}}_{\bullet,0}$ are the mass and angular momentum inflows from the accretion disc to the BH, }which in turn drive the evolution of the BH following the model of \cite{Fiacconi+2018}.

The initial mass of the BH is set to 10$^4 \msun$ and the accretion disc is set to 10$^2 \msun$. The initial accretion rate is set assuming $f^{(\rm init)}_{\rm Edd} = 0.003$. The initial spin parameter of the BH is chosen as $a_{\bullet}=0.7$. This, in turn, allows the determination of the initial angular momentum of both the BH and the accretion disc, based on the masses of the BH, accretion disc, the BH spin parameter, and the initial Eddington fraction. We set the angular momenta of both the BH and the accretion disc to initially be aligned with the angular momentum of the galactic disc.

\begin{figure*}
\includegraphics[width=\textwidth]{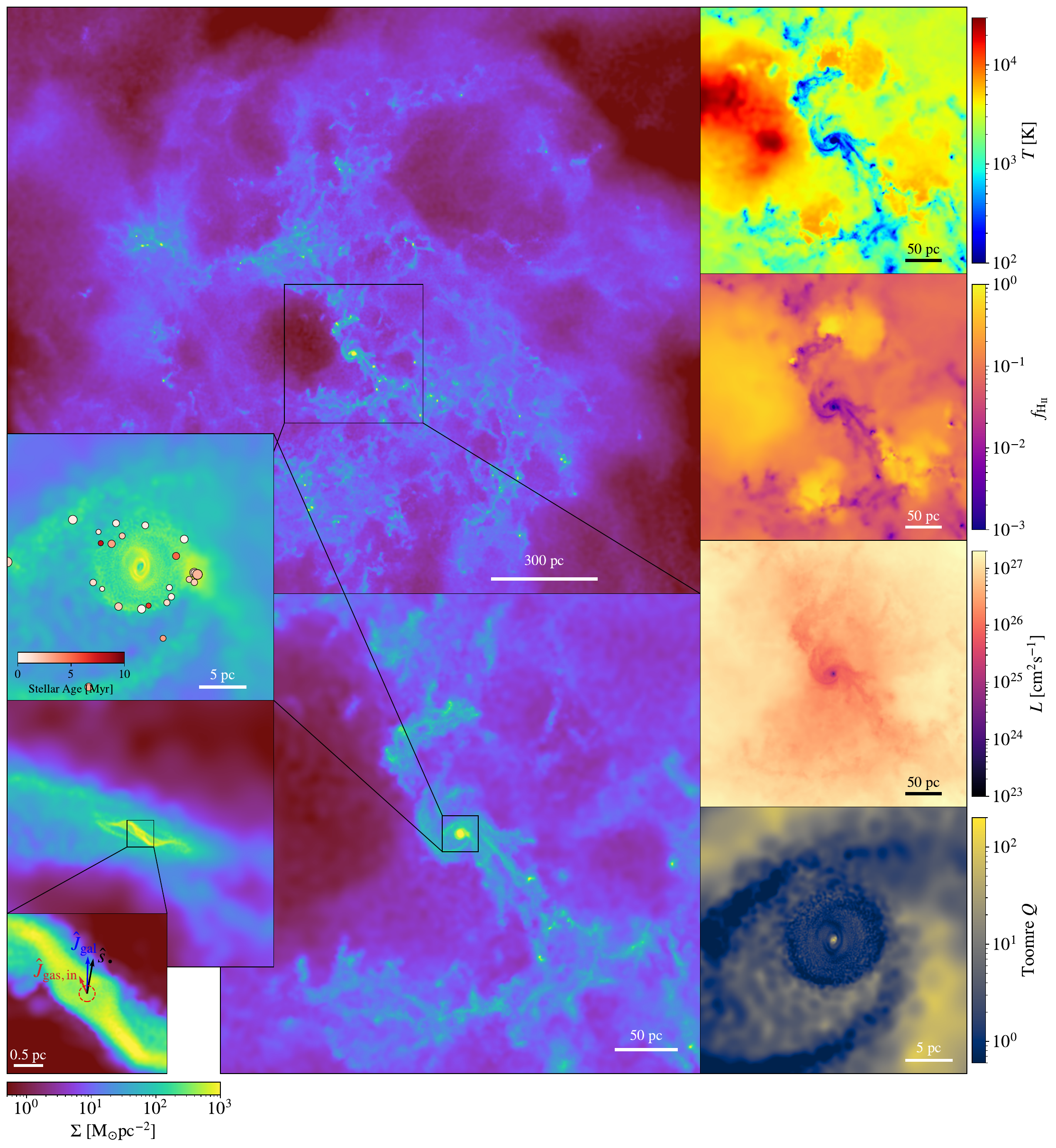}
    \caption{Visualisation of the {\tt NSC-5pc} run at {$t =135$~Myr}. {\it Left}: Gas surface density projections within zoom-in boxes of $2~{\rm kpc}, 400~{\rm pc}, 30~{\rm pc}$ (face-on and edge-on), and $3$~pc on a side. The {\it reddish} dots in one sub-panel represent the locations of young star particles (age $< 10$~Myr), with the dot size indicating the mass of star particles and the colour encoding their age. The stellar mass range is $6.68-37.9 \msun$ at this snapshot. In the bottom sub-panel the {\it red-dashed} circle presents the self-gravity radius of accretion disc, $r_{\rm SG} \sim 0.15$~pc while the {\it red} and {\it blue} arrows show the angular momentum versors of the gas within $r_{\rm SG}$ and galactic disc, respectively, while the {\it black} arrow indicates the BH spin direction. The BH spin evolves due to the gas inflows through the self-gravity radius of the accretion disc.
    {\it Right} (from the {\it top} to the {\it bottom}): maps of density-weighted gas temperature, fraction of H\I\I, density-weighted specific angular momentum, and the Toomre $Q$ parameter. Note that on all scales gas is multiphase, and it exhibits complex morphology and dynamics, with clumpy, filamentary structure and inflowing gas streams on different orbits leading to a warped CND in the centre. For more information, see Section~\ref{sec:overview}.}
    \label{fig:main}
\end{figure*}

\section{Results}
\label{sec3:results}
\subsection{Simulation overview}
\label{sec:overview}

Fig.~\ref{fig:main} presents maps of key gas properties from the {\tt NSC-5pc} run, which simultaneously resolves spatial scales from the CGM ($\approx 200$~kpc) down to the galactic nuclear self-gravitating region ($\sim0.15$~pc). The selected time, $t = 135$~Myr, captures the moment just before the SNe disrupt the nuclear disc, when the circumnuclear gas disc is at its most massive during the simulation. The left panels show the gas surface density of the large-scale galactic disc within a $3$~kpc on a side box, together with zoom-in boxes centred on the BH, with $400~{\rm pc}, 30~{\rm pc}$ (face-on and edge-on), and $3$~pc on a side. The right panels show density-weighted gas temperature, projected H$_{\rm II}$ fraction, $\Sigma_{\rm H_{\rm II}}/(\Sigma_{\rm H_{\rm I}}+\Sigma_{\rm H_{\rm II}})$, density-weighted specific angular momentum and Toomre $Q$ parameter. Coloured dots indicate individual young massive stars with masses above $5\,\msun$, which drive stellar feedback. Multiple stars may overlap at the same position within a single star particle.

The ISM exhibits a thermodynamically complex structure, comprising of cold, dense clumps and filaments, warm, more diffuse gas, and hot, diffuse `voids' shaped by recent stellar feedback. These `voids' vary in size, ranging from a few tens of parsecs to several hundred parsecs. Generally, the ionized gas appears diffuse and hot, while the neutral hydrogen regions trace the cold, dense gas. Regions where the projected ionized fraction is as high as 80\% are driven by recent episodes of massive star formation, followed by photo-ionization, which heats the ISM up to $10^4$~K, as shown in the temperature map. This early feedback mechanism regulates the formation of star clusters and subsequently SNe clustering \citep[e.g.,][]{Smith+2021}. Spatial regions where the temperature exceeds $10^4$~K are due to the recent SN explosions by massive stars. In these regions, remnants of earlier photo-ionization feedback, occurring prior to the SNe, can also be observed.

Gas angular momentum is significantly influenced by SNe, which drive turbulent flows and redistribute angular momentum in the ISM. The multiphase ISM feels the NSC gravitational potential, with orbit crossings generating shocks that efficiently dissipate energy \citep[e.g.,][]{Hopkins+Quataert2011}. Consequently, angular momentum decreases from $\sim 10^{27}~{\rm cm}^2{\rm s}^{-1}$ in the outer region (scales of a few $100$~pc) to $\sim 10^{25}~{\rm cm}^2{\rm s}^{-1}$ in the central region (on $\sim 10$~pc scales). This gas subsequently circularizes around the galactic nucleus, forming a cold, dense circumnuclear disc (CND) with a radius of approximately $7$~pc\footnote{Hereafter, we use the term ``CND'' to refer to the gas disc on the $r\sim0.2-7$~pc scales, distinct from ``accretion disc'', which refers to the subgrid model for the disc near the BH on the $r\lesssim0.2$~pc scale.}. Once a CND forms, the inflowing gas experiences torques from the CND gas, aligning its angular momentum with that of the CND. However, when a large amount of inflowing gas accretes onto the CND, the torque exerted by the CND becomes insufficient to alter the angular momentum of the inflowing gas. As a result, the inflowing gas rotates in a misaligned direction relative to the inner disc, leading to warping of the CND, as shown in the {\it bottom-left} edge-on gas surface density map.

The CND gas at a few parsecs from the BH loses its angular momentum primarily due to the gravitational torques exerted by the NSC and also the torques from the mini-spiral that forms at $r\sim1$~pc, given that the Toomre parameter is low. Consequently, the CND's angular momentum decreases from $10^{25}{\rm cm}^{2}{\rm s}^{-1}$ to below the specific angular momentum of the accretion disc, $L_{\rm d}\sim4\times 10^{23}{\rm cm}^{2}{\rm s}^{-1}$ at $r\sim 0.5$~pc within a few Myr. This circularized gas is then transferred to the accretion disc, altering the mass and angular momentum of the accretion disc and, subsequently, the BH. The {\it bottom-left} panel shows the angular momentum versor of the galactic disc ({\it blue} arrow), of the gas within the self-gravity radius ({\it red} arrow), as well as the spin direction of the BH ({\it black} arrow). Note that these three angular momenta are misaligned, indicating complex and dynamic gas inflows and torque interplay, which we observe throughout the entire galactic evolution (for further details see Section~\ref{sec:warping-disc}). 

The gas surface density of the CND is approximately $10^2-10^3~\msun~{\rm pc}^{-2}$. Along the edge of the CND, the disc becomes Toomre unstable, leading to the formation of tens and even hundreds of young stars orbiting a few pc away from the central BH. Some of these stars explode as SN and are able to (temporarily) halt further star formation. We note that the CND can withstand SNe rates of $0.3$~Myr$^{-1}$ within the $r<10$~pc region (tens of successive SNe during $25$~Myr; for further details see Section~\ref{sec:SFFB}), allowing the BH to continue accreting gas over $150$~Myr. However, the CND can be completely evaporated by a larger SN explosion rate, and it takes $\gtrsim 50$~Myr for a new CND to form from large-scale inflows (for further details see Section~\ref{sec:bh-evolution}).

 \begin{figure*}
\includegraphics[width=0.9\textwidth]{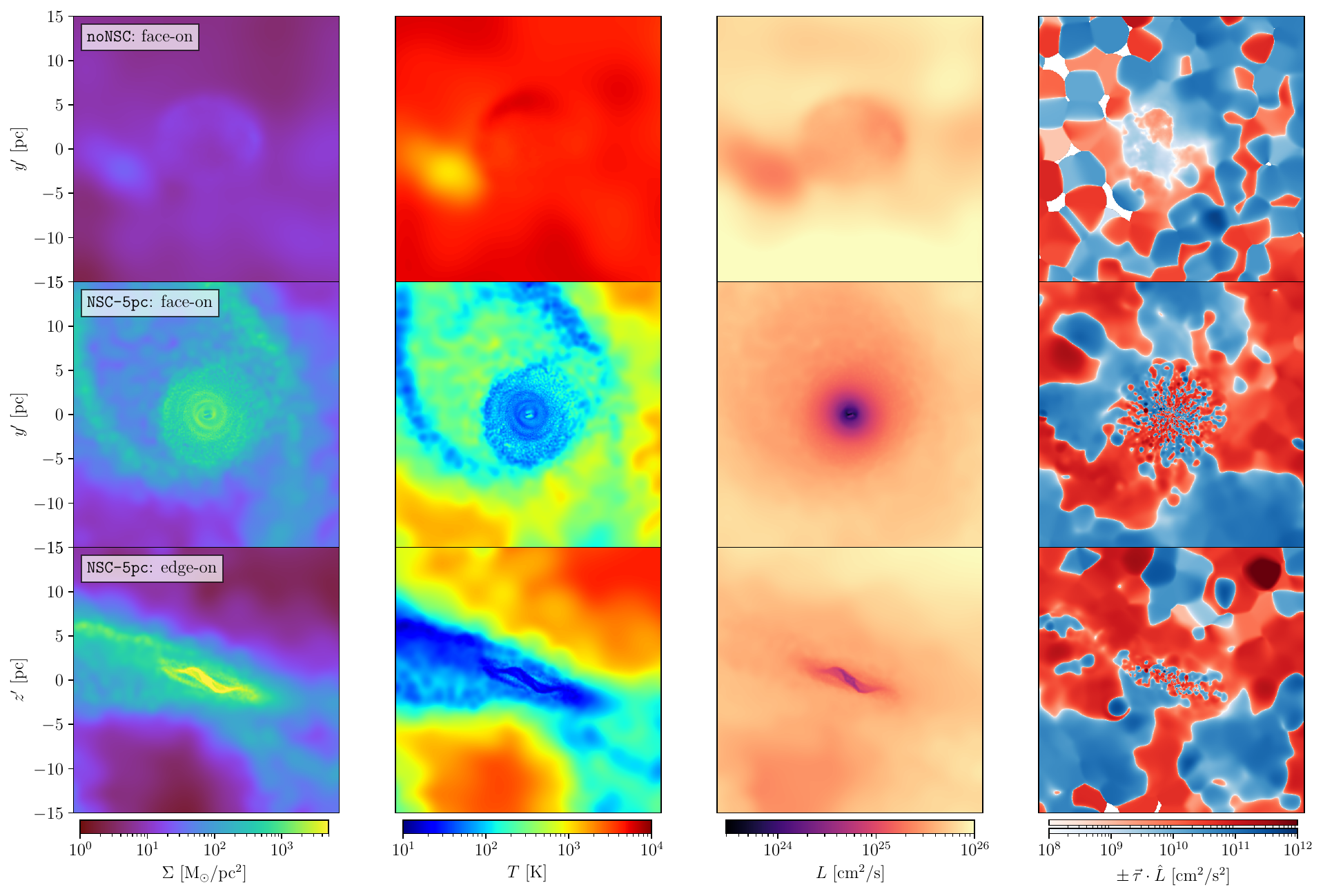}
    \caption{{\it Top two rows:} maps of gas surface density, temperature (density-weighted), specific angular momentum (density-weighted), specific torque (density-weighted) exerted on the gas near the BH at $t= 135$~Myr for {\tt noNSC} and {\tt NSC-5pc} runs. The projection boxes are aligned with the angular momentum vector of the gas and are centred on the BH. {\it Bottom row:} panels show the analogous edge-on maps for {\tt NSC-5pc} run. Central gas and torque properties are significantly influenced by the presence of the NSC. For more information, see Section~\ref{sec:NSCgas-properties}.}
    \label{fig:nsc-projection}
\end{figure*}

\begin{figure*}
\includegraphics[width=0.9\textwidth]{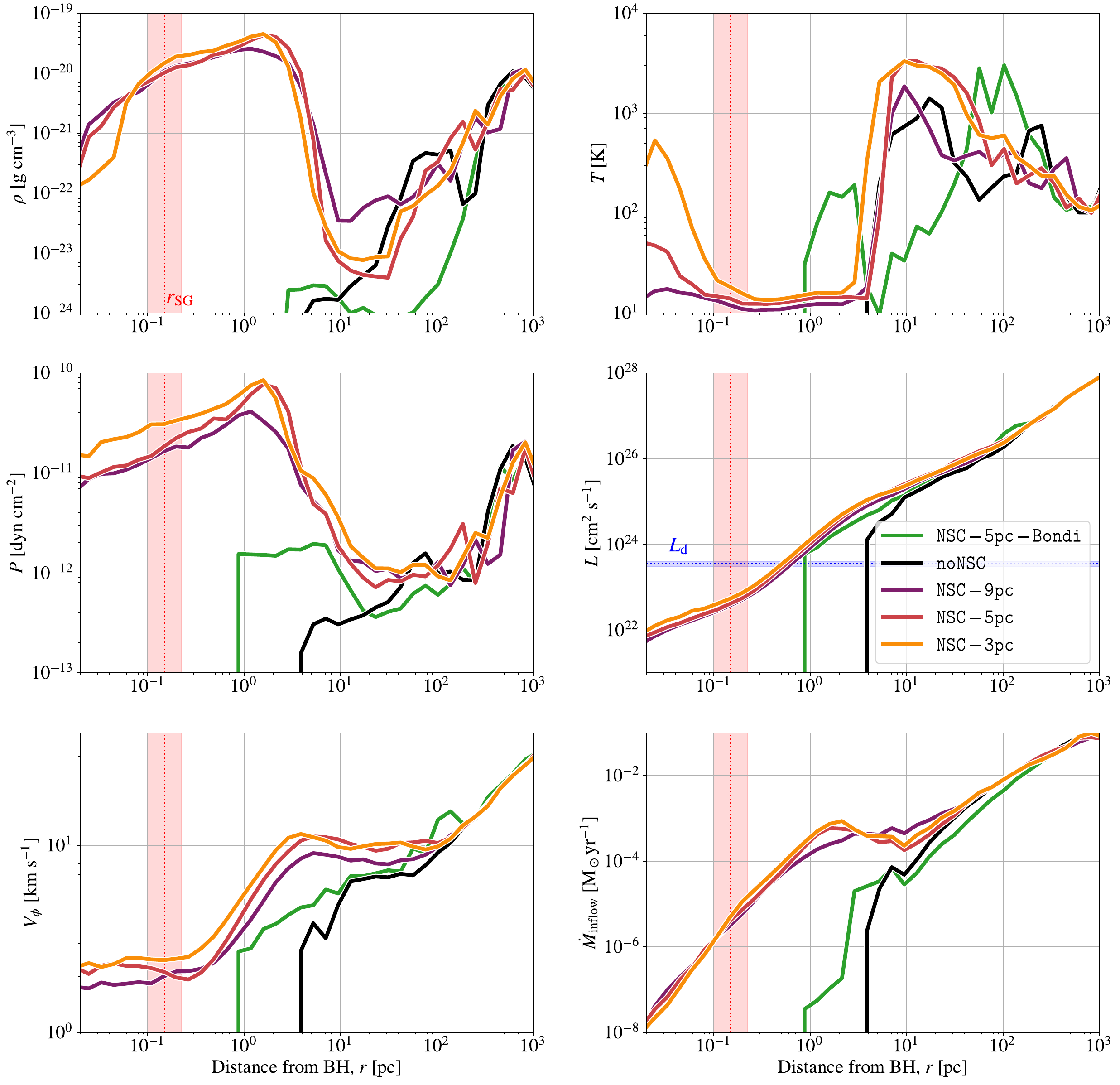}
    \caption{Radial profiles of gas density (density-weighted), temperature (density-weighted), pressure (density-weighted), specific angular momentum (density-weighted), azimuthal velocity (density-weighted) and mass inflow rate. The lines represent the median values during the first $135$~Myr. {\it Red {dotted}} line and {\it red shaded} region indicate the location of $\alpha$-disc self-gravity radius, $r_{\rm SG}$, while the {\it blue dotted} line and {\it blue shaded} region denote the value of the specific angular momentum of accretion disc, $L_{\rm d}$. Mass and angular momentum flux that ultimately may reach the BH depend not only crucially on the numerical resolution, but also on the detailed properties of gas and stars in the innermost region, including the presence of the NSC, as well as on the choice of accretion prescription onto the BH. For more information, see Section~\ref{sec:NSCgas-properties}.}
    \label{fig:nsc-profile}
\end{figure*}

\subsection{The impact of the NSC on the circumnuclear gas}
\label{sec:NSCgas-properties}
In this section, we explore how the presence of an NSC influences the thermodynamic properties of the gas and consequently the mass inflow rate onto the central BH. Fig.~\ref{fig:nsc-projection} shows maps of gas surface density, density-weighted temperature, density-weighted specific angular momentum and density-weighted torque within the central $30$~pc region at $t= 135$~Myr for {\tt noNSC} ({\it top} row; face-on) and {\tt NSC-5pc} ({\it bottom} two rows; face-on and edge-on) runs. We calculate the specific torque on gas cells by directly summing the individual particle specific torques acting on the $i$-th gas cell:
\begin{equation}
{\vec\tau_i}=\sum_{j=1} \vec{r_i}\times\frac{G M_j (\vec{r_j}-\vec{r_i})}{|\vec{r_j}-\vec{r_i}|^3}\,,
\end{equation}
where $j$ runs over all particles within $500$~pc from the BH, $M_j$ is the mass of $j$-th particle, and $\vec{r_i}$ is the position vector of the $i$-th gas particle from BH. This approach allows us to evaluate the influence of the torque exerted by each component (gas, dark matter, and stellar particles for the disc, NSC, and newly formed stars).

In the {\tt noNSC} run, a gas clump with surface density of a few $10 \msun{\rm pc}^{-2}$, is captured by the BH potential. This gas has a temperature of a few times $10^3$ K and a specific angular momentum greater than $10^{25}{\rm cm}^2{\rm s}^{-1}$, i.e., too high to be able to circularize on the $\alpha$-disc. We further note that the gas fails to form any steady structure in the vicinity of the BH in the absence of the NSC. In contrast, the presence of the NSC dramatically alters the gas dynamics. Specifically, in the {\tt NSC-5pc} run, several cold, dense streams lose their angular momentum and circularize onto the nuclear region, resulting in the formation of a thin gaseous disc with a radius of approximately $7$~pc around the BH. Within this disc, the gas surface density reaches very high values of $\gtrsim 500~\msun~{\rm pc}^{-2}$, and the gas is very cold with a temperature of $\lesssim 100$~K. As discussed in the Section~\ref{sec:overview}, when the angular momentum of the larger-scale accreting streams differs from that of the CND -- an occurrence that happens frequently -- the CND becomes warped. In the inner region, complex structures such as multiple mini-spiral arms and mini-bar form at $r \lesssim 3$~pc with the gas in the mini-bar having a particularly low angular momentum. These structures facilitate the  dissipation of the gas angular momentum  via their torques \citep[see e.g.,][]{Shlosman+1989, Hopkins+Quataert2010}. 

To analyse the cause of the gas angular momentum transport, the rightmost panels of Fig.~\ref{fig:nsc-projection} show maps of the density-weighted specific torque acting along the gas angular momentum ($\vec\tau\cdot\hat{L}$) near the BH at $t = 135$~Myr for the {\tt noNSC} and {\tt NSC-5pc} runs. The torque includes contribution from all mass components. In {\tt noNSC}, the torque direction is highly random at $ r\gtrsim5$~pc, and it is mainly exerted by dark matter and gas. In the vicinity of the BH, the torque magnitude is significantly smaller compared to that in {\tt NSC-5pc}. As discussed, the NSC in {\tt NSC-5pc} creates a denser, more structured gas distribution near the BH, resulting in a stronger and more coherent torque. For gas within 5~pc, where the NSC density distribution is most dominant, high-density gas gravitationally interacts with NSC stars, resulting in large-amplitude but finely structured torques within the CND. At an instantaneous level, the spatial regions with $\vec\tau\cdot\hat{L} < 0$ and $\vec\tau\cdot\hat{L} > 0$ are roughly comparable. We discuss total torques and their time evolution in Section~\ref{sec:torque}, where the net trends are revealed.

Fig.~\ref{fig:nsc-profile} shows the radial profiles of key gas properties near the BH across different simulation models. The {\it thick-solid} lines represent time-averaged profiles over the first 135~Myr (with a time step of 1~Myr), spanning from the formation of the CND to its disruption by stellar feedback from in-situ star formation.

First, we focus on the influence of the NSC while keeping the BH accretion model fixed to the '$\alpha$-disc' prescription. All runs exhibit a similar density profile in the $r\gtrsim30$~pc region, but they diverge at smaller radii, where the NSC potential becomes significant. In the absence of the NSC, the time-averaged density profile decreases monotonically toward the BH and drops to negligible levels at $r\lesssim 4$~pc, suggesting a lack of sustained gas inflow. Conversely, in the presence of the NSC, the density rises steeply toward the BH at $r = 10$~pc, marking the onset of the CND. When inflowing gas streams fed from the galactic disc are present, the density outside of the CND ($10~{\rm pc} < r < 100~{\rm pc}$) reaches levels comparable to the outer density of the CND (see Fig.~\ref{fig:nsc-projection}).  However, for most of the time, such inflowing streams are suppressed by intense stellar feedback from young stars near the CND ($r\lesssim 50$~pc), preventing the sustained build-up of dense gas in this region. The density profile peaks at $\gtrsim10^{-20}~{\rm g}~{\rm cm}^{-3}$ at $r = 3$~pc, marking the densest region in the ISM, and slightly decreases as it approaches the BH within the range $r_{\rm SG} < r < 3$~pc. The high-density nature of the CND enables it to withstand SN explosions occurring in the circumnuclear region. At $r < r_{\rm SG}$, the density drops to a few times $10^{-22} {\rm g}~{\rm cm}^{-3}$ as the BH accretes gas from its immediate surroundings. We find no strong trends between runs with different NSC effective radii and the size of the CND that forms; however, more compact NSCs correspond to more diffuse gas in the vicinity of the BH. 

\begin{figure*}
\includegraphics[width=\textwidth]{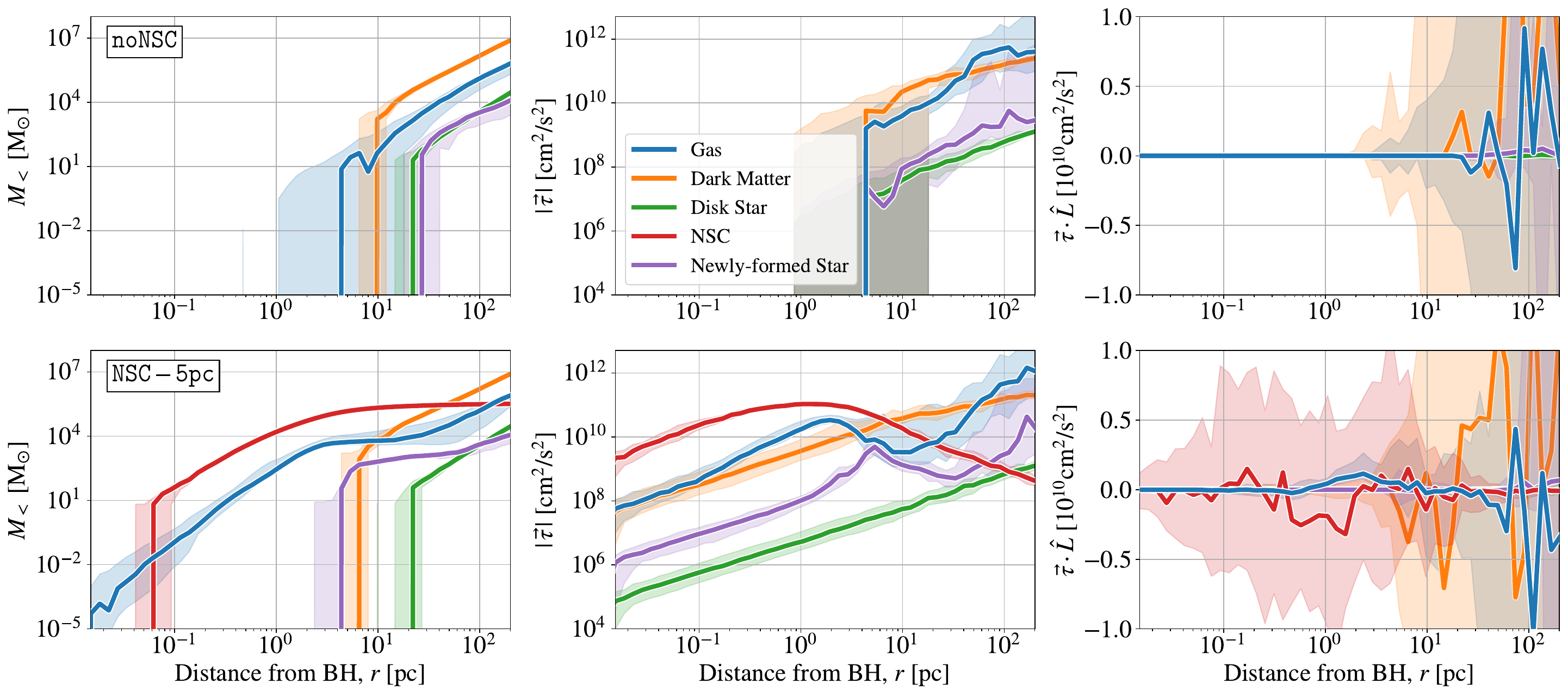}
    \caption{Radial profiles of enclosed mass ({\it left}), the magnitude of specific torque (density-weighted; {\it middle}) and specific torque acting along gas angular momentum, $\vec\tau\cdot\hat{L}$ (density-weighted; {\it right}), exerted on gas by different components. The top and bottom rows show results for the {\tt noNSC} and {\tt NSC-5pc} simulation, respectively. The {\it solid} lines represent the median values during the first 135 Myr, and {\it shaded} region denotes the 16th and 84th percentiles. An NSC is crucial in bringing the gas towards the innermost regions and torquing it down. For more information, see Section~\ref{sec:torque}.}
    \label{fig:torque-prof}
\end{figure*}

In the outer galactic disc ($r>30$~pc), the gas temperature fluctuates between $10^2$ and $10^3$~K due to the formation of cold gas clumps or streams, subsequent star formation, and the heating induced by stellar feedback. In the presence of the NSC, due to the high-density CND that forms, the temperature drops significantly from $\sim10^3$~K to 20~K at the edge of the CND. It then plateaus before rising again as it approaches the self-gravity radius $r_{\rm SG}$. We find a systematic correlation between NSC compactness and CND temperature: a denser NSC corresponds to slightly hotter (and more pressurized) CNDs, with this trend becoming more pronounced within the self-gravity radius. This is attributed to the deeper gravitational potential of a denser NSC, which enhances both the kinetic and thermal energy of the gas (see Fig.~\ref{fig:energy}). The pressure in the galactic disc rises from $10$~pc to $1$~pc, peaking at $\gtrsim8\times10^{-11}~{\rm dyn}~{\rm cm}^{-2}$ for {\tt NSC-3pc} and then decreases in the centre most region as the gas circularizes onto the $\alpha$-disc. 

The specific angular momentum of the gas increases monotonically with $r$. As shown in the Fig.~\ref{fig:nsc-projection}, the specific angular momentum reaches $\sim 10^{25}{\rm cm}^2{\rm s}^{-1}$ at $r=5$~pc, primarily due to the shocks induced by orbit crossings in the presence of the NSC. In the region $r_{\rm SG} < r < 5$~pc, where the CND is located, the slope of the angular momentum profiles becomes slightly steeper, decreasing to $\sim 10^{23}{\rm cm}^2{\rm s}^{-1}$ at $r=r_{\rm SG}$. This value is lower than the specific angular momentum of the $\alpha$ accretion disc, $L_{\rm d} \sim 3\times10^{23}~{\rm cm}^2~{\rm s}^{-1}$ (see Appendix~\ref{sec:appendix-alpha-disc} for the evolution of the accretion disc properties), allowing the gas to effectively circularize onto the accretion disc. The time-averaged profiles of the {\tt noNSC} simulation exhibit similar trends to the runs with an NSC in the outer galactic region; however, they show a clear drop at $r\sim 1$~pc, indicating that negligible gas reaches the BH. 

The {\it bottom-left} panel of Fig.~\ref{fig:nsc-profile} shows the density-weighted gas azimuthal velocity profiles, defined as $V_{\phi}=|\vec{v}-(\vec{v}\cdot\hat{r})\hat{r}|$, for each simulation. The azimuthal velocity is nearly constant at $\sim 2~{\rm km}~{\rm s}^{-1}$ within the self-gravity radius, and further-out it increases with radius, peaking at $8 - 10~{\rm km}~{\rm s}^{-1}$ at $r\sim4$~pc in the presence of the NSC. We note that the entire galactic disc is supersonic, particularly in the CND, where the low temperature ($\lesssim 100$~K) results in a Mach number reaching $\sim 20$ in the presence of an NSC. As expected, a denser NSC leads to higher azimuthal velocities due to the larger enclosed mass, especially within $r < 3$~pc. The radial velocity is of the order of $\sim0.1-1$ km~s$^{-1}$, which is 10-100 times lower than the azimuthal velocity, confirming that the gas is largely rotationally supported. This well-defined gas rotation supports the continuous growth of the accretion disc angular momentum, a topic that will be discussed in Section~\ref{sec:bh-evolution}.

Additionally, to verify that our simulated galaxies exhibit realistic stellar kinematics, within twice the stellar half-mass radius we compute the stellar rotational velocity, $v_{\rm rot,\star}=6.58~{\rm km}~{\rm s}^{-1}$, the velocity dispersion, $\sigma_{v,\star}=14.2 ~{\rm km}~{\rm s}^{-1}$, and the ratio of rotational velocity to velocity dispersion, $v_{\rm rot,\star}/\sigma_{v,\star} \sim 0.46$, which all align well with observations of similarly massive dwarfs as reported by  {\citep[][see also Appendix~\ref{sec:appendix-wlm-rotation} where we directly compare the rotation curve of our simulated dwarf with observational constraints on the WLM dwarf galaxy]{Wheeler+2017}}.

The {\it bottom-right} panel of Fig.~\ref{fig:nsc-profile} shows the mass inflow rate profiles for our simulation suite. To compute the mass flow rate, we measure the net mass flux of gas across concentric spherical shells. The inflow rate increases with radius in the region $10~{\rm pc}<r<1~{\rm kpc}$. In the region $r\sim 2-10~{\rm pc}$, where the CND is located, the inflow rate is higher further in and exhibits a peak at the location of the density maximum. The position and amplitude of this inflow peak have a slight dependence on the NSC density: a denser NSC leads to a higher inflow rate and shifts the peak closer to the BH. This enhanced matter transport inwards at $r\sim2$~pc is driven by the torques exerted by the NSC within the CND, a topic that will be discussed in Section~\ref{sec:torque}. At distances around $r_{\rm SG}$, the average inflow rate is $\lesssim10^{-5} \msun{\rm yr}^{-1}$\footnote{We note here that the inflow rate continues to decrease within $r_{\rm SG}$ as we approach innermost regions. This implies that the inflow rate is likely to depend sensitively on the choice of the spatial region within which it is computed. Here we adopt a numerically robust approach by ensuring that the $r_{\rm SG}$ is well resolved, which may, however, provide an upper bound on the accretion rate onto the $\alpha$-disc. Even higher resolution simulations accurately resolving gas and stellar dynamics and their respective torques will be needed in future to settle this question, which is something we plan to address in our future work.}.

Finally, we compare the $\alpha$-disc models to the Bondi-Hoyle-Lyttleton (Bondi) model, with the BH accretion rate modelled as 
$\dot{M}_{\rm Bondi} = {4 \pi \rho\, {\rm G}^2 M_{\bullet}^2}/{c_{\rm s}^3}$, where $\rho$ and $c_{\rm s}$ represent the density and the sound speed of the gas in the vicinity of the BH, respectively. The model is subject to the same super-Lagrangian refinement conditions, and contains the same NSC as the {\tt NSC-5pc} run. In Fig.~\ref{fig:nsc-profile}, {\it green} lines present the results of the {\tt NSC-5pc-Bondi} run. The Bondi radius is defined as $r_{\rm Bondi}=2{\rm G}M_\bullet/c_{\rm s}^2$. This radius in our simulation is $r_{\rm Bondi}=1-200$~pc and is very well resolved. The time-averaged gas density in the vicinity of the BH is significantly lower than in our fiducial model, particularly within $r<200$~pc, where Bondi accretion occurs. We reiterate that the CND is formed by the NSC potential; however, the CND mass is lower than in the fiducial model due to gas depletion by episodic and excessive BH accretion, which by definition does not take into account gas angular momentum. This makes gas more vulnerable to disruption by stellar feedback in the vicinity of the BH, leading to large fluctuations in the BH accretion rate, which we discuss in Section~\ref{sec:bondi}. Notably, the temperature of the CND is high, reaching $\gtrsim10^2$ K. The elevated temperature of the CND can be explained by two factors. Due to the low density, cooling is inefficient. In addition, under these low-density conditions, the gas maintains a high temperature in order to sustain pressure equilibrium near the BH. Ultimately, this comparison highlights both the importance of numerical resolution and the accurate modelling of accretion processes to understand how BHs acquire their mass.

\begin{figure*}
\centering
\includegraphics[width=\textwidth]{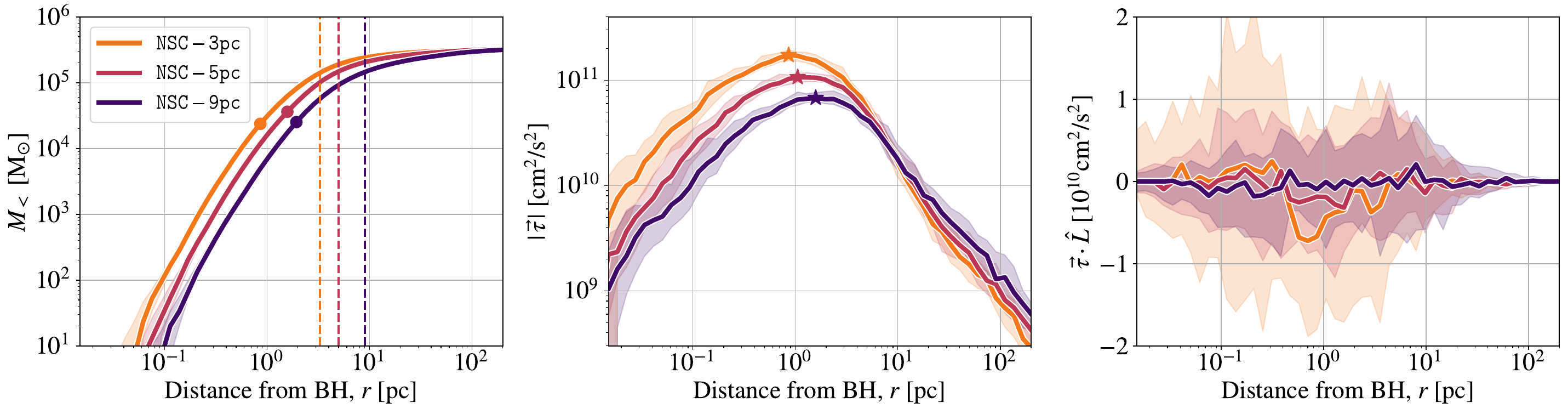}
    \caption{Radial profiles of enclosed mass ({\it left}), the magnitude of specific torque (density-weighted; {\it middle}), and specific torque acting along gas angular momentum, $\vec\tau\cdot\hat{L}$ (density-weighted; {\it right}), exerted on gas by the NSC stars.  The {\it solid} lines represent the median values during the first 135 Myr, and the {\it shaded} region denotes the 16th and 84th percentiles. The {\it dashed-vertical} lines in the {\it left} panel indicate the effective radius of each NSC. The {\it circle} dots indicate the location of maximum value of $|\nabla(M_<(r)/r)|$ in the {\it left} panel, and {\it star} dots indicate the maximum values of the specific torque in {\it middle} panel. For more information, see Section~\ref{sec:torque}. }
    \label{fig:torque-prof-nsc}
\end{figure*}

\subsubsection{Torque properties}
\label{sec:torque}
In this section, we examine the central mass distribution and quantitatively analyse the effects of gravitational torques exerted by each mass component on the gas. Left-hand panels of Fig.~\ref{fig:torque-prof} show the time-averaged radial profiles of enclosed mass, which is split into gas, newly formed stars and stars in the galactic disc, NSC (where present) and dark matter. For $r > 20$~pc, where the influence of the NSC is less significant, the mass distributions of each component are similar between {\tt noNSC} and {\tt NSC-5pc} runs, with dark matter and gas being the dominant contributors to the total mass. However, for $r < 20$~pc, a distinct difference in mass distribution is observed between {\tt noNSC} and {\tt NSC-5pc}, especially for the gas and newly formed stars. In the absence of the NSC, the gas distribution significantly fluctuates compared to {\tt NSC-5pc}, and there is a lack of central gas reservoir, as gas inflows are highly episodic. In the presence of the NSC, its gravitational potential pulls $\sim 6.1\times10^3 \,\mathrm{M_\odot}$ of gas within $r<6$~pc, with the gas distribution extending down to $\lesssim 0.01$~pc. Moreover, thanks to the presence of the NSC, in this central gas-rich region a large number of stars form, preferentially between $2$ and $10$~pc, a topic that will be discussed in Section~\ref{sec:SF-in-CND}.

The middle panels of Fig.~\ref{fig:torque-prof} show the magnitude of the density-weighted specific torque ($\tau$) exerted on the gas by different components, as labelled in the legend. Here, we compute the magnitude of the specific torque for each gas cell and then take the density-weighted average within radial bins. Clearly, for the {\tt noNSC} run, the specific torque exerted on gas can only become apparent at the onset of the gas profile at $r>1$~pc, where the gas distribution starts. The dominant contribution to the torque comes from dark matter and gas self-torque throughout the entire region. As observed in Fig.~\ref{fig:nsc-projection}, the torque magnitude decreases as it approaches the BH, reaching $\lesssim10^{10}$cm$^2$s$^{-2}$ at $r=5$~pc. In the {\tt NSC-5pc} simulation, the most significant contribution to the specific torque comes from the NSC component at $r<8$~pc, which is $5-20$ times greater than the contributions from either gas or dark matter within the region. It reaches a magnitude of 10$^{11}$cm$^2$s$^{-2}$ at $r\sim1$~pc and decreases outwards as the influence of the NSC potential diminishes with increasing radius. Among the mass components in the CND region, the self-torque exerted by gas is the second most dominant. It reaches $\sim3\times10^{10}$cm$^2$s$^{-2}$ at $r\sim2$~pc, where the gas density is the highest and where we observe significant CND warping. We further note, as evidenced in the middle row of Fig.~\ref{fig:nsc-projection}, that the gas in the CND disc has lopsided $m=1$-like structures and an appearance of a mini-bar-like structure within the central pc. The torque exerted by newly formed stars increases and peaks at $6$~pc, near the edge of the CND; however, its contribution remains significantly lower than that of other mass components.

Right-hand panels of Fig.~\ref{fig:torque-prof} show the radial profiles of density-weighted specific torque exerted by each component, but now acting along the gas angular momentum versor ($\vec\tau\cdot\hat{L}$), indicating how the angular momentum of gas is influenced by gravitational torque. In the {\tt noNSC} run, the dark matter and gas components exert a highly fluctuating torque on the gas, suggesting that, in the absence of an NSC, the gas both loses and gains angular momentum through self-gravity and gravitational interactions with the dark matter halo. In the {\tt NSC-5pc} simulation, central $\vec\tau\cdot\hat{L}$ behaviour is more complex. In the region $\gtrsim 1$ to $7$~pc, the time-averaged torque exerted by the gas itself (indicated by the {\it blue} line) is positive, contributing to an increase of the gas angular momentum. Once the CND has formed, misaligned inflows of gas, which have higher gas angular momentum, stream onto and lead to warping of the CND and positive torques. Meanwhile, the net torque exerted by the NSC preferentially acts in the opposite direction to the angular momentum of the inflowing gas and is overall stronger. 

 {This negative torque arises from the dynamical interaction between the rotating, axisymmetric CND and the initially dispersion-supported, spherical NSC. As the two components interact gravitationally, the CND exerts a positive torque on the NSC, transferring angular momentum to it. In return, the wake that the CND generates in the NSC exerts a negative torque on the CND. This mutual angular momentum exchange is analogous to dynamical friction between a rotating disc (or bar) and a spherical dark matter (or stellar) system \citep[see, e.g., the seminal papers by][ {where in our case the CND plays the role of the disc/bar, while the NSC corresponds to the dark matter halo}]{Sellwood+1980, Weinberg+1985, Tremaine+1999}. We emphasize that the axisymmetric, flattened CND induces a wake and a strong quadrupolar torque structure onto the spherically-symmetric NSC, enabling the angular momentum transfer from the CND to the NSC, such that the CND can lose sufficient angular momentum to circularize onto the $\alpha$-disc (see Appendix~\ref{sec:appendix-torque-CND} for more details and a visual of the torque map). Furthermore,}
 we note that the self-torque exerted by gas decreases for $r\lesssim 1.5$~pc and the torque actually becomes negative for $0.4\lesssim r\lesssim 0.8 $~pc, where the mini-bar is located. However, its amplitude remains significantly lower than that of the NSC.

Next, we explore how NSCs of different compactness affect the torque exerted on the gas. Fig.~\ref{fig:torque-prof-nsc} shows the same quantities as Fig.~\ref{fig:torque-prof} but focusing only on the gas-NSC interaction for our 3 different NSC models. We find that the NSC radial mass profiles are very stable and largely remain unchanged over $135$~Myr. As shown in the {\it middle} panel of Fig.~\ref{fig:torque-prof-nsc}, a denser NSC exerts a stronger specific torque, with the the torque peaking closer to the BH at $r\sim0.95, 1.2,$ and $ 1.7$~pc, for the {\tt NSC-3pc}, {\tt NSC-5pc}, and {\tt NSC-9pc} runs, respectively. The positions of maximum torque ({\it star} symbols) correspond well with the spatial location where $|\nabla (M_</r)|$ is maximal ({\it circle} symbols), as $M_</r$ reflects the gravitational potential, and thus, $\nabla (M_</r)$ represents the acceleration vector. 

As observed in Fig.~\ref{fig:nsc-profile}, at the edge of the CND ($r\sim7$~pc), the specific angular momentum of the gas is around $10^{25}{\rm cm}^2{\rm s}^{-1}$. Consequently, the characteristic timescale for the NSC torque to act is $\sim L/\tau\lesssim3$~Myr, which is consistent with our observation that the CND forms and gas inflows toward the BH within $\lesssim3$~Myr from the start of the simulations\footnote{Given the deep gravitational potential of the NSC, the free-fall velocity at $r = 6$~pc is $v_{\rm ff} = \sqrt{2{\rm G}M_</r} \sim16~{\rm km}~{\rm s}^{-1}$, which corresponds to a free-fall timescale of approximately $0.40$~Myr. However, the gas has a significant azimuthal velocity component, $V_{\phi} > 10~{\rm km}~{\rm s}^{-1}$ at $r = 6$~pc (see Fig.~\ref{fig:nsc-profile}), requiring angular momentum transport via torques in order to be accreted onto the BH. Therefore, the relevant timescale for circularization should be $L/\tau\lesssim 3$~Myr.}. A denser NSC exerts a stronger torque and promotes faster formation of the CND, which in turn influences the time evolution of the BH accretion, a topic that will be discussed in Section~\ref{sec:bh-evolution}. For $r>10$~pc, this trend reverses: a compact NSC exerts weaker torques, as its mass distribution becomes less significant farther away from the centre. As shown in the {\it right} panel, the NSC torque predominantly acts on the gas in the opposite direction of the gas angular momentum, reducing it, particularly in the region $r\sim 0.7-7$~pc. Within the innermost region ($r\lesssim0.7$~pc), the torque mostly fluctuates in sign, without clear trends. 

As mentioned earlier, the CND gas loses its angular momentum due to the torque exerted by the NSC, which, in turn, increases the angular momentum of the NSC. For example, in the {\tt NSC-3pc} simulation the specific angular momentum of the NSC within $r=6$~pc initially measures approximately $3.15\times10^{22}{\rm cm}^2{\rm s}^{-1}$ and increases to a median value of approximately $6.97\times10^{22}{\rm cm}^2{\rm s}^{-1}$ over 135~Myr.
The median value of the NSC's specific angular momentum increase depends on its density, with more compact NSCs leading to greater gas inflow, increasing by 125\%, 25.1\%, and 12.4\% for {\tt NSC-3pc}, {\tt NSC-5pc}, and {\tt NSC-9pc}, respectively. We measured $v_{\rm rot}/\sigma$ of our simulated NSCs to be $\sim0.0143-0.023$, which is small but still comparable to observed NSCs, ranging from 0.02 to 0.3 within their effective radii \citep[see Fig. 2, e.g., FCC177, FCC277 and FCC310 in][]{Lyubenova+Tsatsi2019}. We note that the specific angular momentum of the NSCs is two orders of magnitude smaller than that of the CND (see Fig. \ref{fig:nsc-profile}). Even if the CND were associated with a more rotation-supported NSC (say, with a tenfold higher $v_{\rm rot}/\sigma$), the specific angular momentum of the simulated NSC would still be much smaller than that of the CND, and it would still able to exert a strong enough negative torque on the CND.

Finally, to investigate how the CND rotation may induce NSC ellipticity, $\epsilon$, we compute the mass tensor of the NSCs for $r<6$~pc following \citep[see e.g.][]{Despali+2014},
\begin{equation}    
M_{\alpha\beta}=\frac{1}{N}\sum r_{i,\alpha} r_{i,\beta}, 
\end{equation} 
where $r_i$ is the position vector of $i$-th particle, and the summation is carried out for $r<r_{\rm eff}$. Diagonalizing the mass tensor yields eigenvalues $l_i$ ($i=1,2,3;~l_1\geq l_2\geq l_3$), from which the ellipticity is given by $e=({\lambda_1-\lambda_3})/{2\Lambda}$, where $\lambda_i=\sqrt{l_i}$, $\Lambda=\sum{\lambda_i}$. Our results show that the NSC ellipticity is strongly influenced by the CND rotation. For {\tt NSC-3pc}, ellipticity increases from $\epsilon=0.00206$ to $0.0103$, while for {\tt NSC-5pc} and {\tt NSC-9pc}, it increases from $0.00261$ to $0.00638$, and from $0.00259$ to $0.00378$, respectively. The observed ellipticity of NSCs with masses of $M_{\rm NSC}=10^5-10^6 \msun$ is reported to be approximately $0.02-0.4$ \citep[for e.g.,][]{Lyubenova+Tsatsi2019, Hoyer+2023}, which exceeds those in our simulations. This likely arises because our initial NSCs are highly isotropic and we are simulating a dwarf galaxy in isolation over a limited period of time. Nonetheless, our findings demonstrate that CND formation can significantly enhance the rotation and ellipticity of the host NSC. We further note that the correlation between NSC density and both angular momentum transport and orbital ellipticity aligns with the behaviour of $\vec{\tau} \cdot \hat{L}$. In the low-density {\tt NSC-9pc} case, although the torque is weaker, the gas has intrinsically lower angular momentum due to the shallower potential and lower kinetic energy, allowing for effective circularization.

\begin{figure*}
\centering
\includegraphics[width=\textwidth]{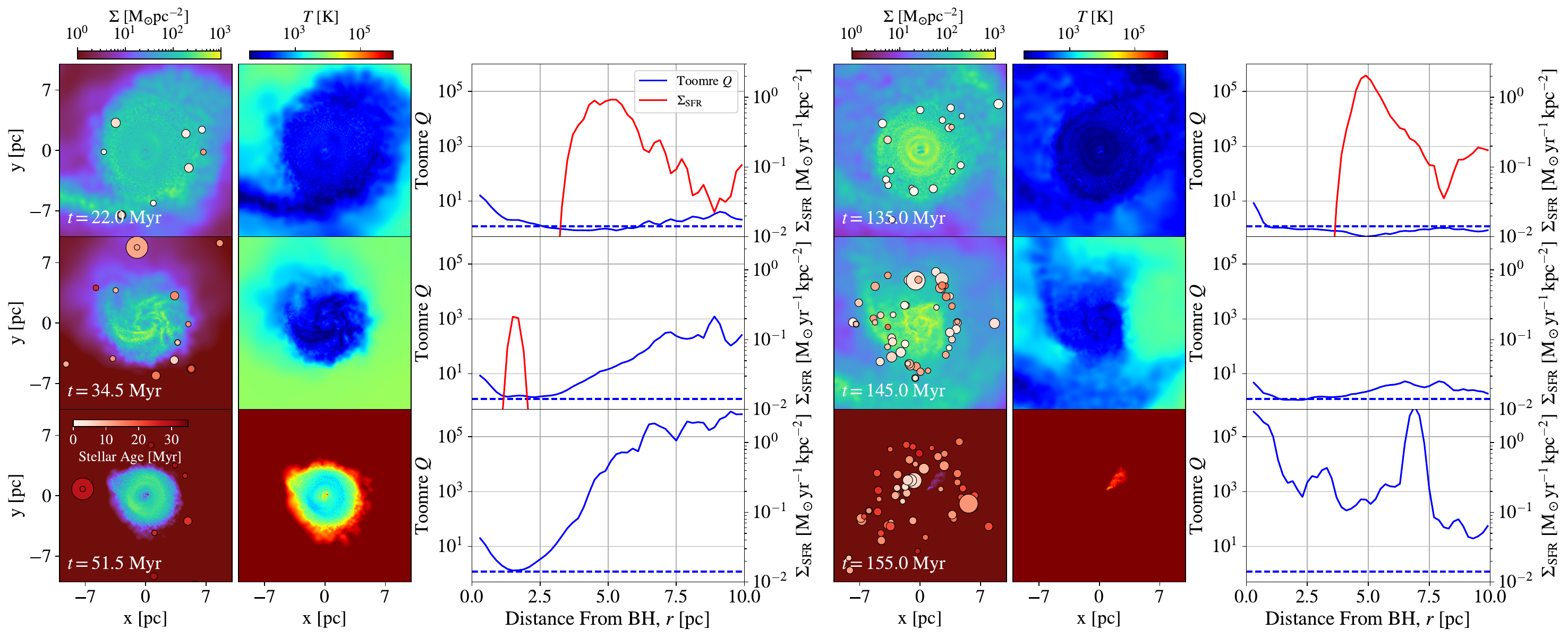}
    \caption{Time evolution of the CND in {\tt NSC-5pc} due to feedback from massive stars and SN explosions. The first three columns illustrate the formation and evolution of the `first' generation of stars in the CND, while the subsequent three columns show the `second' generation. The panels display maps of gas surface density ({\it left} columns), density-weighted temperature ({\it middle} columns), as well as radial profiles of the Toomre $Q$ parameter ({\it blue-solid} lines) and the surface density of SFR ({\it red-solid} lines) in the {\it right} columns. {\it Blue-dashed} lines show the threshold value of the Toomre parameter, $Q=1.2$. The coloured dots represent the locations of individual young massive stars responsible for stellar feedback  (age $< 35$~Myr; mass $>5\,\msun$), with the dot size indicating the stellar masses and the colour encoding their age. Across the simulated time span, the stellar mass of massive stars ranges from $5.4$ to $78.4\, \msun$. The CND survives SN explosions from the first stellar generation and keeps feeding the central BH, but becomes Toomre stable, preventing further in-situ star formation. At a later time, large-scale gas inflows replenish the CND, which becomes Toomre unstable again, and drives a large SFR peak, which ultimately leads to the entire CND dispersal. For more information, see Section~\ref{sec:SF-in-CND}.}
    \label{fig:SNe-proj}
\end{figure*}

\begin{figure*}
\centering
\includegraphics[width=\textwidth]{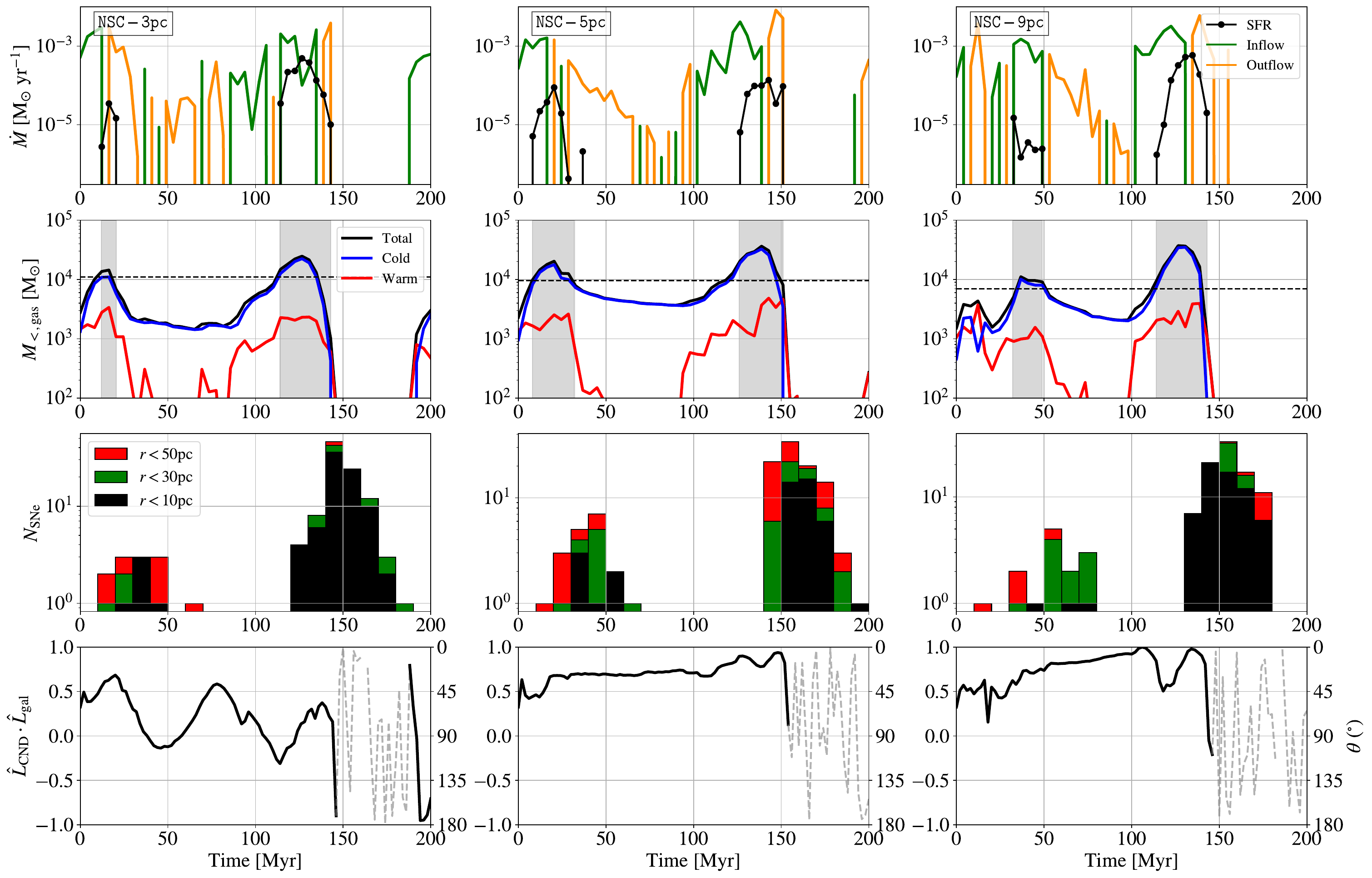}
    \caption{The time evolution of SFR in central 10~pc region as well as the gas inflow and outflow rates at $r=10$~pc ({\it top} row); enclosed gas mass, also split by gas temperature ($r < 10$~pc; {\it second} row); number of SNe within different spatial regions ({\it third} row); and angular momentum direction of CND compared to the galactic disc ($\hat{L}_{\rm CND}\cdot\hat{L}_{\rm gal}$; {\it bottom} row). Gas is categorized as `cold' or `warm' based on a temperature threshold of $100$ K. The {\it black-dashed} horizontal lines in the {\it second}-row panels indicate the mass threshold at which star formation begins in CNDs, while the shaded region represents the period when SFR is non-zero. The angular momentum of CND is density-weighted. It is shown with a solid line when the total enclosed mass ($M_{\rm <,gas}$) exceeds $10^3\msun$ and with a light grey dashed line otherwise. 
    The CNDs in the different NSC runs survive the SNe after the first stellar generation; however, after the second stellar generation, the CNDs are evaporated by the SNe in the circumnuclear region. For more information, see Section~\ref{sec:SF-in-CND}.
    }
    \label{fig:sfr-evol}
\end{figure*} 

\subsection{Star formation in the circumnuclear region}
\label{sec:SF-in-CND}
\subsubsection{How does in-situ star formation feedback affect CND properties?}

NSCs are typically thought to be composed of old stars, however, recent studies have confirmed the presence of multiple stellar populations and ongoing star formation \citep[e.g.,][]{Paudel+Yoon2020, Fahrion+2022, Fahrion+2024}. Therefore, in this section, we investigate gravitational instability in the circumnuclear region and its role in triggering star formation. Fig.~\ref{fig:SNe-proj} illustrates the evolution of the CND in the {\tt NSC-5pc} run, displaying gas surface density and temperature maps alongside radial profiles of the Toomre $Q$ parameter and SFR surface density. The left-hand set of panels depicts the formation and evolution of the initial stellar generation, while the right-hand set of panels shows the second generation. Individual young massive stars with masses above $5\,\msun$ and ages less than $35$~Myr are shown as coloured dots. This age threshold approximately corresponds to the lifetime of an $8\,\msun$ star. 

At $t=22$~Myr, within the dense, cold CND, the gas becomes gravitationally unstable at $r\simeq4-7$~pc (see also Fig.~\ref{fig:main}). In regions where $Q \lesssim 1.2$, young stars form at a rate reaching $\Sigma_{\rm SFR}\simeq1\msun{\rm yr}^{-1}{\rm kpc}^{-2}$. As illustrated with the dots colour-coded by stellar age, at $t=22$~Myr massive stars form and rotate in the CND within the Toomre-unstable region, emitting UV photons that ionize and heat the surrounding CND gas. After their short lifetime on the main sequence, these massive stars explode as SNe, partially disrupting the CND and sweeping away gas streams from the circumnuclear region (see maps at $t=34.5$ Myr). This strong stellar feedback in the CND renders the circumnuclear region for $r>6$~pc highly stable, with $Q$ $ \gtrsim 10^2$. These SNe explosions not only heat the circumnuclear region, but also generate shocks within the CND, compressing the gas toward the BH and shifting the minimum Toomre Q closer to the centre.
This process triggers a significant yet short-lived boost in both in-situ star formation (see panels at $t=34.5$~Myr with $\Sigma_{\rm SFR}\sim0.2\msun{\rm yr}^{-1}{\rm kpc}^{-2}$ at $r\sim2$~pc) and BH accretion. Despite the disruption, the CND withstands the SNe of the first stellar generation (see panels at $t=51.5$~Myr, which show the circumnuclear region after the final SN have exploded from this generation), although it becomes Toomre-stable and further star formation ceases. This allows the BH to stably accrete gas from the CND.

Around $50$~Myr after the previous SN explosions ($t\approx100$~Myr), dense gas streams from the outer regions are captured by the NSC potential and accreted onto the CND, triggering the formation of a second stellar generation. By $t = 135$~Myr, the large gas inflow and subsequent gravitational instability lead to the formation of a large number of young massive stars.  At this time, nearly the entire CND region becomes Toomre-unstable, with the SFR surface density reaching $\Sigma_{\rm SFR} \gtrsim 2 \msun \, {\rm yr}^{-1} \, {\rm kpc}^{-2}$ at $r \sim 5$ pc. Similar to the first stellar generation, the feedback from young massive stars disrupts the CND. However, by $t = 155$~Myr, SNe from the second-generation stars completely evaporate the CND. As a result, inflow to the accretion disc is terminated. Due to CND's diffuse and hot state, the Toomre $Q$ reaches values as high as $Q > 10^5$ at $r \sim 1$ pc, rendering the entire circumnuclear region stable.  

\begin{figure*}
\centering
\includegraphics[width=0.9\textwidth]{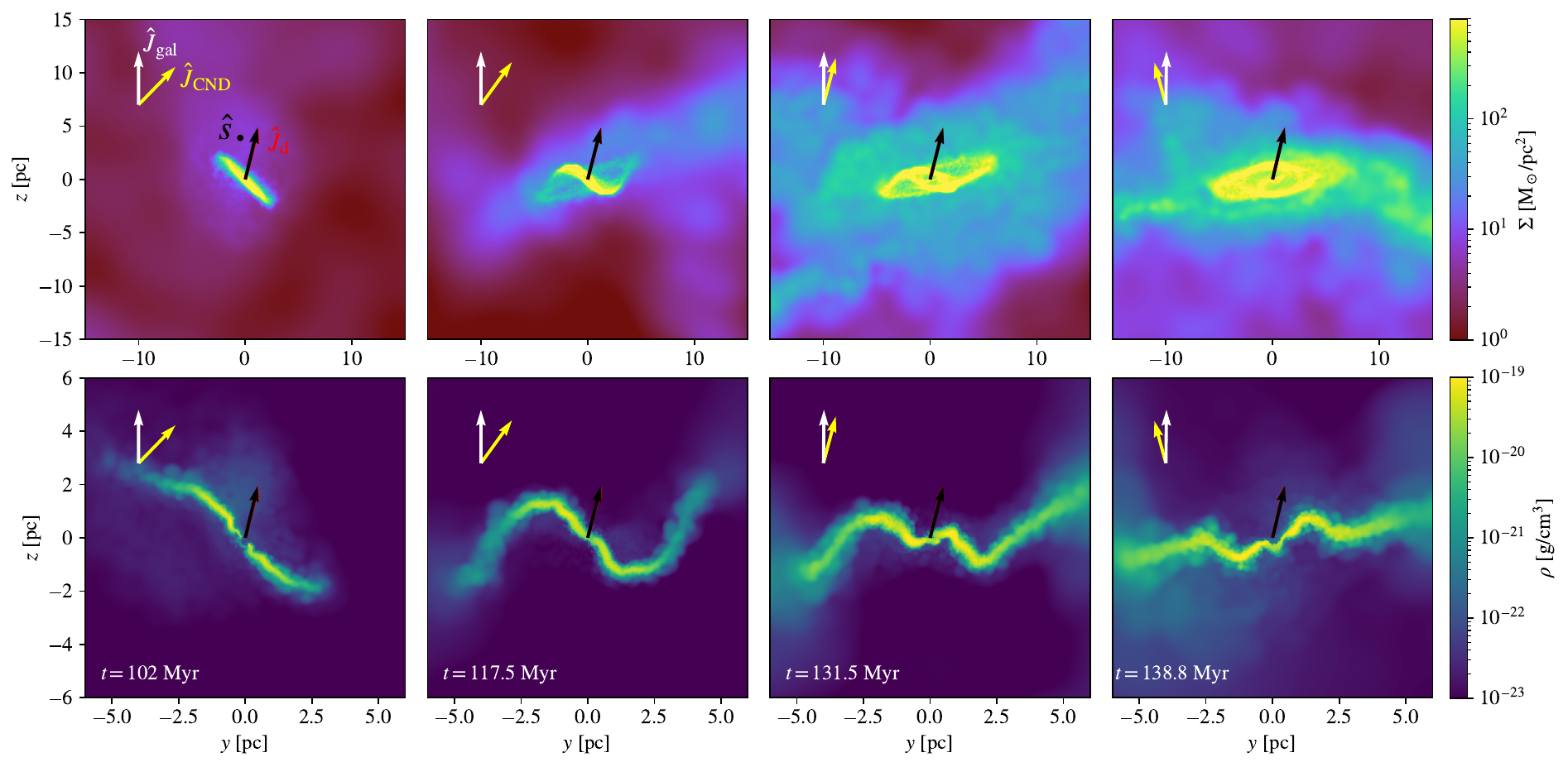}
    \caption{Gas density projections ({\it top}) and slice plots ({\it bottom}) in the nuclear region at four different times in {\tt NSC-5pc} run. The projection box is $30$~pc in size and centred on the BH, while the slice plot is taken on a very thin plane of $12$~pc in length that contains the BH. The arrows indicate the direction of angular momentum of the galactic disc ({\it white}), CND ({\it yellow}), accretion disc ({\it red}) and BH spin ({\it black}). The inflowing gas has a different angular momentum direction compared to the CND, causing the CND to gradually reorient itself from the outer edge toward the centre and forming a warped disc, which is re-aligning with the angular momentum of the inflowing gas. The inner disc undergoes rapid precession. For further details, see Section~\ref{sec:warping-disc}.} 
    \label{fig:warping-disc-gas}
\end{figure*}

\begin{figure*}
\centering
\includegraphics[width=0.76\textwidth]{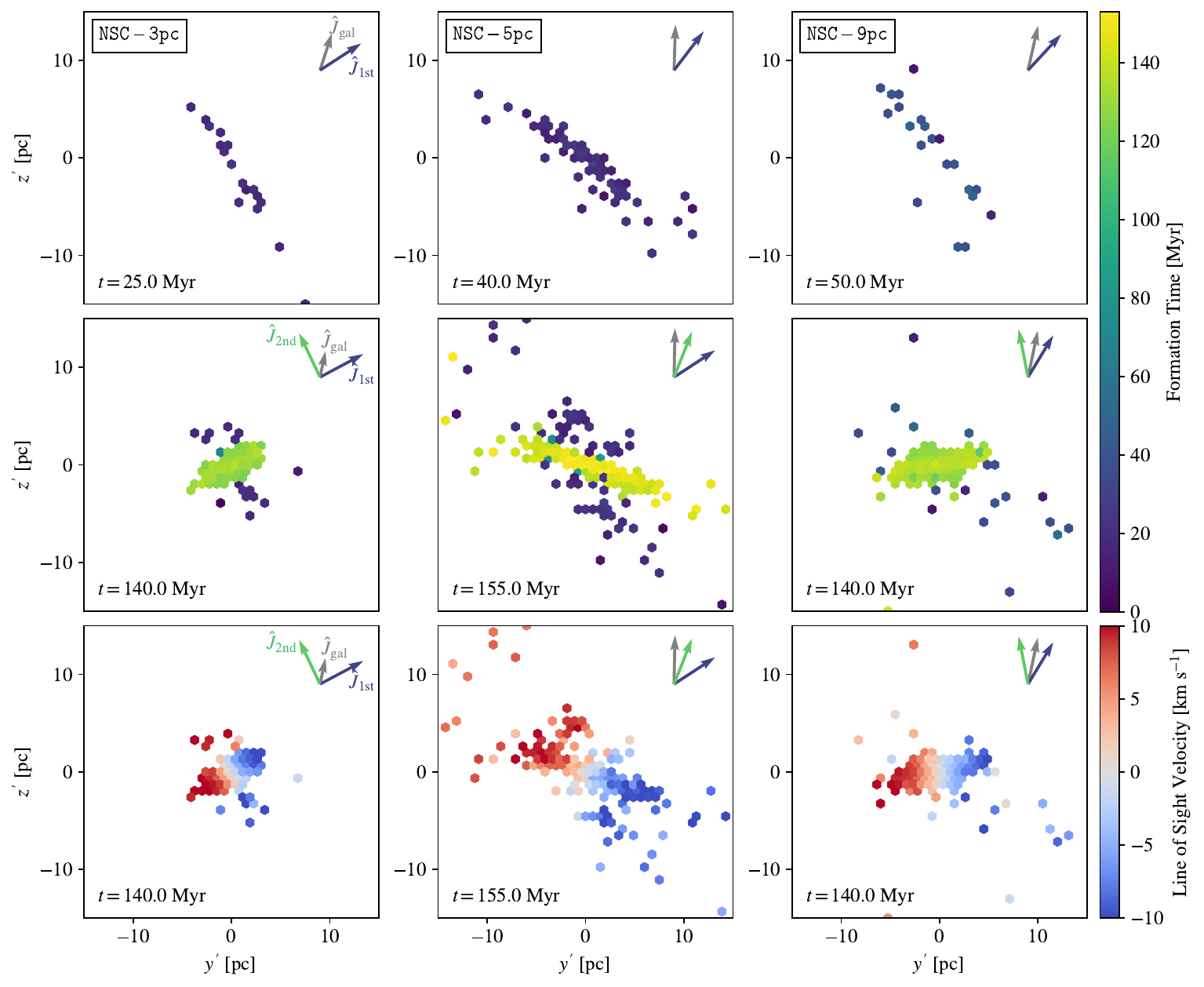}
    \caption{The projections of newly formed {\it star particles} in the nuclear region at the completion of the first ({\it top}) and second ({\it middle} and {\it bottom}) stellar generations for the three different NSC runs. The colour indicates the formation time ({\it top} and {\it middle}) of each stellar particle and its line-of-sight velocity ({\it bottom}). The $z$-axis of the projection box is aligned with the cross product of the angular momentum of the first and second stellar generation for clarity. The {\it top right} arrows indicate the angular momentum direction of the galactic disc ({\it grey}), the stellar disc of the first generation ({\it navy}) and the second generation ({\it green}). 
    The CND changes its direction over time, resulting in each stellar generation having a different rotational plane. For more information, see Section~\ref{sec:warping-disc}.} 
    \label{fig:warping-disc-star}
\end{figure*}

In Fig.~\ref{fig:sfr-evol}, we present the time evolution of the SFR, inflow/outflow rate ({\it top} row), and enclosed gas mass ({\it second} row) within $10$~pc of the BH for simulations with NSCs of different effective radii. The gas mass is divided into cold ($T<100$~K) and warm gas ($T>100$~K). The CND gas mass fluctuates between $10^3-10^4\msun$, which is slightly lower than values reported in the literature \citep[$10^4-10^6 \msun$; for e.g.,][]{Schartmann+2018, Dinh+2021, Solanki+2023}, where CNDs are modelled in more massive host systems. NSCs with different densities do not significantly affect the SFR in the CND or the enclosed gas mass. As observed in Fig.~\ref{fig:main}, gas accretes onto the CND primarily in the form of dense streams, where it loses its internal energy through cooling. Within $r<10$~pc, cold gas dominates, comprising approximately $80-99$\% of the total gas mass. We find that when the total gas mass exceeds $\sim10^4\msun$ ({\it black dashed} line in the second row panels), the CND becomes unstable and begins forming stars. We note that this limit increases in denser NSCs, which is attributed to stronger tidal forces (larger second derivatives of the potential; see also Appendix~\ref{sec:appendix-energy-profile}), leading to enhanced gravitational instability.

In the CND, star formation occurs through multiple cycles, with two distinct episodes observed over a span of $200$~Myr (at around $t \sim 10-50$~Myr and $t\sim 110-150$~Myr), each lasting $\lesssim25$~Myr. As shown in Fig.~\ref{fig:SNe-proj}, the SNe from the first generation stars do not completely evaporate the CND, allowing it to persist for $\sim140$~Myr in all three runs. The highly dense nature of the CND ($\Sigma_{\rm gas}\gtrsim10^3 \msun~{\rm pc}^{-2}$; see Fig.~\ref{fig:nsc-projection}) enables it to withstand tens of SNe explosions, with an event rate of at least $0.3~{\rm Myr}^{-1}$ within $r<10$~pc and $0.7~{\rm Myr}^{-1}$ within $r<50$~pc \citep[see also e.g., ][]{Shi+2023}. During the second stellar generation, the SFR in all three runs reaches $\geq10^{-4}\msun~{\rm yr}^{-1}$, which is about 10\% of the SFR in the entire galactic disc (see also Fig.~\ref{fig:CGM-profile}). The SNe from the second stellar generation evaporate the CND, causing the central region to lose more than $90\%$ of its gas. The {\it third} row of Fig.~\ref{fig:sfr-evol} shows the number of SN events in time bins ($N_{\rm bins} = 20$; $\Delta t=10$~Myr) within $10$~pc, $30$~pc, and $50$~pc from the BH. The CND gas evaporates when the number of SN events becomes sufficiently high, i.e. around 20 per $10$~Myrs. 

The total stellar mass in the nucleus ($r\le10$~pc) is summarized in Table~\ref{tab:runs}, with the mass fraction of newly formed stars in the nucleus relative to the total stellar mass being 1.22\%, 0.538\%, and 1.39\% for the {\tt NSC-3pc}, {\tt NSC-5pc} and {\tt NSC-9pc} runs, respectively. We find that the in-situ star formation is influenced by the conditions of the ISM within $50$~pc of the BH. Notably, the SFR in the {\tt NSC-5pc} run is relatively lower than in the other runs. As a result of early stellar feedback and SNe explosions near the CND ($r>10$~pc) during the early phase of the second stellar generation (see the {\it red} bar), the hot gas mass increases at $t=140$~Myr. Even in the absence of SNe within $r<10$~pc, this rise in hot gas mass prevents gravitational instability and suppresses SFR in the CND. Furthermore, in the simulation without an NSC, star formation does not occur within $r<10$~pc of the BH over the $200$~Myr period.

The bottom row of Fig.~\ref{fig:sfr-evol} shows the evolution of the angular momentum of the CND over $200$~Myr. We calculate the density-weighted angular momentum of the CND, $\vec{L}_{\rm CND}$ within $r<10$~pc and compute the inner product of its versor with that of the galactic disc ($\hat{L}_{\rm CND}\cdot\hat{L}_{\rm gal}$) to estimate the alignment of the CND relative to the galactic disc. Gas inflows into the vicinity of the BH and forms the CND within $\lesssim3$~Myr of the beginning of the simulation, after which the CND rotation direction changes due to both inflows and outflows. Typically, the CND does not manage to align with the galactic disc, oscillating at angles of approximately $\lesssim45^\circ$ in the {\tt NSC-5pc} and {\tt NSC-9pc} runs, while in the {\tt NSC-3pc} run, it exhibits a slightly broader range of misalignment of $30-110^\circ$. For the {\tt NSC-3pc} run, the SNe within the CND induce stronger outflows compared to the other runs, reaching rates of $10^{-5}\msun{\rm yr}^{-1}$ at $t=20$~Myr, rapidly shifting the CND's direction from $\sim40^\circ$ to $\sim100^\circ$. The subsequent inflowing gas alters the CND's orientation, causing back-and-forth variations by more than $\gtrsim40^\circ$ for $t=10-140$~Myr. We note that this variation in CND rotation leads to warping within the CND, which will be discussed in the following section.

\subsection{Circumnuclear disc and galactic disc misalignment and warping}
\label{sec:warping-disc}

The {\it top} panels of Fig.~\ref{fig:warping-disc-gas} show a sequence of gas surface densities in the nuclear region at four different times for the {\tt NSC-5pc} run. The angular momentum directions of the galactic disc ({\it white}), CND ({\it yellow}), accretion disc ({\it red}) and BH spin ({\it black}) are indicated with arrows. At $t = 102$~Myr, the angular momentum direction of the CND deviates by approximately $40^\circ$ from that of the galactic disc. Once the CND forms, the torque exerted by the CND aligns the inflowing gas to the CND, as shown in Fig.~\ref{fig:torque-prof}. However, when a large amount of inflowing gas, rotating along a different axis, is accreted onto the CND, it continues to be misaligned with respect to the CND's inner regions (see the second and third panels in Fig.~\ref{fig:warping-disc-gas}). This leads to the formation of a warped CND. Over time, we observe that the inner disc gradually aligns with the outer disc due to gas depletion by the BH and the torque exerted by the outer disc. Specifically, by $t=138.8$~Myr, the inner disc becomes aligned with the outer disc. 

As shown in the {\it bottom} panels of Fig.~\ref{fig:warping-disc-gas}, where central gas density slices are displayed, the rotation of the inner region ($r \lesssim 2$~pc) of the CND is rapid and chaotic, exhibiting precession. This precession is induced by the relative motion of the BH with respect to the CND, which leads to mini-spirals and multiple circumnuclear rings within the CND. The direction of both the BH spin and the angular momentum of the accretion disc remains nearly unchanged over 30~Myr. We note that the specific angular momentum of the inflowing gas, $L_{\rm in}$, oscillates around a level that is approximately 10\% of that of the $\alpha$-accretion disc (which will be discussed in detail later). The mass accretion rate is $\dot{M}_{\rm in} \sim 0.01~M_{\rm d}~{\rm Myr}^{-1}$, resulting in a 0.1\% difference in total angular momentum over 1~Myr. The detailed evolution of the accretion disc and BH spin, as well as the strong alignment between them, will be discussed in Section~\ref{sec:bh-evolution}.

\begin{figure*}
\includegraphics[width=\textwidth]{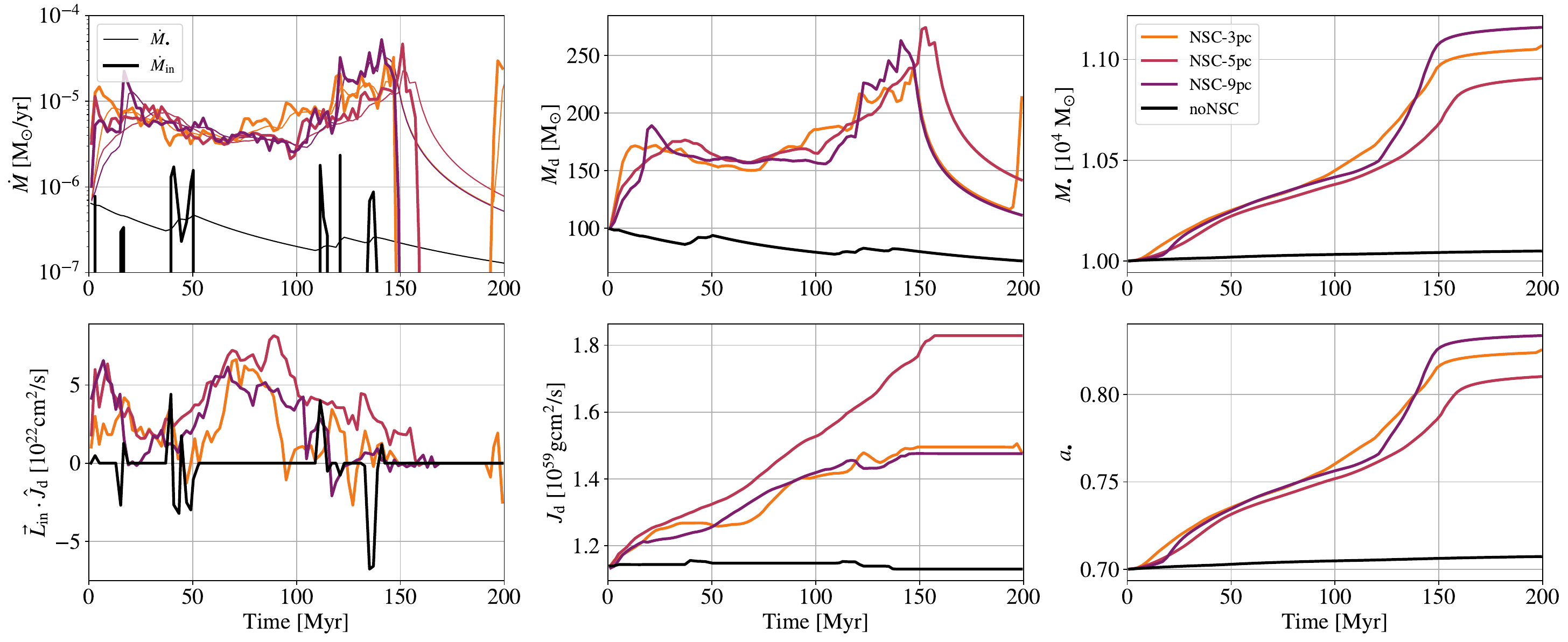}
    \caption{The mass ({\it top}) and angular momentum ({\it bottom}) evolution of inflowing gas ({\it left}), accretion disc ({\it middle}) and BH ({\it right}) for runs without and with NSCs. Thick lines in the {\it top left} panel represent the inflowing mass rate to the accretion disc, while the thin lines represent the BH accretion rate. The {\it bottom left} panel shows the specific angular momentum of inflowing gas, aligned with the angular momentum of the accretion disc. Inclusion of an NSC leads to a sustained BH mass growth and spin-up. For more information, see Section~\ref{sec:adspin}.
    }
    \label{fig:bh-ad-evol}
\end{figure*}

This significant change in the angular momentum of the CND results in distinct rotational characteristics for different stellar generations. Fig.~\ref{fig:warping-disc-star} shows projections of newly formed stars in the nuclear region at the completion of the first and second stellar generations for our different NSC runs. For the first stellar generation, spanning $t \sim 10-50$~Myr, the angular momentum of the newly formed stars deviates by $52^\circ$, $41^\circ$, and $40^\circ$ from that of the galactic disc in the {\tt NSC-3pc}, {\tt NSC-5pc}, and {\tt NSC-9pc} runs, respectively. However, for $t \sim100-140$~Myr, the inflowing gas alters the rotation direction of the CND (see the bottom panels of Fig.~\ref{fig:sfr-evol}), resulting in the formation of young stars aligned with the reoriented CND. As shown in the second and third rows of Fig.~\ref{fig:warping-disc-star}, the young stars rotate in a plane which is clearly distinct from that of the older stellar disc. Notably, in the {\tt NSC-3pc} run, the rotation directions of the first and second generations differ by nearly $90^\circ$.  Additionally, we observe that the stellar disc formed by the second generation of stars is itself warped in the {\tt NSC-3pc} and {\tt NSC-9pc} runs. This suggests that different stellar generations can exhibit complex and distinct kinematic properties in the nuclear region. Interestingly, this configuration is reminiscent of the Milky Way nuclear star cluster, where two stellar discs -- one rotating clockwise and the other counter-clockwise -- have been observed around Sagittarius A* \citep[for e.g.,][]{Genzel+2010}.

\subsection{Evolution of the black hole}
\label{sec:bh-evolution}
\subsubsection{Black hole and accretion disc evolution}
\label{sec:adspin}

We now analyse how larger-scale mass and angular momentum flows influence BH and $\alpha$-accretion disc evolution.  Fig.~\ref{fig:bh-ad-evol} shows mass and angular momentum (or spin) evolution of the inflowing gas onto the accretion disc ({\it left} column), accretion disc ({\it middle} column) and BH ({\it right} column) in simulations with and without an NSC. In the {\tt noNSC} run, gas intermittently flows towards the accretion disc, with a mass inflow rate ranging between $10^{-6}$ and $10^{-7}\msun~{\rm yr}^{-1}$. In contrast, the inflow is considerably more sustained and higher in the presence of an NSC, oscillating around $10^{-5}\msun~{\rm yr}^{-1}$. This inflow rate is consistent with the time-averaged radial inflow profiles observed near the self-gravity radius, $r_{\rm SG}\sim1.5$~pc, as shown in Fig.~\ref{fig:nsc-profile}.

At the beginning of the simulation, we observe that the CND forms slightly faster in the run with a denser NSC, due to the stronger torque (see Fig.~\ref{fig:torque-prof-nsc}). This initially results in a higher inflow rate, which is then regulated by stellar feedback from the young stars in the CND. After the feedback-induced regulation, the CND becomes dynamically stable, and the BH accretion persists uninterrupted until $t\sim140$~Myr in all three simulations with an NSC, at which point the CND is completely evaporated by SNe formed in situ. Notably, initially SN explosions enhance the mass inflow rate to the accretion disc, leading to sharp increases at $t\sim20$~Myr and $t\sim120-150$~Myr. This is because SNe rapidly compress the surrounding gas, generating shocks near the BH that concentrate the gas and facilitate efficient mass inflow onto the accretion disc\footnote{While SNe may contribute to angular momentum cancellation, their dominant effect is to transport pre-existing low angular momentum gas from $r \lesssim 1$~pc (see Fig.~\ref{fig:nsc-profile}) toward the accretion disc’s self-gravity radius.}.

When the inflow rate exceeds the BH accretion rate, the accretion disc mass increases, and vice versa, as shown in the {\it middle} panel. The initial mass of the accretion disc is set to 100~$\msun$. In the presence of an NSC, the accretion disc mass increases up to $150 \,\msun$, at which point the BH accretion rate becomes comparable to the inflowing mass rate during the first $20$~Myr, establishing a stable equilibrium until $t=100$~Myr, when a large inflow of material from the ISM occurs (see Fig.~\ref{fig:sfr-evol}). As previously shown, bursty star formation and subsequent SNe at around $t=140-150$~Myr evaporate the nuclear gas, causing the accretion disc mass to decrease as it is accreted onto the BH. In the absence of an NSC, episodic inflows lead to slight increases in the accretion disc mass, but overall, the accretion disc mass is reduced due to BH accretion\footnote{ {In low accretion rate regimes, especially in the {\tt noNSC} run, a thick disc model may provide a more realistic description \citep[see e.g.][]{Koudmani+2024}. However, we apply the $\alpha$-disc model uniformly across all runs to enable a fair comparison of NSC effects. A more detailed treatment of accretion disc physics will be explored in future work.}}.

By comparing $\dot{M}_{\rm in}$ ({\it top-left} panel) and ${M}_{\rm d}$ ({\it top-middle} panel) in Fig.~\ref{fig:bh-ad-evol}, we find that the BH accretion rate approximately follows the evolutionary trend of the accretion disc (see also Equation~(2) in \cite{Fiacconi+2018}). Over $200$~Myr, the BH mass increases by $10\%$ in the presence of an NSC, compared to only a $0.5\%$ increase in its absence (see also Table~\ref{tab:runs}). However, no clear correlation is observed between the NSC density and the growth of the accretion disc or BH mass. The evolution of the accretion disc and BH is instead determined by both the state of the CND and stellar feedback in the circumnuclear region. While the BH growth in the presence of the NSC is modest, we emphasize here that our dwarf galaxy simulations are performed both in approximate equilibrium and in isolation, and we would expect that the fresh gas supply from the cosmic web and environmental perturbations, especially at high redshift \citep{Smith+2019}, may lead to higher BH fuelling.

The {\it bottom-left} panel shows the evolution of the specific angular momentum of the inflowing gas, which influences the accretion disc's angular momentum. The angular momentum amplitude oscillates around $5\times10^{22}{\rm cm}^2{\rm s}^{-1}$ across the different runs. In the absence of an NSC, the direction of the inflowing gas shows random variation, resulting in minimal changes to the accretion disc's angular momentum, $J_{\rm d}$ over the entire $200$~Myr. In contrast, the presence of an NSC enables coherently aligned gas accretion for $\sim140$~Myr. This originates from the rotationally-supported CND formed in the presence of the NSC (see Fig.~\ref{fig:nsc-profile}), leading to an increase in the angular momentum of the accretion disc, as shown in the {\it bottom-middle} panel. Notably, in the {\tt NSC-5pc} simulation, the inflowing gas remains aligned with the rotation of the accretion disc throughout the $140$~Myr period, resulting in a 63\% increase in angular momentum, whereas other runs with an NSC show increases of up to $36\%$.

Finally, the \textit{bottom-right} panel shows the evolution of the spin parameter of the BH. The Bardeen-Petterson torques drive alignment between the accretion disc and the BH spin, ensuring that in the setups considered here, the accretion disc always corotates with the BH such that matter accretion onto the BH drives the growth of the spin parameter in a manner analogous to the mass growth of the BH \citep[see Fig.~\ref{fig:warping-disc-gas} and][]{Fiacconi+2018}. Since our BH mass is small ($M_\bullet\sim10^4 \,\msun$) and the accretion disc typically dominates the total angular momentum budget (see Fig.~\ref{fig:appendix-bh-evolution}), the time-scale of the \cite{Bardeen+Petterson1975} effect is on the order of a few Myr. Although the angular momentum of the accretion disc changes over time due to gas inflow, the BH and the disc repeatedly re-align. The prolonged coherent matter accretion drives the growth of the spin parameter, increasing it from an initial value of 0.7 to $\gtrsim0.8$ in the presence of the NSC. However, in the absence of an NSC and the supply of gas for accretion that it promotes, the BH spin remains nearly constant, staying at $\lesssim0.71$.

\subsubsection{Bondi-Hoyle-Lyttleton accretion vs. $\alpha$-disc accretion model}
\label{sec:bondi}

\begin{figure}
\centering	\includegraphics[width=0.45\textwidth]{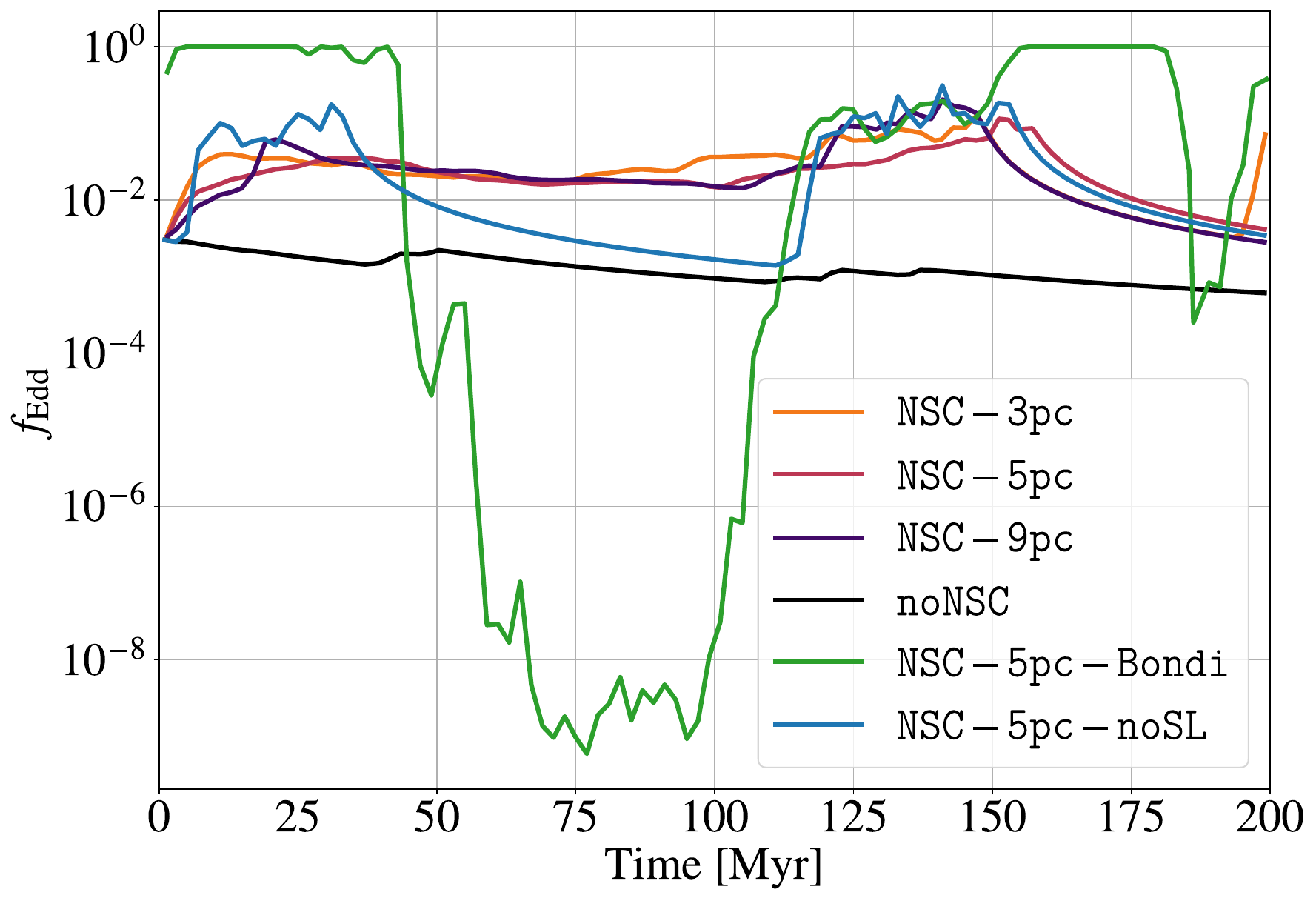}
    \caption{Time evolution of BH's Eddington fraction for different simulation models. In the Bondi accretion model BH growth is overestimated as gas angular momentum is not taken into account, whereas the $\alpha$-disc model shows much more moderate growth, which is boosted in the presence of a NSC. For more information, see Section~\ref{sec:bondi}.}
    \label{fig:bondi-vs-adspin}
\end{figure}

Fig.~\ref{fig:bondi-vs-adspin} shows the BH Eddington fractions for different accretion schemes and different NSC properties that we have explored in this study. Consistent with the findings from Fig.~\ref{fig:bh-ad-evol}, the Eddington fraction in the fiducial model (i.e. with the `$\alpha$-disc' scheme) sensitively depends on the presence of an NSC: with an NSC the Eddington fraction is $ \gtrsim 10^{-2}$, while in the absence of an NSC it is about one order of magnitude lower, $\sim 10^{-3}$. We further note that the `$\alpha$-disc' simulation with a NSC but without super-Lagrangian refinement ({\tt NSC-5pc-noSL}) while qualitatively similar to {\tt NSC-5pc}, exhibits clear quantitative differences: the gas in the vicinity of the BH within the CND cannot reach as high densities, and hence is unable to withstand strong stellar feedback which leads to its disruption. As a result, gas inflow to the accretion disc ceases, resulting in only two distinct episodes of BH accretion in this run. This result highlights the need for both very high resolution and realistic physical modelling to be able to retain feedback-resistant material and track BH fuelling with greater accuracy. 

Specifically, this point is further underscored when we turn to the analysis of the Bondi accretion model. In the {\tt NSC-5pc-Bondi} run, the BH reaches the Eddington limit ($f_{\rm Edd}=1$) at $t<40$~Myr and $t=155-180$~Myr, but in between it drops as low as $10^{-9}$. We note that the initially high BH accretion rate in the Bondi model inhibits the formation of a dense CND, which slightly delays star formation in the CND (both for the first and second generation). At $t\sim50$~Myr, SNe in the circumnuclear region completely evaporate the CND, which is less massive and more diffuse than the CND in the fiducial models. At $t\sim100$~Myr, the NSC accretes gas again and forms a new CND. With the Bondi model, the BH mass increases by $460\%$ over $200$~Myr, while the $\alpha$-disc accretion model shows only a $10\%$ increase for the same simulation setup (see Table~\ref{tab:runs}). The Bondi accretion model neglects the angular momentum of the inflowing gas, clearly leading to an overestimation of the BH accretion rate. However, it should be noted that, in the presence of AGN feedback, the strong dependence of Bondi accretion on gas density and sound speed makes the BH growth self-regulating, thereby ultimately suppressing very large BH growth we observe here.

\section{Discussion}
\label{sec4:Discussion}
In this study, we take advantage of very high-resolution simulations of individual dwarf galaxies where individual star formation is resolved. Furthermore, by accurately resolving gas on the scale of the accretion disc self-gravity radius with super-Lagrangian refinement, we investigate how the multiphase ISM feeds the BH and influences its spin, both with and without an NSC. 

A study that is closest to ours is by \citet{Partmann+2025}, who explored BH accretion and nuclear star formation with NSCs in dwarf galaxy systems using pc-scale resolution simulations. Comparison to our results provides useful insights into the role of NSCs and accretion models. In general, we find a broad agreement with their findings, which is encouraging and firmly underscores the presence of an NSC can promote nuclear gas inflows and support the growth of the central BH in simulations of dwarf galaxies (and likely in more massive systems as well). More specifically, their study found that the explosion of nuclear SNe evaporates all the gas within the central $10$~pc, limiting BH accretion to short periods of only a few tens of Myr. Our simulations having a higher spatial and mass resolution near the BH due to super-Lagrangian refinement, capture the formation of a CND - a much denser (reaching $\Sigma_{\rm CND}\gtrsim10^3\msun~{\rm pc}^2$) and colder ($T_{\rm CND}\sim10^2$~K) structure. This structure can remain largely intact despite SN feedback, surviving SNe rates of $\sim 0.3$~Myr$^{-1}$ in the $r < 10$~pc region (and $0.7$~Myr$^{-1}$ in $r < 50$~pc; see Fig.\ref{fig:sfr-evol}), thereby allowing the host BH to accrete stably over long periods (i.e. of the order of 140~Myr).

This result aligns with the findings of \cite{Shi+2023}, who investigated BH accretion within massive star-forming gas clumps and reported that surface densities exceeding $10^3\msun~{\rm pc}^{-2}$ can persist throughout the episodes of strong stellar feedback. However, \cite{Shi+2023} did not observe the formation of a fast-rotating CND, as they did not model a centrally concentrated gravitational potential such as the one provided by the NSC in our simulations. Similarly, \cite{Schartmann+2018} studied star formation in CNDs with radii of several hundred parsecs. They found that the CND undergoes strong fragmentation, forming massive clumps due to gravitational instability and intense nuclear starburst. In our simulation, the CND forms within a lower-mass galaxy, leading to a CND mass that is 100 times lower. Moreover, the presence of an NSC leads to a steeper potential gradient compared to the bulge, enhancing the rotation of the CND. This enhanced rotation suppresses severe fragmentation, distinguishing our findings from previous studies of star formation in CNDs.

\begin{figure}
\includegraphics[width=0.5\textwidth]{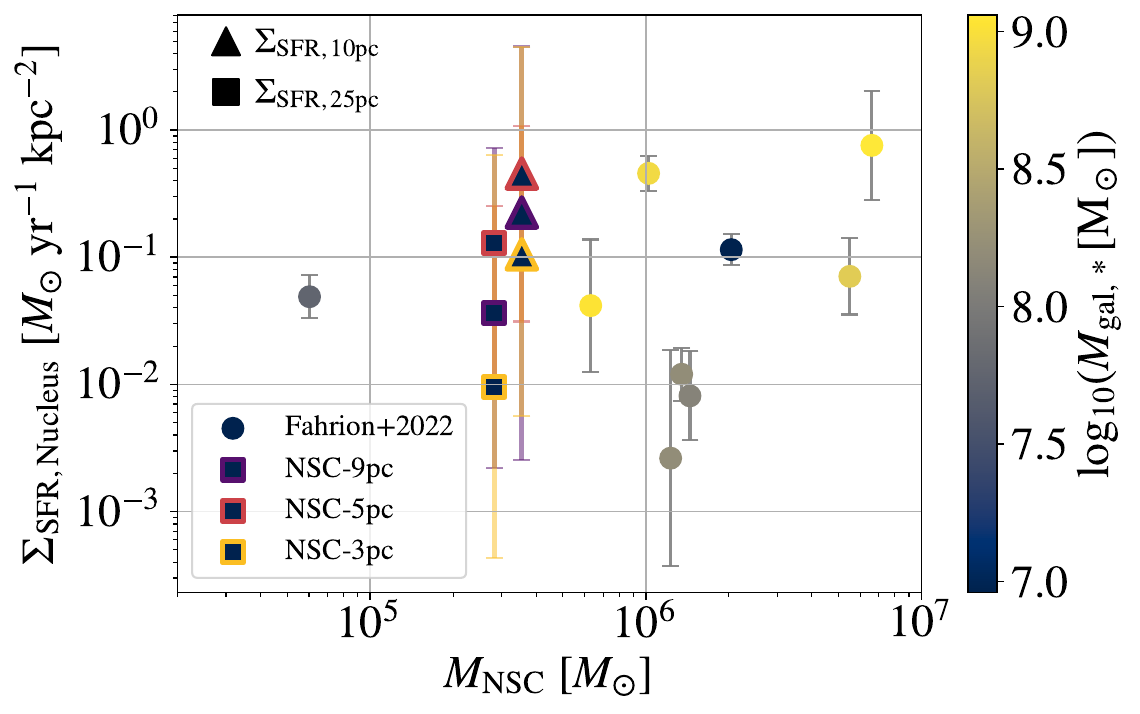}
\caption{The SFR surface density in nuclear region vs. NSC mass for this work and observations in nucleated late-type dwarf galaxies from \citep[][{\it circle}]{Fahrion+2022}. The colour of symbols represents the stellar mass of the host galaxy. The {\it triangle} and {\it square} symbols represent the median value of the SFR surface density for $t=0-200$~Myr within spheres of $r=10$~pc and $25$~pc, respectively, which correspond to the half of FWHM values estimated based on the maximum and minimum resolutions, following \citet{Fahrion+2022}. The error bars associated with {\it triangle} and {\it square} symbols indicate the $5$th and $95$th percentiles of the distribution evaluated over $t=0-200$~Myr. Note that both {\it triangle} and {\it square} symbols correspond to the same NSC mass, $M_{\rm NSC}= 3.16\times10^5\msun$, but are slightly offset for clarity. We find that the simulated SFR surface density matches well the observations, both in terms of NSC masses as well as host galaxy stellar mass. For more information, see Section~\ref{sec4:Discussion}.}
\label{fig:SFR-comparison}
\end{figure}

To further understand whether our simulated CNDs and the in-situ SFR they sustain are realistic, we compare with recent observations of dwarf galaxies. \citet{Fahrion+2022} studied the SFR surface density in the circumnuclear regions of nucleated late-type dwarf galaxies using integral-field spectroscopy data obtained with the Multi Unit Spectroscopic Explorer (MUSE). Fig.~\ref{fig:SFR-comparison} compares the time-averaged (median) SFR surface density within the full width at half maximum (FWHM) and the mass of NSCs between our work and that of \citet{Fahrion+2022}. We calculate the SFR surface density over $200$~Myr with a timestep of $10$~Myr. The maximum and minimum FWHM of NSC stars in our simulations are estimated based on the minimum and maximum spatial resolutions of the observed samples in \citet[][4.8 and $20$~pc]{Fahrion+2022}, resulting in FWHM values of $20$~pc and $50$~pc, respectively. All three different NSCs share the same FWHM values for these resolutions.

The SFR surface densities in the simulations range from $\sim10^{-2}$ to $0.4 \,\msun$~yr$^{-1}$~kpc$^{-2}$, comparable to the observed data. Notably, despite having the smallest total stellar mass in the central region within the simulation sample, {\tt NSC-5pc} exhibits the highest median SFR surface density\footnote{However, a distinct correlation between NSC mass and either SFR surface density or the stellar mass of the host galaxy is not present in the observational sample. Similarly, no strict correlation is found between the SFR and NSC density.}. This can be attributed to its steadier SFR evolution, which is regulated by SN feedback in the outer regions ($r>10$~pc; see Fig.~\ref{fig:sfr-evol}). In contrast, simulations such as {\tt NSC-3pc} and {\tt NSC-9pc} experience more bursty star formation. Moreover, we note that approximately 1\% of the total mass of newly formed stars originates in the nuclear region when an NSC is present (see Table~\ref{tab:runs}). This ongoing star formation contributes to the gradual growth of NSCs, which increase in mass by $1-2\%$ over 200~Myr through in-situ star formation. In contrast, {\tt noNSC} shows no star formation in the vicinity of the BH within $r < 10$~pc, further emphasizing the localized nature of star formation driven by NSCs.

In Section~\ref{sec:warping-disc}, we have analysed a persistent misalignment between the CND and the galactic disc, including a warped CND, as well as a misalignment between CND and the inner accretion disc and BH spin.
This finding is consistent with the widely observed finding that the `spin axis' as inferred from AGN jet orientations, shows misalignment with the angular momentum of gas on an observable ``macroscopic'' scale within galaxies. \citet{Kinney+2000} reported an angular offset between the accretion disc and the galactic disc in a sample of Seyfert galaxies and discussed the possibility that this offset arises from the inflowing gas onto the accretion disc being tilted or from the gravitational potential of an NSC surrounding the BH being misaligned. The warped disc in the CND or accretion disc can be probed using (Mega)Maser observations via VLBI.
For instance, in NGC 4258, a thin gas disc within the sup-parsec region of the nucleus is found to be substantially warped \citep[e.g.,][]{Miyoshi+1995,Baan+2022,Liang+2024}. MUSE observations of M87 reveal a $400~{\rm km}{\rm s}^{-1}$ biconical outflow aligned with the jet, and a sub-arcsecond ($\sim5$~pc) ionized gas disc that is significantly warped. The innermost part of the disc has a position angle nearly perpendicular to the jet axis, indicating strong kinematic misalignment on small scales \citep[e.g.,][]{Jeter+2021, Osorno+2023}. The distinct kinematic behaviour of different stellar generations in nuclear regions is reminiscent of the central $0.5$~pc of Sgr A$^{*}$ region, where two counter-rotating stellar discs of similar ages have been observed \citep[e.g.,][]{Paumard+2006,Bartko+2009, Genzel+2010,Mastrobuono-Battisti+2019}.

\subsection{Caveats and future work}
In this section, we highlight some caveats and future developments for our model. In this study, for the sake of simplicity and to isolate the impact of stellar feedback, we focus on the BH accretion and do not include any BH feedback, such as jets, wind or radiation, which can affect the CND evolution and ISM in the circumnuclear region.  {Such feedback may drive mass-loaded outflows that starve the nucleus of fresh fuel, particularly in the shallow potential wells of dwarf galaxies \citep[e.g.,][]{Koudmani+2022, Sharma+2023}.} In future work, we plan to include BH-driven jet and wind feedback fully coupled to the BH spin to explore this important issue \citep[e.g.,][]{Talbot+2022, Talbot+2024}. 

 {In addition, our simulations do not model cosmological context. However, to somewhat mitigate this caveat, we calibrate the thermodynamical properties of the CGM based on cosmological zoom-in simulations of dwarf galaxies \citep{Koudmani+2022} to mimic the likely CGM inflow rates and model more realistic galactic outflow propagation. We also note that major mergers are extremely rare in such very low mass systems; for instance, \citet{Koudmani+2022} report only one merger with a stellar mass ratio greater than 1:100 at $z=4$ \citep[see also][]{Martin+2021}. None the less, fully cosmological zoom-in simulations of dwarf galaxies with NSC will be crucial to better understand the likely BH fuelling in these systems, which is something we plan to tackle in our future work. Such cosmological simulations, including AGN feedback, will be able to shed light on the possible link between the morphological/kinematical structure of dwarfs (with and without NSCs) and AGN activity \citep[see e.g.][for an observational study]{Kimbrell+2021} as well as on the relationship between the BH mass and NSC mass \citep[see e.g.][]{Greene+2020} across cosmic time.}

Also, our simulations do not account for magnetic fields, which may alter the dynamics of the CND and subsequent BH accretion. For example, \cite{Hopkins+2024a} showed that magnetic fields may exert a strong negative torque on the gas, leading to the formation of a thick and dense CND. 

Another important accretion channel for IMBHs is tidal disruption events (TDEs) of stars, especially in dense stellar systems such as NSC \citep[e.g.,][]{Zubovas+2019, Lee+2023, Rantala+2024, Chang+2025}. Using observational constraints on stellar density profiles in star clusters and BH masses, \cite{Chang+2025} estimated the TDE rate to be $10^{-6}-10^{-4}~{\rm yr}^{-1}$ for BHs with masses in the range of $10^4-10^5~\msun$, which becomes comparable to gas accretion rates in our simulation. Our simulation models the NSC using a single particle mass ($6~M_\odot$) with a softened gravitational force ($\epsilon_{\rm NSC} = 0.175$~pc), which implies that we are unable to properly capture gravitational dynamics on scales smaller than $\epsilon_{\rm NSC}$. Consequently, both the relaxation process and mass segregation in NSCs cannot be resolved within the scope of our current work. To achieve a more accurate assessment of BH growth, we plan to incorporate the TDE channel into our BH accretion modelling and implement direct $N$-body calculations with a more reasonable stellar mass distribution.

In the super-Lagrangian refinement region, we explicitly prevent star formation in gas cells with masses below $0.08 \msun$ in order to avoid the formation of excessively low-mass star particles. However, as shown in Fig.~\ref{fig:SNe-proj}, the Toomre-unstable region is located at $r>0.3$~pc, where larger gas cells are located, and we have confirmed that all gas cells with $m_{\rm cell} < 0.08 \msun$ would have anyway been ineligible for star formation via our standard prescription (i.e. the local Jeans mass was always more than 8 times the cell mass) throughout the entire simulation. 

\section{Conclusions}
\label{sec5:conclusion}
In this study we perform a high-resolution {\it MandelZoom} simulation suite considering an isolated dwarf galaxy at a mean baryonic mass resolution of $20\,\msun$ and peak spatial resolution of $\lesssim0.01$~pc. The dwarf galaxy is composed of a central IMBH with a mass of $10^4\, \msun$, a gaseous and stellar disc with a combined baryonic mass of $\sim 7.8 \times 10^7 \, \msun$, and a DM halo. The total mass of the galaxy is $10^{10}\,\msun$ and it is embedded within an extended CGM.

To model a realistic ISM, we explicitly sample stars according to an IMF and track the individual stellar evolution, incorporating stellar feedback mechanisms such as photo-ionization, photoelectric heating and individual SN explosions. Furthermore, our simulation suite explores dwarf galaxies both with and without NSCs, covering a range of NSC densities. By employing a super-Lagrangian refinement approach, we attain a spatial resolution of $\lesssim0.01$~pc near the BH. This enables us to reliably capture the detailed hydrodynamic and gravitational interactions from the galactic-disc-scale down to the self-gravitating region of the $\alpha$-accretion disc, where we track BH mass and spin evolution assuming the standard Shakura-Sunyaev accretion disc model. The combination of super-Lagrangian refinement, resolved individual stars, and detailed BH accretion modelling allows us to resolve key processes governing angular momentum and mass transfer onto the IMBH, providing critical insights into its long-term growth and thermodynamic state on the surrounding environment. We summarize our main results as follows:
\begin{itemize}
\item Multiphase ISM structure within the galactic disc comprises of cold-dense clumps, filaments, warm medium and hot-diffuse voids, which are shaped by stellar feedback. In the presence of an NSC, the multiphase gas is effectively captured and circularized around the galactic nucleus, forming a cold, dense CND with a radius of $\lesssim7$~pc. The CND has a mass of $10^3-10^4 \msun$, a very high surface density of $\sim10^3\msun~{\rm pc}^{-2}$ and is mainly composed of cold gas with $\lesssim10^2$~K. This CND is highly turbulent, with a Mach number of $\sim20$, due to shocks induced by circumnuclear structures, such as mini-spirals, mini-bar and circumnuclear rings, as well as in-situ star formation.

\item The gas in the CND constantly loses its angular momentum primarily due to the torque exerted by the NSC at $r\sim1$ pc and also due to circumnuclear bars, spirals and rings at $r\lesssim1$ pc. This leads to effective gas circularization onto the $\alpha$-accretion disc and ultimately transfer of mass and angular momentum onto the BH. The BH growth is nonetheless moderate, leading to $\sim 10\%$ increase with respect to the inital BH mass after $200$~Myrs, but it is sustained at about $\sim 0.01$ of the Eddington rate, with peaks up to $0.1$, which correspond to bolometric luminosities of $L_{\rm bol}\sim1.2\times10^{40}-4.2\times10^{41}~{\rm erg}~{\rm s}^{-1}$.

\item When the CND becomes Toomre unstable, young stars form in the CND, and the subsequent SNe heat the CND and halt successive star formation. The star formation episodes last $\lesssim25$~Myr, resulting in multiple stellar generations in the nucleus. Due to its highly dense nature, the CND can withstand a SNe rate of $0.3$~Myr$^{-1}$ with tens of successive SNe explosions within $r\lesssim 10$~pc, which allows the BH to continue accreting gas over $150$~Myr.

\item The CND is misaligned with respect to the galactic disc and changes its rotational axis on a timescale of $\gtrsim 20$~Myrs. Large inflows of fresh ISM gas further lead to CND warping. As stars form within the CND and its rotation axis evolves over time, different stellar generations move within distinct rotational planes and are warped as well.

\item The mass of the $\alpha$-accretion disc fluctuates around $1-2\%$ of the BH mass, due to a balance between inflows from the ISM and BH accretion. For our IMBH mass regime, the accretion disc dominates the angular momentum budget, which results in the alignment of the BH spin to the accretion disc. This allows the BH spin parameter $a_{\bullet}$ to steadily increase due to prolonged coherent accretion in the presence of an NSC. 

\item In the absence of an NSC, the gas fails to form any stable structure in the vicinity of the BH, with low mass flux flows having random angular momentum sporadically reaching the accretion disc. As a result, both the accretion disc and BH accretion are largely insignificant and much reduced compared to the simulations with NSCs. This indicates that NSCs are crucial in understanding IMBH growth in dwarf galaxies.

\item When comparing the $\alpha$-accretion disc model with the traditionally used Bondi model, we find that the latter leads to a very different accretion history and significantly faster BH growth, resulting in a final BH mass that is a factor of $\sim 4$ higher. Since the Bondi model assumes radial symmetry, neglects the angular momentum of accreting gas, and does not differentiate between gas phases, it ultimately leads to a systematic overestimation of BH accretion.
\end{itemize}

Our simulations highlight the need to study BH fuelling within a resolved, multi-phase ISM at sub-pc resolution. We demonstrate how the complex interplay between stellar feedback, NSCs and dense (warped) CNDs can both lead to nuclear in-situ star formation and sustained BH growth. Exploring such setups, where key BH feedback channels are accounted for as well, in full cosmological simulations is becoming possible. It will ultimately allow us to shed light on the build-up of the elusive high redshift black hole population, which is currently one of the most puzzling astrophysical quests.    

\section*{Acknowledgements} 
We would like to thank Antti Rantala, Christian Partmann, Hagai Perets and Massimo Dotti for useful discussions and helpful comments during the development of this manuscript.  {We furthermore thank the referee for constructive comments on the manuscript.} E.S., D.S. and M.A.B. acknowledge support from the Science and Technology Facilities Council (STFC), grant code ST/W000997/1. M.A.B. acknowledges support from a UKRI Stephen Hawking Fellowship (EP/X04257X/1). S.K. has been supported by an 1851 Research Fellowship from the Royal Commission for the Exhibition of 1851 and a Junior Research Fellowship from St Catharine's College, Cambridge. This work was performed using the following facilities: the Cambridge Service for Data Driven Discovery (CSD3) operated by the University of Cambridge Research Computing Service (www.csd3.cam.ac.uk), provided by Dell EMC and Intel using Tier2 funding from the Engineering and Physical Sciences Research Council (capital grant EP/P020259/1), and DiRAC funding from the STFC (www.dirac.ac.uk), as well as the DiRAC@Durham facility operated by the Institute for Computational Cosmology on behalf of the STFC DiRAC HPC Facility (www.dirac.ac.uk). The equipment was funded by BEIS capital funding via STFC capital grants ST/P002293/1, ST/R002371/1 and ST/S002502/1, Durham University, and STFC operations grant ST/R000832/1. DiRAC is part of the National e-Infrastructure.

\section*{Data Availability}

The data underlying this article will be shared on reasonable request to the corresponding author.



\bibliographystyle{mnras}
\bibliography{ref} 

\begin{thebibliography}{}
\makeatletter
\relax
\def\mn@urlcharsother{\let\do\@makeother \do\$\do\&\do\#\do\^\do\_\do\%\do\~}
\def\mn@doi{\begingroup\mn@urlcharsother \@ifnextchar [ {\mn@doi@}
  {\mn@doi@[]}}
\def\mn@doi@[#1]#2{\def\@tempa{#1}\ifx\@tempa\@empty \href
  {http://dx.doi.org/#2} {doi:#2}\else \href {http://dx.doi.org/#2} {#1}\fi
  \endgroup}
\def\mn@eprint#1#2{\mn@eprint@#1:#2::\@nil}
\def\mn@eprint@arXiv#1{\href {http://arxiv.org/abs/#1} {{\tt arXiv:#1}}}
\def\mn@eprint@dblp#1{\href {http://dblp.uni-trier.de/rec/bibtex/#1.xml}
  {dblp:#1}}
\def\mn@eprint@#1:#2:#3:#4\@nil{\def\@tempa {#1}\def\@tempb {#2}\def\@tempc
  {#3}\ifx \@tempc \@empty \let \@tempc \@tempb \let \@tempb \@tempa \fi \ifx
  \@tempb \@empty \def\@tempb {arXiv}\fi \@ifundefined
  {mn@eprint@\@tempb}{\@tempb:\@tempc}{\expandafter \expandafter \csname
  mn@eprint@\@tempb\endcsname \expandafter{\@tempc}}}

\bibitem[\protect\citeauthoryear{{Adamo} et~al.,}{{Adamo}
  et~al.}{2024}]{Adamo+2024}
{Adamo} A.,  et~al., 2024, \mn@doi [\nat] {10.1038/s41586-024-07703-7}, \href
  {https://ui.adsabs.harvard.edu/abs/2024Natur.632..513A} {632, 513}

\bibitem[\protect\citeauthoryear{{Agarwal} \& {Milosavljevi{\'c}}}{{Agarwal} \&
  {Milosavljevi{\'c}}}{2011}]{Agarwal+Milosavljevic2011}
{Agarwal} M.,  {Milosavljevi{\'c}} M.,  2011, \mn@doi [\apj]
  {10.1088/0004-637X/729/1/35}, \href
  {https://ui.adsabs.harvard.edu/abs/2011ApJ...729...35A} {729, 35}

\bibitem[\protect\citeauthoryear{{Antonini}, {Barausse}  \& {Silk}}{{Antonini}
  et~al.}{2015}]{Antonini+2015}
{Antonini} F.,  {Barausse} E.,   {Silk} J.,  2015, \mn@doi [\apj]
  {10.1088/0004-637X/812/1/72}, \href
  {https://ui.adsabs.harvard.edu/abs/2015ApJ...812...72A} {812, 72}

\bibitem[\protect\citeauthoryear{{Arca Sedda}, {Kamlah}, {Spurzem}, {Rizzuto},
  {Naab}, {Giersz}  \& {Berczik}}{{Arca Sedda} et~al.}{2023}]{Arca_Sedda+2023}
{Arca Sedda} M.,  {Kamlah} A. W.~H.,  {Spurzem} R.,  {Rizzuto} F.~P.,  {Naab}
  T.,  {Giersz} M.,   {Berczik} P.,  2023, \mn@doi [\mnras]
  {10.1093/mnras/stad2292}, \href
  {https://ui.adsabs.harvard.edu/abs/2023MNRAS.526..429A} {526, 429}

\bibitem[\protect\citeauthoryear{{Baan}, {An}, {Henkel}, {Imai}, {Kostenko}  \&
  {Sobolev}}{{Baan} et~al.}{2022}]{Baan+2022}
{Baan} W.~A.,  {An} T.,  {Henkel} C.,  {Imai} H.,  {Kostenko} V.,   {Sobolev}
  A.,  2022, \mn@doi [Nature Astronomy] {10.1038/s41550-022-01706-y}, \href
  {https://ui.adsabs.harvard.edu/abs/2022NatAs...6..976B} {6, 976}

\bibitem[\protect\citeauthoryear{{Bardeen} \& {Petterson}}{{Bardeen} \&
  {Petterson}}{1975}]{Bardeen+Petterson1975}
{Bardeen} J.~M.,  {Petterson} J.~A.,  1975, \mn@doi [\apjl] {10.1086/181711},
  \href {https://ui.adsabs.harvard.edu/abs/1975ApJ...195L..65B} {195, L65}

\bibitem[\protect\citeauthoryear{{Barnes} \& {Hernquist}}{{Barnes} \&
  {Hernquist}}{1991}]{Barnes+Hernquist1991}
{Barnes} J.~E.,  {Hernquist} L.~E.,  1991, \mn@doi [\apjl] {10.1086/185978},
  \href {https://ui.adsabs.harvard.edu/abs/1991ApJ...370L..65B} {370, L65}

\bibitem[\protect\citeauthoryear{{Bartko} et~al.,}{{Bartko}
  et~al.}{2009}]{Bartko+2009}
{Bartko} H.,  et~al., 2009, \mn@doi [\apj] {10.1088/0004-637X/697/2/1741},
  \href {https://ui.adsabs.harvard.edu/abs/2009ApJ...697.1741B} {697, 1741}

\bibitem[\protect\citeauthoryear{Beckmann et~al.,}{Beckmann
  et~al.}{2023}]{Beckmann+2023}
Beckmann R.~S.,  et~al., 2023, \mn@doi [\mnras] {10.1093/mnras/stad1795}, \href
  {https://ui.adsabs.harvard.edu/abs/2022arXiv221113614B} {527, 10867}

\bibitem[\protect\citeauthoryear{{Begelman} \& {Rees}}{{Begelman} \&
  {Rees}}{1978}]{Begelman+Rees1978}
{Begelman} M.~C.,  {Rees} M.~J.,  1978, \mn@doi [\mnras]
  {10.1093/mnras/185.4.847}, \href
  {https://ui.adsabs.harvard.edu/abs/1978MNRAS.185..847B} {185, 847}

\bibitem[\protect\citeauthoryear{{Bourne}, {Fiacconi}, {Sijacki}, {Piotrowska}
  \& {Koudmani}}{{Bourne} et~al.}{2024}]{Bourne+2024}
{Bourne} M.~A.,  {Fiacconi} D.,  {Sijacki} D.,  {Piotrowska} J.~M.,
  {Koudmani} S.,  2024, \mn@doi [\mnras] {10.1093/mnras/stae2143}, \href
  {https://ui.adsabs.harvard.edu/abs/2024MNRAS.534.3448B} {534, 3448}

\bibitem[\protect\citeauthoryear{{Bressan}, {Marigo}, {Girardi}, {Salasnich},
  {Dal Cero}, {Rubele}  \& {Nanni}}{{Bressan} et~al.}{2012}]{Bressan+2012}
{Bressan} A.,  {Marigo} P.,  {Girardi} L.,  {Salasnich} B.,  {Dal Cero} C.,
  {Rubele} S.,   {Nanni} A.,  2012, \mn@doi [\mnras]
  {10.1111/j.1365-2966.2012.21948.x}, \href
  {https://ui.adsabs.harvard.edu/abs/2012MNRAS.427..127B} {427, 127}

\bibitem[\protect\citeauthoryear{{Bustamante} \& {Springel}}{{Bustamante} \&
  {Springel}}{2019}]{Bustamante+Springel2019}
{Bustamante} S.,  {Springel} V.,  2019, \mn@doi [\mnras]
  {10.1093/mnras/stz2836}, \href
  {https://ui.adsabs.harvard.edu/abs/2019MNRAS.490.4133B} {490, 4133}

\bibitem[\protect\citeauthoryear{{Bykov}, {Gilfanov}  \& {Sunyaev}}{{Bykov}
  et~al.}{2024}]{Bykov+2024}
{Bykov} S.~D.,  {Gilfanov} M.~R.,   {Sunyaev} R.~A.,  2024, \mn@doi [\mnras]
  {10.1093/mnras/stad3355}, \href
  {https://ui.adsabs.harvard.edu/abs/2024MNRAS.527.1962B} {527, 1962}

\bibitem[\protect\citeauthoryear{{Campanelli}, {Lousto}, {Zlochower}  \&
  {Merritt}}{{Campanelli} et~al.}{2007}]{Campanelli+2007}
{Campanelli} M.,  {Lousto} C.~O.,  {Zlochower} Y.,   {Merritt} D.,  2007,
  \mn@doi [\prl] {10.1103/PhysRevLett.98.231102}, \href
  {https://ui.adsabs.harvard.edu/abs/2007PhRvL..98w1102C} {98, 231102}

\bibitem[\protect\citeauthoryear{{Capuzzo-Dolcetta}}{{Capuzzo-Dolcetta}}{1993}]{Capuzzo-Dolcetta1993}
{Capuzzo-Dolcetta} R.,  1993, \mn@doi [\apj] {10.1086/173189}, \href
  {https://ui.adsabs.harvard.edu/abs/1993ApJ...415..616C} {415, 616}

\bibitem[\protect\citeauthoryear{{Capuzzo-Dolcetta} \&
  {Miocchi}}{{Capuzzo-Dolcetta} \&
  {Miocchi}}{2008}]{Capuzzo-Dolcetta+Miocchi2008}
{Capuzzo-Dolcetta} R.,  {Miocchi} P.,  2008, \mn@doi [\apj] {10.1086/588017},
  \href {https://ui.adsabs.harvard.edu/abs/2008ApJ...681.1136C} {681, 1136}

\bibitem[\protect\citeauthoryear{{Carlsten}, {Greene}, {Beaton}  \&
  {Greco}}{{Carlsten} et~al.}{2022}]{Carlsten+2022}
{Carlsten} S.~G.,  {Greene} J.~E.,  {Beaton} R.~L.,   {Greco} J.~P.,  2022,
  \mn@doi [\apj] {10.3847/1538-4357/ac457e}, \href
  {https://ui.adsabs.harvard.edu/abs/2022ApJ...927...44C} {927, 44}

\bibitem[\protect\citeauthoryear{{Chang}, {Dai}, {Pfister}, {Kar Chowdhury}  \&
  {Natarajan}}{{Chang} et~al.}{2025}]{Chang+2025}
{Chang} J. N.~Y.,  {Dai} L.,  {Pfister} H.,  {Kar Chowdhury} R.,   {Natarajan}
  P.,  2025, \mn@doi [\apjl] {10.3847/2041-8213/adace7}, \href
  {https://ui.adsabs.harvard.edu/abs/2025ApJ...980L..22C} {980, L22}

\bibitem[\protect\citeauthoryear{{Chieffi} \& {Limongi}}{{Chieffi} \&
  {Limongi}}{2004}]{Chieffi+2004}
{Chieffi} A.,  {Limongi} M.,  2004, \mn@doi [\apj] {10.1086/392523}, \href
  {https://ui.adsabs.harvard.edu/abs/2004ApJ...608..405C} {608, 405}

\bibitem[\protect\citeauthoryear{{Cui} et~al.,}{{Cui} et~al.}{2023}]{Cui+2023}
{Cui} Y.,  et~al., 2023, \mn@doi [\nat] {10.1038/s41586-023-06479-6}, \href
  {https://ui.adsabs.harvard.edu/abs/2023Natur.621..711C} {621, 711}

\bibitem[\protect\citeauthoryear{{Curtis} \& {Sijacki}}{{Curtis} \&
  {Sijacki}}{2015}]{Curtis+Sijacki2015}
{Curtis} M.,  {Sijacki} D.,  2015, \mn@doi [\mnras] {10.1093/mnras/stv2246},
  \href {https://ui.adsabs.harvard.edu/abs/2015MNRAS.454.3445C} {454, 3445}

\bibitem[\protect\citeauthoryear{{Curtis} \& {Sijacki}}{{Curtis} \&
  {Sijacki}}{2016}]{Curtis+Sijacki2016}
{Curtis} M.,  {Sijacki} D.,  2016, \mn@doi [\mnras] {10.1093/mnras/stw1944},
  \href {https://ui.adsabs.harvard.edu/abs/2016MNRAS.463...63C} {463, 63}

\bibitem[\protect\citeauthoryear{{Davis} et~al.,}{{Davis}
  et~al.}{2020}]{Davis+2020}
{Davis} T.~A.,  et~al., 2020, \mn@doi [\mnras] {10.1093/mnras/staa1567}, \href
  {https://ui.adsabs.harvard.edu/abs/2020MNRAS.496.4061D} {496, 4061}

\bibitem[\protect\citeauthoryear{{Davis}, {Graham}, {Soria}, {Jin},
  {Karachentsev}, {Karachentseva}  \& {D'Onghia}}{{Davis}
  et~al.}{2024}]{Davis+2024}
{Davis} B.~L.,  {Graham} A.~W.,  {Soria} R.,  {Jin} Z.,  {Karachentsev} I.~D.,
  {Karachentseva} V.~E.,   {D'Onghia} E.,  2024, \mn@doi [\apj]
  {10.3847/1538-4357/ad55eb}, \href
  {https://ui.adsabs.harvard.edu/abs/2024ApJ...971..123D} {971, 123}

\bibitem[\protect\citeauthoryear{{Despali}, {Giocoli}  \& {Tormen}}{{Despali}
  et~al.}{2014}]{Despali+2014}
{Despali} G.,  {Giocoli} C.,   {Tormen} G.,  2014, \mn@doi [\mnras]
  {10.1093/mnras/stu1393}, \href
  {https://ui.adsabs.harvard.edu/abs/2014MNRAS.443.3208D} {443, 3208}

\bibitem[\protect\citeauthoryear{{Dinh}, {Salas}, {Morris}  \& {Naoz}}{{Dinh}
  et~al.}{2021}]{Dinh+2021}
{Dinh} C.~K.,  {Salas} J.~M.,  {Morris} M.~R.,   {Naoz} S.,  2021, \mn@doi
  [\apj] {10.3847/1538-4357/ac185b}, \href
  {https://ui.adsabs.harvard.edu/abs/2021ApJ...920...79D} {920, 79}

\bibitem[\protect\citeauthoryear{{Dong-P{\'a}ez}, {Volonteri}, {Beckmann},
  {Dubois}, {Trebitsch}, {Mangiagli}, {Vergani}  \& {Webb}}{{Dong-P{\'a}ez}
  et~al.}{2023}]{Dong-Paez+2023}
{Dong-P{\'a}ez} C.~A.,  {Volonteri} M.,  {Beckmann} R.~S.,  {Dubois} Y.,
  {Trebitsch} M.,  {Mangiagli} A.,  {Vergani} S.~D.,   {Webb} N.~A.,  2023,
  \mn@doi [\aap] {10.1051/0004-6361/202346295}, \href
  {https://ui.adsabs.harvard.edu/abs/2023A&A...673A.120D} {673, A120}

\bibitem[\protect\citeauthoryear{{Dotti}, {Colpi}, {Pallini}, {Perego}  \&
  {Volonteri}}{{Dotti} et~al.}{2013}]{Dotti+2013}
{Dotti} M.,  {Colpi} M.,  {Pallini} S.,  {Perego} A.,   {Volonteri} M.,  2013,
  \mn@doi [\apj] {10.1088/0004-637X/762/2/68}, \href
  {https://ui.adsabs.harvard.edu/abs/2013ApJ...762...68D} {762, 68}

\bibitem[\protect\citeauthoryear{{Dubois}, {Volonteri}  \& {Silk}}{{Dubois}
  et~al.}{2014}]{Dubois+2014}
{Dubois} Y.,  {Volonteri} M.,   {Silk} J.,  2014, \mn@doi [\mnras]
  {10.1093/mnras/stu373}, \href
  {https://ui.adsabs.harvard.edu/abs/2014MNRAS.440.1590D} {440, 1590}

\bibitem[\protect\citeauthoryear{{Dubois} et~al.,}{{Dubois}
  et~al.}{2021}]{Dubois+2021}
{Dubois} Y.,  et~al., 2021, \mn@doi [\aap] {10.1051/0004-6361/202039429}, \href
  {https://ui.adsabs.harvard.edu/abs/2021A&A...651A.109D} {651, A109}

\bibitem[\protect\citeauthoryear{{Emerick}, {Bryan}  \& {Mac Low}}{{Emerick}
  et~al.}{2019}]{Emerick+2019}
{Emerick} A.,  {Bryan} G.~L.,   {Mac Low} M.-M.,  2019, \mn@doi [\mnras]
  {10.1093/mnras/sty2689}, \href
  {https://ui.adsabs.harvard.edu/abs/2019MNRAS.482.1304E} {482, 1304}

\bibitem[\protect\citeauthoryear{{Emsellem}, {Renaud}, {Bournaud}, {Elmegreen},
  {Combes}  \& {Gabor}}{{Emsellem} et~al.}{2015}]{Emsellem+2015}
{Emsellem} E.,  {Renaud} F.,  {Bournaud} F.,  {Elmegreen} B.,  {Combes} F.,
  {Gabor} J.~M.,  2015, \mn@doi [\mnras] {10.1093/mnras/stu2209}, \href
  {https://ui.adsabs.harvard.edu/abs/2015MNRAS.446.2468E} {446, 2468}

\bibitem[\protect\citeauthoryear{{Escala}}{{Escala}}{2007}]{Escala2007}
{Escala} A.,  2007, \mn@doi [\apj] {10.1086/523092}, \href
  {https://ui.adsabs.harvard.edu/abs/2007ApJ...671.1264E} {671, 1264}

\bibitem[\protect\citeauthoryear{{Fahrion} et~al.,}{{Fahrion}
  et~al.}{2021}]{Fahrion+2021}
{Fahrion} K.,  et~al., 2021, \mn@doi [\aap] {10.1051/0004-6361/202140644},
  \href {https://ui.adsabs.harvard.edu/abs/2021A&A...650A.137F} {650, A137}

\bibitem[\protect\citeauthoryear{{Fahrion} et~al.,}{{Fahrion}
  et~al.}{2022}]{Fahrion+2022}
{Fahrion} K.,  et~al., 2022, \mn@doi [\aap] {10.1051/0004-6361/202244932},
  \href {https://ui.adsabs.harvard.edu/abs/2022A&A...667A.101F} {667, A101}

\bibitem[\protect\citeauthoryear{{Fahrion} et~al.,}{{Fahrion}
  et~al.}{2024}]{Fahrion+2024}
{Fahrion} K.,  et~al., 2024, \mn@doi [\aap] {10.1051/0004-6361/202449629},
  \href {https://ui.adsabs.harvard.edu/abs/2024A&A...687A..83F} {687, A83}

\bibitem[\protect\citeauthoryear{{Farrell}, {Webb}, {Barret}, {Godet}  \&
  {Rodrigues}}{{Farrell} et~al.}{2009}]{Farrell+2009}
{Farrell} S.~A.,  {Webb} N.~A.,  {Barret} D.,  {Godet} O.,   {Rodrigues} J.~M.,
   2009, \mn@doi [\nat] {10.1038/nature08083}, \href
  {https://ui.adsabs.harvard.edu/abs/2009Natur.460...73F} {460, 73}

\bibitem[\protect\citeauthoryear{{Fedrigo}, {Cattorini}, {Giacomazzo}  \&
  {Colpi}}{{Fedrigo} et~al.}{2024}]{Fedrigo+2024}
{Fedrigo} G.,  {Cattorini} F.,  {Giacomazzo} B.,   {Colpi} M.,  2024, \mn@doi
  [\prd] {10.1103/PhysRevD.109.103024}, \href
  {https://ui.adsabs.harvard.edu/abs/2024PhRvD.109j3024F} {109, 103024}

\bibitem[\protect\citeauthoryear{{Ferland} et~al.,}{{Ferland}
  et~al.}{2013}]{Ferland+2013}
{Ferland} G.~J.,  et~al., 2013, \mn@doi [\rmxaa] {10.48550/arXiv.1302.4485},
  \href {https://ui.adsabs.harvard.edu/abs/2013RMxAA..49..137F} {49, 137}

\bibitem[\protect\citeauthoryear{{Ferrarese} \& {Merritt}}{{Ferrarese} \&
  {Merritt}}{2000}]{Ferrarese+Merritt2000}
{Ferrarese} L.,  {Merritt} D.,  2000, \mn@doi [\apjl] {10.1086/312838}, \href
  {https://ui.adsabs.harvard.edu/abs/2000ApJ...539L...9F} {539, L9}

\bibitem[\protect\citeauthoryear{{Fiacconi}, {Sijacki}  \&
  {Pringle}}{{Fiacconi} et~al.}{2018}]{Fiacconi+2018}
{Fiacconi} D.,  {Sijacki} D.,   {Pringle} J.~E.,  2018, \mn@doi [\mnras]
  {10.1093/mnras/sty893}, \href
  {https://ui.adsabs.harvard.edu/abs/2018MNRAS.477.3807F} {477, 3807}

\bibitem[\protect\citeauthoryear{{Freitag}, {Rasio}  \& {Baumgardt}}{{Freitag}
  et~al.}{2006}]{Freitag+2006}
{Freitag} M.,  {Rasio} F.~A.,   {Baumgardt} H.,  2006, \mn@doi [\mnras]
  {10.1111/j.1365-2966.2006.10095.x}, \href
  {https://ui.adsabs.harvard.edu/abs/2006MNRAS.368..121F} {368, 121}

\bibitem[\protect\citeauthoryear{{Fujii}, {Wang}, {Tanikawa}, {Hirai}  \&
  {Saitoh}}{{Fujii} et~al.}{2024}]{Fujii+2024}
{Fujii} M.~S.,  {Wang} L.,  {Tanikawa} A.,  {Hirai} Y.,   {Saitoh} T.~R.,
  2024, \mn@doi [Science] {10.1126/science.adi4211}, \href
  {https://ui.adsabs.harvard.edu/abs/2024Sci...384.1488F} {384, 1488}

\bibitem[\protect\citeauthoryear{{Fujimoto} et~al.,}{{Fujimoto}
  et~al.}{2024}]{Fujimoto+2024}
{Fujimoto} S.,  et~al., 2024, \mn@doi [Nature Astronomy in press]
  {10.48550/arXiv.2402.18543}, \href
  {https://ui.adsabs.harvard.edu/abs/2024arXiv240218543F} {p. arXiv:2402.18543}

\bibitem[\protect\citeauthoryear{{Genzel}, {Eisenhauer}  \&
  {Gillessen}}{{Genzel} et~al.}{2010}]{Genzel+2010}
{Genzel} R.,  {Eisenhauer} F.,   {Gillessen} S.,  2010, \mn@doi [Reviews of
  Modern Physics] {10.1103/RevModPhys.82.3121}, \href
  {https://ui.adsabs.harvard.edu/abs/2010RvMP...82.3121G} {82, 3121}

\bibitem[\protect\citeauthoryear{{Georgiev}, {Hilker}, {Puzia}, {Goudfrooij}
  \& {Baumgardt}}{{Georgiev} et~al.}{2009}]{Georgiev+2009b}
{Georgiev} I.~Y.,  {Hilker} M.,  {Puzia} T.~H.,  {Goudfrooij} P.,   {Baumgardt}
  H.,  2009, \mn@doi [\mnras] {10.1111/j.1365-2966.2009.14776.x}, \href
  {https://ui.adsabs.harvard.edu/abs/2009MNRAS.396.1075G} {396, 1075}

\bibitem[\protect\citeauthoryear{{Georgiev}, {B{\"o}ker}, {Leigh},
  {L{\"u}tzgendorf}  \& {Neumayer}}{{Georgiev} et~al.}{2016}]{Georgiev+2016}
{Georgiev} I.~Y.,  {B{\"o}ker} T.,  {Leigh} N.,  {L{\"u}tzgendorf} N.,
  {Neumayer} N.,  2016, \mn@doi [\mnras] {10.1093/mnras/stw093}, \href
  {https://ui.adsabs.harvard.edu/abs/2016MNRAS.457.2122G} {457, 2122}

\bibitem[\protect\citeauthoryear{{Gerosa}, {Veronesi}, {Lodato}  \&
  {Rosotti}}{{Gerosa} et~al.}{2015}]{Gerosa+15}
{Gerosa} D.,  {Veronesi} B.,  {Lodato} G.,   {Rosotti} G.,  2015, \mn@doi
  [\mnras] {10.1093/mnras/stv1214}, \href
  {https://ui.adsabs.harvard.edu/abs/2015MNRAS.451.3941G} {451, 3941}

\bibitem[\protect\citeauthoryear{{Gnedin}, {Ostriker}  \& {Tremaine}}{{Gnedin}
  et~al.}{2014}]{Gnedin+2014}
{Gnedin} O.~Y.,  {Ostriker} J.~P.,   {Tremaine} S.,  2014, \mn@doi [\apj]
  {10.1088/0004-637X/785/1/71}, \href
  {https://ui.adsabs.harvard.edu/abs/2014ApJ...785...71G} {785, 71}

\bibitem[\protect\citeauthoryear{{Gonz{\'a}lez Prieto}, {Weatherford},
  {Fragione}, {Kremer}  \& {Rasio}}{{Gonz{\'a}lez Prieto}
  et~al.}{2024}]{Gonzalez-Prieto+2024}
{Gonz{\'a}lez Prieto} E.,  {Weatherford} N.~C.,  {Fragione} G.,  {Kremer} K.,
  {Rasio} F.~A.,  2024, \mn@doi [\apj] {10.3847/1538-4357/ad43d6}, \href
  {https://ui.adsabs.harvard.edu/abs/2024ApJ...969...29G} {969, 29}

\bibitem[\protect\citeauthoryear{{Greene} \& {Ho}}{{Greene} \&
  {Ho}}{2007}]{Greene+Ho2007}
{Greene} J.~E.,  {Ho} L.~C.,  2007, \mn@doi [\apj] {10.1086/522082}, \href
  {https://ui.adsabs.harvard.edu/abs/2007ApJ...670...92G} {670, 92}

\bibitem[\protect\citeauthoryear{{Greene}, {Strader}  \& {Ho}}{{Greene}
  et~al.}{2020}]{Greene+2020}
{Greene} J.~E.,  {Strader} J.,   {Ho} L.~C.,  2020, \mn@doi [\araa]
  {10.1146/annurev-astro-032620-021835}, \href
  {https://ui.adsabs.harvard.edu/abs/2020ARA&A..58..257G} {58, 257}

\bibitem[\protect\citeauthoryear{{G{\"u}ltekin} et~al.,}{{G{\"u}ltekin}
  et~al.}{2022}]{Gultekin+2022}
{G{\"u}ltekin} K.,  et~al., 2022, \mn@doi [\mnras] {10.1093/mnras/stac2608},
  \href {https://ui.adsabs.harvard.edu/abs/2022MNRAS.516.6123G} {516, 6123}

\bibitem[\protect\citeauthoryear{{Haardt} \& {Madau}}{{Haardt} \&
  {Madau}}{2012}]{Haardt+Madau2012}
{Haardt} F.,  {Madau} P.,  2012, \mn@doi [\apj] {10.1088/0004-637X/746/2/125},
  \href {https://ui.adsabs.harvard.edu/abs/2012ApJ...746..125H} {746, 125}

\bibitem[\protect\citeauthoryear{{H{\"a}berle} et~al.,}{{H{\"a}berle}
  et~al.}{2024}]{Haberle+2024}
{H{\"a}berle} M.,  et~al., 2024, \mn@doi [\nat] {10.1038/s41586-024-07511-z},
  \href {https://ui.adsabs.harvard.edu/abs/2024Natur.631..285H} {631, 285}

\bibitem[\protect\citeauthoryear{{H{\"a}ring} \& {Rix}}{{H{\"a}ring} \&
  {Rix}}{2004}]{Haring+Rix2004}
{H{\"a}ring} N.,  {Rix} H.-W.,  2004, \mn@doi [\apjl] {10.1086/383567}, \href
  {https://ui.adsabs.harvard.edu/abs/2004ApJ...604L..89H} {604, L89}

\bibitem[\protect\citeauthoryear{{Hernquist}}{{Hernquist}}{1989}]{Hernquist1989}
{Hernquist} L.,  1989, \mn@doi [\nat] {10.1038/340687a0}, \href
  {https://ui.adsabs.harvard.edu/abs/1989Natur.340..687H} {340, 687}

\bibitem[\protect\citeauthoryear{{Hernquist}}{{Hernquist}}{1990}]{Hernquist1990}
{Hernquist} L.,  1990, \mn@doi [\apj] {10.1086/168845}, \href
  {https://ui.adsabs.harvard.edu/abs/1990ApJ...356..359H} {356, 359}

\bibitem[\protect\citeauthoryear{{Hopkins} \& {Quataert}}{{Hopkins} \&
  {Quataert}}{2010}]{Hopkins+Quataert2010}
{Hopkins} P.~F.,  {Quataert} E.,  2010, \mn@doi [\mnras]
  {10.1111/j.1365-2966.2010.17064.x}, \href
  {https://ui.adsabs.harvard.edu/abs/2010MNRAS.407.1529H} {407, 1529}

\bibitem[\protect\citeauthoryear{{Hopkins} \& {Quataert}}{{Hopkins} \&
  {Quataert}}{2011}]{Hopkins+Quataert2011}
{Hopkins} P.~F.,  {Quataert} E.,  2011, \mn@doi [\mnras]
  {10.1111/j.1365-2966.2011.18542.x}, \href
  {https://ui.adsabs.harvard.edu/abs/2011MNRAS.415.1027H} {415, 1027}

\bibitem[\protect\citeauthoryear{{Hopkins} et~al.,}{{Hopkins}
  et~al.}{2024}]{Hopkins+2024a}
{Hopkins} P.~F.,  et~al., 2024, \mn@doi [The Open Journal of Astrophysics]
  {10.21105/astro.2310.04506}, \href
  {https://ui.adsabs.harvard.edu/abs/2024OJAp....7E..19H} {7, 19}

\bibitem[\protect\citeauthoryear{{Hoyer}, {Neumayer}, {Georgiev}, {Seth}  \&
  {Greene}}{{Hoyer} et~al.}{2021}]{Hoyer+2021}
{Hoyer} N.,  {Neumayer} N.,  {Georgiev} I.~Y.,  {Seth} A.~C.,   {Greene} J.~E.,
   2021, \mn@doi [\mnras] {10.1093/mnras/stab2277}, \href
  {https://ui.adsabs.harvard.edu/abs/2021MNRAS.507.3246H} {507, 3246}

\bibitem[\protect\citeauthoryear{{Hoyer}, {Neumayer}, {Seth}, {Georgiev}  \&
  {Greene}}{{Hoyer} et~al.}{2023}]{Hoyer+2023}
{Hoyer} N.,  {Neumayer} N.,  {Seth} A.~C.,  {Georgiev} I.~Y.,   {Greene} J.~E.,
   2023, \mn@doi [\mnras] {10.1093/mnras/stad220}, \href
  {https://ui.adsabs.harvard.edu/abs/2023MNRAS.520.4664H} {520, 4664}

\bibitem[\protect\citeauthoryear{{Hu{\v{s}}ko}, {Lacey}, {Schaye}, {Nobels}  \&
  {Schaller}}{{Hu{\v{s}}ko} et~al.}{2024}]{Husko+2024}
{Hu{\v{s}}ko} F.,  {Lacey} C.~G.,  {Schaye} J.,  {Nobels} F. S.~J.,
  {Schaller} M.,  2024, \mn@doi [\mnras] {10.1093/mnras/stad3548}, \href
  {https://ui.adsabs.harvard.edu/abs/2024MNRAS.527.5988H} {527, 5988}

\bibitem[\protect\citeauthoryear{{Jeter} \& {Broderick}}{{Jeter} \&
  {Broderick}}{2021}]{Jeter+2021}
{Jeter} B.,  {Broderick} A.~E.,  2021, \mn@doi [\apj]
  {10.3847/1538-4357/abda3d}, \href
  {https://ui.adsabs.harvard.edu/abs/2021ApJ...908..139J} {908, 139}

\bibitem[\protect\citeauthoryear{{Katz}, {Sijacki}  \& {Haehnelt}}{{Katz}
  et~al.}{2015}]{Katz+2015}
{Katz} H.,  {Sijacki} D.,   {Haehnelt} M.~G.,  2015, \mn@doi [\mnras]
  {10.1093/mnras/stv1048}, \href
  {https://ui.adsabs.harvard.edu/abs/2015MNRAS.451.2352K} {451, 2352}

\bibitem[\protect\citeauthoryear{{Kim}, {L{\'o}pez}, {Jonker}, {Ho}  \&
  {Im}}{{Kim} et~al.}{2020}]{Kim+2020}
{Kim} M.,  {L{\'o}pez} K.~M.,  {Jonker} P.~G.,  {Ho} L.~C.,   {Im} M.,  2020,
  \mn@doi [\mnras] {10.1093/mnrasl/slaa011}, \href
  {https://ui.adsabs.harvard.edu/abs/2020MNRAS.493L..76K} {493, L76}

\bibitem[\protect\citeauthoryear{{Kimbrell}, {Reines}, {Schutte}, {Greene}  \&
  {Geha}}{{Kimbrell} et~al.}{2021}]{Kimbrell+2021}
{Kimbrell} S.~J.,  {Reines} A.~E.,  {Schutte} Z.,  {Greene} J.~E.,   {Geha} M.,
   2021, \mn@doi [\apj] {10.3847/1538-4357/abec40}, \href
  {https://ui.adsabs.harvard.edu/abs/2021ApJ...911..134K} {911, 134}

\bibitem[\protect\citeauthoryear{{King}}{{King}}{1966}]{King+1966}
{King} I.~R.,  1966, \mn@doi [\aj] {10.1086/109857}, \href
  {https://ui.adsabs.harvard.edu/abs/1966AJ.....71...64K} {71, 64}

\bibitem[\protect\citeauthoryear{{King} \& {Pringle}}{{King} \&
  {Pringle}}{2006}]{King+Pringle2006}
{King} A.~R.,  {Pringle} J.~E.,  2006, \mn@doi [\mnras]
  {10.1111/j.1745-3933.2006.00249.x}, \href
  {https://ui.adsabs.harvard.edu/abs/2006MNRAS.373L..90K} {373, L90}

\bibitem[\protect\citeauthoryear{{King}, {Lubow}, {Ogilvie}  \&
  {Pringle}}{{King} et~al.}{2005}]{King+2005}
{King} A.~R.,  {Lubow} S.~H.,  {Ogilvie} G.~I.,   {Pringle} J.~E.,  2005,
  \mn@doi [\mnras] {10.1111/j.1365-2966.2005.09378.x}, \href
  {https://ui.adsabs.harvard.edu/abs/2005MNRAS.363...49K} {363, 49}

\bibitem[\protect\citeauthoryear{{Kinney}, {Schmitt}, {Clarke}, {Pringle},
  {Ulvestad}  \& {Antonucci}}{{Kinney} et~al.}{2000}]{Kinney+2000}
{Kinney} A.~L.,  {Schmitt} H.~R.,  {Clarke} C.~J.,  {Pringle} J.~E.,
  {Ulvestad} J.~S.,   {Antonucci} R.~R.~J.,  2000, \mn@doi [\apj]
  {10.1086/309016}, \href
  {https://ui.adsabs.harvard.edu/abs/2000ApJ...537..152K} {537, 152}

\bibitem[\protect\citeauthoryear{{Kormendy} \& {Ho}}{{Kormendy} \&
  {Ho}}{2013}]{Kormendy+Ho2013}
{Kormendy} J.,  {Ho} L.~C.,  2013, \mn@doi [\araa]
  {10.1146/annurev-astro-082708-101811}, \href
  {https://ui.adsabs.harvard.edu/abs/2013ARA&A..51..511K} {51, 511}

\bibitem[\protect\citeauthoryear{{Koudmani}, {Sijacki}  \& {Smith}}{{Koudmani}
  et~al.}{2022}]{Koudmani+2022}
{Koudmani} S.,  {Sijacki} D.,   {Smith} M.~C.,  2022, \mn@doi [\mnras]
  {10.1093/mnras/stac2252}, \href
  {https://ui.adsabs.harvard.edu/abs/2022MNRAS.516.2112K} {516, 2112}

\bibitem[\protect\citeauthoryear{{Koudmani}, {Somerville}, {Sijacki}, {Bourne},
  {Jiang}  \& {Profit}}{{Koudmani} et~al.}{2024}]{Koudmani+2024}
{Koudmani} S.,  {Somerville} R.~S.,  {Sijacki} D.,  {Bourne} M.~A.,  {Jiang}
  Y.-F.,   {Profit} K.,  2024, \mn@doi [\mnras] {10.1093/mnras/stae1422}, \href
  {https://ui.adsabs.harvard.edu/abs/2024MNRAS.532...60K} {532, 60}

\bibitem[\protect\citeauthoryear{{Kroupa}}{{Kroupa}}{2001}]{Kroupa2001}
{Kroupa} P.,  2001, \mn@doi [\mnras] {10.1046/j.1365-8711.2001.04022.x}, \href
  {https://ui.adsabs.harvard.edu/abs/2001MNRAS.322..231K} {322, 231}

\bibitem[\protect\citeauthoryear{{Krumholz} \& {Tan}}{{Krumholz} \&
  {Tan}}{2007}]{Krumholz+Tan2007}
{Krumholz} M.~R.,  {Tan} J.~C.,  2007, \mn@doi [\apj] {10.1086/509101}, \href
  {https://ui.adsabs.harvard.edu/abs/2007ApJ...654..304K} {654, 304}

\bibitem[\protect\citeauthoryear{{Laine}, {Shlosman}, {Knapen}  \&
  {Peletier}}{{Laine} et~al.}{2002}]{Laine+2002}
{Laine} S.,  {Shlosman} I.,  {Knapen} J.~H.,   {Peletier} R.~F.,  2002, \mn@doi
  [\apj] {10.1086/323964}, \href
  {https://ui.adsabs.harvard.edu/abs/2002ApJ...567...97L} {567, 97}

\bibitem[\protect\citeauthoryear{{Lanz} \& {Hubeny}}{{Lanz} \&
  {Hubeny}}{2003}]{Lanz+2003}
{Lanz} T.,  {Hubeny} I.,  2003, \mn@doi [\apjs] {10.1086/374373}, \href
  {https://ui.adsabs.harvard.edu/abs/2003ApJS..146..417L} {146, 417}

\bibitem[\protect\citeauthoryear{{Lee}, {Kim}  \& {Oh}}{{Lee}
  et~al.}{2023}]{Lee+2023}
{Lee} S.,  {Kim} J.-h.,   {Oh} B.~K.,  2023, \mn@doi [\apj]
  {10.3847/1538-4357/aca75d}, \href
  {https://ui.adsabs.harvard.edu/abs/2023ApJ...943...77L} {943, 77}

\bibitem[\protect\citeauthoryear{{Levine}, {Gnedin}  \& {Hamilton}}{{Levine}
  et~al.}{2010}]{Levine+2010}
{Levine} R.,  {Gnedin} N.~Y.,   {Hamilton} A. J.~S.,  2010, \mn@doi [\apj]
  {10.1088/0004-637X/716/2/1386}, \href
  {https://ui.adsabs.harvard.edu/abs/2010ApJ...716.1386L} {716, 1386}

\bibitem[\protect\citeauthoryear{{Liang} et~al.,}{{Liang}
  et~al.}{2024}]{Liang+2024}
{Liang} F.-H.,  et~al., 2024, \mn@doi [\mnras] {10.1093/mnras/stad3675}, \href
  {https://ui.adsabs.harvard.edu/abs/2024MNRAS.527.9343L} {527, 9343}

\bibitem[\protect\citeauthoryear{{Lodato} \& {Pringle}}{{Lodato} \&
  {Pringle}}{2006}]{Lodato+Pringle2006}
{Lodato} G.,  {Pringle} J.~E.,  2006, \mn@doi [\mnras]
  {10.1111/j.1365-2966.2006.10194.x}, \href
  {https://ui.adsabs.harvard.edu/abs/2006MNRAS.368.1196L} {368, 1196}

\bibitem[\protect\citeauthoryear{{Loose}, {Kruegel}  \& {Tutukov}}{{Loose}
  et~al.}{1982}]{Loose+1982}
{Loose} H.~H.,  {Kruegel} E.,   {Tutukov} A.,  1982, \aap, \href
  {https://ui.adsabs.harvard.edu/abs/1982A&A...105..342L} {105, 342}

\bibitem[\protect\citeauthoryear{{Lupi}, {Haardt}, {Dotti}, {Fiacconi}, {Mayer}
   \& {Madau}}{{Lupi} et~al.}{2016}]{Lupi+2016}
{Lupi} A.,  {Haardt} F.,  {Dotti} M.,  {Fiacconi} D.,  {Mayer} L.,   {Madau}
  P.,  2016, \mn@doi [\mnras] {10.1093/mnras/stv2877}, \href
  {https://ui.adsabs.harvard.edu/abs/2016MNRAS.456.2993L} {456, 2993}

\bibitem[\protect\citeauthoryear{{Lyu}, {Zhang}, {Paudel}, {Li}, {Tang},
  {Chen}, {Kong}  \& {Peng}}{{Lyu} et~al.}{2025}]{lyu+2025}
{Lyu} W.,  {Zhang} H.-X.,  {Paudel} S.,  {Li} T.,  {Tang} Y.,  {Chen} G.,
  {Kong} X.,   {Peng} E.~W.,  2025, \mn@doi [\apj] {10.3847/1538-4357/ad9b0d},
  \href {https://ui.adsabs.harvard.edu/abs/2024arXiv241203132L} {979, 85}

\bibitem[\protect\citeauthoryear{{Lyubenova} \& {Tsatsi}}{{Lyubenova} \&
  {Tsatsi}}{2019}]{Lyubenova+Tsatsi2019}
{Lyubenova} M.,  {Tsatsi} A.,  2019, \mn@doi [\aap]
  {10.1051/0004-6361/201833954}, \href
  {https://ui.adsabs.harvard.edu/abs/2019A&A...629A..44L} {629, A44}

\bibitem[\protect\citeauthoryear{{Maciejewski}}{{Maciejewski}}{2004}]{Maciejewski2004}
{Maciejewski} W.,  2004, \mn@doi [\mnras] {10.1111/j.1365-2966.2004.08254.x},
  \href {https://ui.adsabs.harvard.edu/abs/2004MNRAS.354..892M} {354, 892}

\bibitem[\protect\citeauthoryear{{Magorrian} et~al.,}{{Magorrian}
  et~al.}{1998}]{Magorrian+1998}
{Magorrian} J.,  et~al., 1998, \mn@doi [\aj] {10.1086/300353}, \href
  {https://ui.adsabs.harvard.edu/abs/1998AJ....115.2285M} {115, 2285}

\bibitem[\protect\citeauthoryear{{Maiolino} et~al.,}{{Maiolino}
  et~al.}{2024}]{Maiolino+2024}
{Maiolino} R.,  et~al., 2024, \mn@doi [\nat] {10.1038/s41586-024-07052-5},
  \href {https://ui.adsabs.harvard.edu/abs/2024Natur.627...59M} {627, 59}

\bibitem[\protect\citeauthoryear{{Martin} et~al.,}{{Martin}
  et~al.}{2021}]{Martin+2021}
{Martin} G.,  et~al., 2021, \mn@doi [\mnras] {10.1093/mnras/staa3443}, \href
  {https://ui.adsabs.harvard.edu/abs/2021MNRAS.500.4937M} {500, 4937}

\bibitem[\protect\citeauthoryear{{Mastrobuono-Battisti}, {Perets},
  {Gualandris}, {Neumayer}  \& {Sippel}}{{Mastrobuono-Battisti}
  et~al.}{2019}]{Mastrobuono-Battisti+2019}
{Mastrobuono-Battisti} A.,  {Perets} H.~B.,  {Gualandris} A.,  {Neumayer} N.,
  {Sippel} A.~C.,  2019, \mn@doi [\mnras] {10.1093/mnras/stz3004}, \href
  {https://ui.adsabs.harvard.edu/abs/2019MNRAS.490.5820M} {490, 5820}

\bibitem[\protect\citeauthoryear{{Mayer}, {Fiacconi}, {Bonoli}, {Quinn},
  {Ro{\v{s}}kar}, {Shen}  \& {Wadsley}}{{Mayer} et~al.}{2015}]{Mayer+2015}
{Mayer} L.,  {Fiacconi} D.,  {Bonoli} S.,  {Quinn} T.,  {Ro{\v{s}}kar} R.,
  {Shen} S.,   {Wadsley} J.,  2015, \mn@doi [\apj]
  {10.1088/0004-637X/810/1/51}, \href
  {https://ui.adsabs.harvard.edu/abs/2015ApJ...810...51M} {810, 51}

\bibitem[\protect\citeauthoryear{{McKinney}, {Tchekhovskoy}  \&
  {Blandford}}{{McKinney} et~al.}{2012}]{McKinney+2012}
{McKinney} J.~C.,  {Tchekhovskoy} A.,   {Blandford} R.~D.,  2012, \mn@doi
  [\mnras] {10.1111/j.1365-2966.2012.21074.x}, \href
  {https://ui.adsabs.harvard.edu/abs/2012MNRAS.423.3083M} {423, 3083}

\bibitem[\protect\citeauthoryear{{Mezcua}}{{Mezcua}}{2017}]{Mezcua2017}
{Mezcua} M.,  2017, \mn@doi [International Journal of Modern Physics D]
  {10.1142/S021827181730021X}, \href
  {https://ui.adsabs.harvard.edu/abs/2017IJMPD..2630021M} {26, 1730021}

\bibitem[\protect\citeauthoryear{{Milosavljevi{\'c}}}{{Milosavljevi{\'c}}}{2004}]{Milosavljevic2004}
{Milosavljevi{\'c}} M.,  2004, \mn@doi [\apjl] {10.1086/420696}, \href
  {https://ui.adsabs.harvard.edu/abs/2004ApJ...605L..13M} {605, L13}

\bibitem[\protect\citeauthoryear{{Miyoshi}, {Moran}, {Herrnstein}, {Greenhill},
  {Nakai}, {Diamond}  \& {Inoue}}{{Miyoshi} et~al.}{1995}]{Miyoshi+1995}
{Miyoshi} M.,  {Moran} J.,  {Herrnstein} J.,  {Greenhill} L.,  {Nakai} N.,
  {Diamond} P.,   {Inoue} M.,  1995, \mn@doi [\nat] {10.1038/373127a0}, \href
  {https://ui.adsabs.harvard.edu/abs/1995Natur.373..127M} {373, 127}

\bibitem[\protect\citeauthoryear{{Neumayer}, {Seth}  \& {B{\"o}ker}}{{Neumayer}
  et~al.}{2020}]{Neumayer+2020}
{Neumayer} N.,  {Seth} A.,   {B{\"o}ker} T.,  2020, \mn@doi [\aapr]
  {10.1007/s00159-020-00125-0}, \href
  {https://ui.adsabs.harvard.edu/abs/2020A&ARv..28....4N} {28, 4}

\bibitem[\protect\citeauthoryear{{Osorno}, {Nagar}, {Richtler}, {Humire},
  {Gebhardt}  \& {Gultekin}}{{Osorno} et~al.}{2023}]{Osorno+2023}
{Osorno} J.,  {Nagar} N.,  {Richtler} T.,  {Humire} P.,  {Gebhardt} K.,
  {Gultekin} K.,  2023, \mn@doi [\aap] {10.1051/0004-6361/202346549}, \href
  {https://ui.adsabs.harvard.edu/abs/2023A&A...679A..37O} {679, A37}

\bibitem[\protect\citeauthoryear{{Pacucci}, {Nguyen}, {Carniani}, {Maiolino}
  \& {Fan}}{{Pacucci} et~al.}{2023}]{Pacucci+2023}
{Pacucci} F.,  {Nguyen} B.,  {Carniani} S.,  {Maiolino} R.,   {Fan} X.,  2023,
  \mn@doi [\apjl] {10.3847/2041-8213/ad0158}, \href
  {https://ui.adsabs.harvard.edu/abs/2023ApJ...957L...3P} {957, L3}

\bibitem[\protect\citeauthoryear{{Pakmor}, {Springel}, {Bauer}, {Mocz},
  {Munoz}, {Ohlmann}, {Schaal}  \& {Zhu}}{{Pakmor} et~al.}{2016}]{Pakmor+2016}
{Pakmor} R.,  {Springel} V.,  {Bauer} A.,  {Mocz} P.,  {Munoz} D.~J.,
  {Ohlmann} S.~T.,  {Schaal} K.,   {Zhu} C.,  2016, \mn@doi [\mnras]
  {10.1093/mnras/stv2380}, \href
  {https://ui.adsabs.harvard.edu/abs/2016MNRAS.455.1134P} {455, 1134}

\bibitem[\protect\citeauthoryear{{Papaloizou} \& {Pringle}}{{Papaloizou} \&
  {Pringle}}{1983}]{Papaloizou+Pringle1983}
{Papaloizou} J.~C.~B.,  {Pringle} J.~E.,  1983, \mn@doi [\mnras]
  {10.1093/mnras/202.4.1181}, \href
  {https://ui.adsabs.harvard.edu/abs/1983MNRAS.202.1181P} {202, 1181}

\bibitem[\protect\citeauthoryear{{Partmann}, {Naab}, {Lah{\'e}n}, {Rantala},
  {Hirschmann}, {Hislop}, {Petersson}  \& {Johansson}}{{Partmann}
  et~al.}{2025}]{Partmann+2025}
{Partmann} C.,  {Naab} T.,  {Lah{\'e}n} N.,  {Rantala} A.,  {Hirschmann} M.,
  {Hislop} J.~M.,  {Petersson} J.,   {Johansson} P.~H.,  2025, \mn@doi [\mnras]
  {10.1093/mnras/staf002}, \href
  {https://ui.adsabs.harvard.edu/abs/2025MNRAS.537..956P} {537, 956}

\bibitem[\protect\citeauthoryear{{Paudel} \& {Yoon}}{{Paudel} \&
  {Yoon}}{2020}]{Paudel+Yoon2020}
{Paudel} S.,  {Yoon} S.-J.,  2020, \mn@doi [\apjl] {10.3847/2041-8213/aba6ed},
  \href {https://ui.adsabs.harvard.edu/abs/2020ApJ...898L..47P} {898, L47}

\bibitem[\protect\citeauthoryear{{Paumard} et~al.,}{{Paumard}
  et~al.}{2006}]{Paumard+2006}
{Paumard} T.,  et~al., 2006, \mn@doi [\apj] {10.1086/503273}, \href
  {https://ui.adsabs.harvard.edu/abs/2006ApJ...643.1011P} {643, 1011}

\bibitem[\protect\citeauthoryear{{Pechetti} et~al.,}{{Pechetti}
  et~al.}{2022}]{Pechetti+2022}
{Pechetti} R.,  et~al., 2022, \mn@doi [\apj] {10.3847/1538-4357/ac339f}, \href
  {https://ui.adsabs.harvard.edu/abs/2022ApJ...924...48P} {924, 48}

\bibitem[\protect\citeauthoryear{{Peebles}}{{Peebles}}{1972}]{Peebles1972}
{Peebles} P.~J.~E.,  1972, \mn@doi [General Relativity and Gravitation]
  {10.1007/BF00755923}, \href
  {https://ui.adsabs.harvard.edu/abs/1972GReGr...3...63P} {3, 63}

\bibitem[\protect\citeauthoryear{{Peirani} et~al.,}{{Peirani}
  et~al.}{2024}]{Peirani+2024}
{Peirani} S.,  et~al., 2024, \mn@doi [\aap] {10.1051/0004-6361/202349101},
  \href {https://ui.adsabs.harvard.edu/abs/2024A&A...686A.233P} {686, A233}

\bibitem[\protect\citeauthoryear{{Portaluri}, {Corsini}, {Morelli}, {Hartmann},
  {Dalla Bont{\`a}}, {Debattista}  \& {Pizzella}}{{Portaluri}
  et~al.}{2013}]{Portaluri+2013}
{Portaluri} E.,  {Corsini} E.~M.,  {Morelli} L.,  {Hartmann} M.,  {Dalla
  Bont{\`a}} E.,  {Debattista} V.~P.,   {Pizzella} A.,  2013, \mn@doi [\mnras]
  {10.1093/mnras/stt738}, \href
  {https://ui.adsabs.harvard.edu/abs/2013MNRAS.433..434P} {433, 434}

\bibitem[\protect\citeauthoryear{{Portegies Zwart} \& {McMillan}}{{Portegies
  Zwart} \& {McMillan}}{2002}]{Portegies_Zwart+McMillan2002}
{Portegies Zwart} S.~F.,  {McMillan} S. L.~W.,  2002, \mn@doi [\apj]
  {10.1086/341798}, \href
  {https://ui.adsabs.harvard.edu/abs/2002ApJ...576..899P} {576, 899}

\bibitem[\protect\citeauthoryear{{Poulain} et~al.,}{{Poulain}
  et~al.}{2025}]{Poulain+2025}
{Poulain} M.,  et~al., 2025, \mn@doi [\nat] {10.1038/s41586-025-08783-9}, \href
  {https://ui.adsabs.harvard.edu/abs/2025arXiv250406749P} {640, 902}

\bibitem[\protect\citeauthoryear{{Rahmati}, {Pawlik}, {Rai{\v{c}}evi{\'c}}  \&
  {Schaye}}{{Rahmati} et~al.}{2013}]{Rahmati+2013}
{Rahmati} A.,  {Pawlik} A.~H.,  {Rai{\v{c}}evi{\'c}} M.,   {Schaye} J.,  2013,
  \mn@doi [\mnras] {10.1093/mnras/stt066}, \href
  {https://ui.adsabs.harvard.edu/abs/2013MNRAS.430.2427R} {430, 2427}

\bibitem[\protect\citeauthoryear{{Rantala}, {Naab}  \& {Lah{\'e}n}}{{Rantala}
  et~al.}{2024}]{Rantala+2024}
{Rantala} A.,  {Naab} T.,   {Lah{\'e}n} N.,  2024, \mn@doi [\mnras]
  {10.1093/mnras/stae1413}, \href
  {https://ui.adsabs.harvard.edu/abs/2024MNRAS.531.3770R} {531, 3770}

\bibitem[\protect\citeauthoryear{{Read}, {Iorio}, {Agertz}  \&
  {Fraternali}}{{Read} et~al.}{2016}]{Read+2016}
{Read} J.~I.,  {Iorio} G.,  {Agertz} O.,   {Fraternali} F.,  2016, \mn@doi
  [\mnras] {10.1093/mnras/stw1876}, \href
  {https://ui.adsabs.harvard.edu/abs/2016MNRAS.462.3628R} {462, 3628}

\bibitem[\protect\citeauthoryear{{Reines}}{{Reines}}{2022}]{Reines2022}
{Reines} A.~E.,  2022, \mn@doi [Nature Astronomy] {10.1038/s41550-021-01556-0},
  \href {https://ui.adsabs.harvard.edu/abs/2022NatAs...6...26R} {6, 26}

\bibitem[\protect\citeauthoryear{{Reines} \& {Volonteri}}{{Reines} \&
  {Volonteri}}{2015}]{Reines+Volonteri2015}
{Reines} A.~E.,  {Volonteri} M.,  2015, \mn@doi [\apj]
  {10.1088/0004-637X/813/2/82}, \href
  {https://ui.adsabs.harvard.edu/abs/2015ApJ...813...82R} {813, 82}

\bibitem[\protect\citeauthoryear{{R{\'e}my-Ruyer} et~al.,}{{R{\'e}my-Ruyer}
  et~al.}{2014}]{Remy-Ruyer+2014}
{R{\'e}my-Ruyer} A.,  et~al., 2014, \mn@doi [\aap]
  {10.1051/0004-6361/201322803}, \href
  {https://ui.adsabs.harvard.edu/abs/2014A&A...563A..31R} {563, A31}

\bibitem[\protect\citeauthoryear{{Rennehan}, {Babul}, {Moa}  \&
  {Dav{\'e}}}{{Rennehan} et~al.}{2024}]{Rennehan+2024}
{Rennehan} D.,  {Babul} A.,  {Moa} B.,   {Dav{\'e}} R.,  2024, \mn@doi [\mnras]
  {10.1093/mnras/stae1785}, \href
  {https://ui.adsabs.harvard.edu/abs/2024MNRAS.532.4793R} {532, 4793}

\bibitem[\protect\citeauthoryear{{Rizzuto}, {Naab}, {Rantala}, {Johansson},
  {Ostriker}, {Stone}, {Liao}  \& {Irodotou}}{{Rizzuto}
  et~al.}{2023}]{Rizzuto+2023}
{Rizzuto} F.~P.,  {Naab} T.,  {Rantala} A.,  {Johansson} P.~H.,  {Ostriker}
  J.~P.,  {Stone} N.~C.,  {Liao} S.,   {Irodotou} D.,  2023, \mn@doi [\mnras]
  {10.1093/mnras/stad734}, \href
  {https://ui.adsabs.harvard.edu/abs/2023MNRAS.521.2930R} {521, 2930}

\bibitem[\protect\citeauthoryear{{Sacchi}, {Bogdan}, {Chadayammuri}  \&
  {Ricarte}}{{Sacchi} et~al.}{2024}]{Sacchi+2024}
{Sacchi} A.,  {Bogdan} A.,  {Chadayammuri} U.,   {Ricarte} A.,  2024, \mn@doi
  [\apj] {10.3847/1538-4357/ad684e}, \href
  {https://ui.adsabs.harvard.edu/abs/2024arXiv240601707S} {974, 14}

\bibitem[\protect\citeauthoryear{{Sala}, {Valentini}, {Biffi}  \&
  {Dolag}}{{Sala} et~al.}{2024}]{Sala+2024}
{Sala} L.,  {Valentini} M.,  {Biffi} V.,   {Dolag} K.,  2024, \mn@doi [\aap]
  {10.1051/0004-6361/202348925}, \href
  {https://ui.adsabs.harvard.edu/abs/2024A&A...685A..92S} {685, A92}

\bibitem[\protect\citeauthoryear{{S{\'a}nchez-Janssen}
  et~al.,}{{S{\'a}nchez-Janssen} et~al.}{2019}]{Sanchez-Janssen+2019}
{S{\'a}nchez-Janssen} R.,  et~al., 2019, \mn@doi [\apj]
  {10.3847/1538-4357/aaf4fd}, \href
  {https://ui.adsabs.harvard.edu/abs/2019ApJ...878...18S} {878, 18}

\bibitem[\protect\citeauthoryear{{Sanchez}, {Reines}, {Bogd{\'a}n}  \&
  {Kraft}}{{Sanchez} et~al.}{2024}]{Sanchez+2024}
{Sanchez} A.~A.,  {Reines} A.~E.,  {Bogd{\'a}n} {\'A}.,   {Kraft} R.~P.,  2024,
  \mn@doi [\apj] {10.3847/1538-4357/ad7588}, \href
  {https://ui.adsabs.harvard.edu/abs/2024ApJ...974....3S} {974, 3}

\bibitem[\protect\citeauthoryear{{Santoro} et~al.,}{{Santoro}
  et~al.}{2022}]{Santoro+2022}
{Santoro} F.,  et~al., 2022, \mn@doi [\aap] {10.1051/0004-6361/202141907},
  \href {https://ui.adsabs.harvard.edu/abs/2022A&A...658A.188S} {658, A188}

\bibitem[\protect\citeauthoryear{{Schartmann}, {Mould}, {Wada}, {Burkert},
  {Durr{\'e}}, {Behrendt}, {Davies}  \& {Burtscher}}{{Schartmann}
  et~al.}{2018}]{Schartmann+2018}
{Schartmann} M.,  {Mould} J.,  {Wada} K.,  {Burkert} A.,  {Durr{\'e}} M.,
  {Behrendt} M.,  {Davies} R.~I.,   {Burtscher} L.,  2018, \mn@doi [\mnras]
  {10.1093/mnras/stx2381}, \href
  {https://ui.adsabs.harvard.edu/abs/2018MNRAS.473..953S} {473, 953}

\bibitem[\protect\citeauthoryear{{Schinnerer}, {B{\"o}ker}, {Emsellem}  \&
  {Downes}}{{Schinnerer} et~al.}{2007}]{Schinnerer+2007}
{Schinnerer} E.,  {B{\"o}ker} T.,  {Emsellem} E.,   {Downes} D.,  2007, \mn@doi
  [\aap] {10.1051/0004-6361:20066711}, \href
  {https://ui.adsabs.harvard.edu/abs/2007A&A...462L..27S} {462, L27}

\bibitem[\protect\citeauthoryear{{Schnittman}}{{Schnittman}}{2007}]{Schnittman2007}
{Schnittman} J.~D.,  2007, \mn@doi [\apjl] {10.1086/522203}, \href
  {https://ui.adsabs.harvard.edu/abs/2007ApJ...667L.133S} {667, L133}

\bibitem[\protect\citeauthoryear{{Sellwood}}{{Sellwood}}{1980}]{Sellwood+1980}
{Sellwood} J.~A.,  1980, \aap, \href
  {https://ui.adsabs.harvard.edu/abs/1980A&A....89..296S} {89, 296}

\bibitem[\protect\citeauthoryear{{Shakura} \& {Sunyaev}}{{Shakura} \&
  {Sunyaev}}{1973}]{Shakura+Sunyaev1973}
{Shakura} N.~I.,  {Sunyaev} R.~A.,  1973, \aap, \href
  {https://ui.adsabs.harvard.edu/abs/1973A&A....24..337S} {24, 337}

\bibitem[\protect\citeauthoryear{{Sharma}, {Brooks}, {Tremmel}, {Bellovary}  \&
  {Quinn}}{{Sharma} et~al.}{2023}]{Sharma+2023}
{Sharma} R.~S.,  {Brooks} A.~M.,  {Tremmel} M.,  {Bellovary} J.,   {Quinn}
  T.~R.,  2023, \mn@doi [\apj] {10.3847/1538-4357/ace046}, \href
  {https://ui.adsabs.harvard.edu/abs/2023ApJ...957...16S} {957, 16}

\bibitem[\protect\citeauthoryear{{Shi}, {Kremer}, {Grudi{\'c}},
  {Gerling-Dunsmore}  \& {Hopkins}}{{Shi} et~al.}{2023}]{Shi+2023}
{Shi} Y.,  {Kremer} K.,  {Grudi{\'c}} M.~Y.,  {Gerling-Dunsmore} H.~J.,
  {Hopkins} P.~F.,  2023, \mn@doi [\mnras] {10.1093/mnras/stac3245}, \href
  {https://ui.adsabs.harvard.edu/abs/2023MNRAS.518.3606S} {518, 3606}

\bibitem[\protect\citeauthoryear{{Shin}, {Kim}  \& {Oh}}{{Shin}
  et~al.}{2021}]{Shin+2021}
{Shin} E.-J.,  {Kim} J.-H.,   {Oh} B.~K.,  2021, \mn@doi [\apj]
  {10.3847/1538-4357/abffd0}, \href
  {https://ui.adsabs.harvard.edu/abs/2021ApJ...917...12S} {917, 12}

\bibitem[\protect\citeauthoryear{{Shlosman}, {Frank}  \& {Begelman}}{{Shlosman}
  et~al.}{1989}]{Shlosman+1989}
{Shlosman} I.,  {Frank} J.,   {Begelman} M.~C.,  1989, \mn@doi [\nat]
  {10.1038/338045a0}, \href
  {https://ui.adsabs.harvard.edu/abs/1989Natur.338...45S} {338, 45}

\bibitem[\protect\citeauthoryear{{Sijacki}, {Springel}  \&
  {Haehnelt}}{{Sijacki} et~al.}{2009}]{Sijacki+2009}
{Sijacki} D.,  {Springel} V.,   {Haehnelt} M.~G.,  2009, \mn@doi [\mnras]
  {10.1111/j.1365-2966.2009.15452.x}, \href
  {https://ui.adsabs.harvard.edu/abs/2009MNRAS.400..100S} {400, 100}

\bibitem[\protect\citeauthoryear{{Smith}}{{Smith}}{2021}]{Smith2021}
{Smith} M.~C.,  2021, \mn@doi [\mnras] {10.1093/mnras/stab291}, \href
  {https://ui.adsabs.harvard.edu/abs/2021MNRAS.502.5417S} {502, 5417}

\bibitem[\protect\citeauthoryear{{Smith} et~al.,}{{Smith}
  et~al.}{2017}]{Smith+2017}
{Smith} B.~D.,  et~al., 2017, \mn@doi [\mnras] {10.1093/mnras/stw3291}, \href
  {https://ui.adsabs.harvard.edu/abs/2017MNRAS.466.2217S} {466, 2217}

\bibitem[\protect\citeauthoryear{{Smith}, {Sijacki}  \& {Shen}}{{Smith}
  et~al.}{2018}]{Smith+2018}
{Smith} M.~C.,  {Sijacki} D.,   {Shen} S.,  2018, \mn@doi [\mnras]
  {10.1093/mnras/sty994}, \href
  {https://ui.adsabs.harvard.edu/abs/2018MNRAS.478..302S} {478, 302}

\bibitem[\protect\citeauthoryear{{Smith}, {Sijacki}  \& {Shen}}{{Smith}
  et~al.}{2019}]{Smith+2019}
{Smith} M.~C.,  {Sijacki} D.,   {Shen} S.,  2019, \mn@doi [\mnras]
  {10.1093/mnras/stz599}, \href
  {https://ui.adsabs.harvard.edu/abs/2019MNRAS.485.3317S} {485, 3317}

\bibitem[\protect\citeauthoryear{{Smith}, {Bryan}, {Somerville}, {Hu},
  {Teyssier}, {Burkhart}  \& {Hernquist}}{{Smith} et~al.}{2021}]{Smith+2021}
{Smith} M.~C.,  {Bryan} G.~L.,  {Somerville} R.~S.,  {Hu} C.-Y.,  {Teyssier}
  R.,  {Burkhart} B.,   {Hernquist} L.,  2021, \mn@doi [\mnras]
  {10.1093/mnras/stab1896}, \href
  {https://ui.adsabs.harvard.edu/abs/2021MNRAS.506.3882S} {506, 3882}

\bibitem[\protect\citeauthoryear{{Solanki}, {Ressler}, {Murchikova}, {Stone}
  \& {Morris}}{{Solanki} et~al.}{2023}]{Solanki+2023}
{Solanki} S.,  {Ressler} S.~M.,  {Murchikova} L.,  {Stone} J.~M.,   {Morris}
  M.~R.,  2023, \mn@doi [\apj] {10.3847/1538-4357/acdb6f}, \href
  {https://ui.adsabs.harvard.edu/abs/2023ApJ...953...22S} {953, 22}

\bibitem[\protect\citeauthoryear{{Spitzer} \& {Hart}}{{Spitzer} \&
  {Hart}}{1971}]{Spitzer+Hart1971}
{Spitzer} Jr. L.,  {Hart} M.~H.,  1971, \mn@doi [\apj] {10.1086/150855}, \href
  {https://ui.adsabs.harvard.edu/abs/1971ApJ...164..399S} {164, 399}

\bibitem[\protect\citeauthoryear{{Spitzer} \& {Stone}}{{Spitzer} \&
  {Stone}}{1967}]{Spitzer+Stone1967}
{Spitzer} Jr. L.,  {Stone} M.~E.,  1967, \mn@doi [\apj] {10.1086/149032}, \href
  {https://ui.adsabs.harvard.edu/abs/1967ApJ...147..519S} {147, 519}

\bibitem[\protect\citeauthoryear{{Springel}}{{Springel}}{2010}]{Springel+2010}
{Springel} V.,  2010, \mn@doi [\mnras] {10.1111/j.1365-2966.2009.15715.x},
  \href {https://ui.adsabs.harvard.edu/abs/2010MNRAS.401..791S} {401, 791}

\bibitem[\protect\citeauthoryear{{Springel}, {Di Matteo}  \&
  {Hernquist}}{{Springel} et~al.}{2005}]{Springel+2005}
{Springel} V.,  {Di Matteo} T.,   {Hernquist} L.,  2005, \mn@doi [\mnras]
  {10.1111/j.1365-2966.2005.09238.x}, \href
  {https://ui.adsabs.harvard.edu/abs/2005MNRAS.361..776S} {361, 776}

\bibitem[\protect\citeauthoryear{{Stone}, {K{\"u}pper}  \& {Ostriker}}{{Stone}
  et~al.}{2017}]{Stone+2017}
{Stone} N.~C.,  {K{\"u}pper} A. H.~W.,   {Ostriker} J.~P.,  2017, \mn@doi
  [\mnras] {10.1093/mnras/stx097}, \href
  {https://ui.adsabs.harvard.edu/abs/2017MNRAS.467.4180S} {467, 4180}

\bibitem[\protect\citeauthoryear{{Talbot}, {Bourne}  \& {Sijacki}}{{Talbot}
  et~al.}{2021}]{Talbot+2021}
{Talbot} R.~Y.,  {Bourne} M.~A.,   {Sijacki} D.,  2021, \mn@doi [\mnras]
  {10.1093/mnras/stab804}, \href
  {https://ui.adsabs.harvard.edu/abs/2021MNRAS.504.3619T} {504, 3619}

\bibitem[\protect\citeauthoryear{{Talbot}, {Sijacki}  \& {Bourne}}{{Talbot}
  et~al.}{2022}]{Talbot+2022}
{Talbot} R.~Y.,  {Sijacki} D.,   {Bourne} M.~A.,  2022, \mn@doi [\mnras]
  {10.1093/mnras/stac1566}, \href
  {https://ui.adsabs.harvard.edu/abs/2022MNRAS.514.4535T} {514, 4535}

\bibitem[\protect\citeauthoryear{{Talbot}, {Sijacki}  \& {Bourne}}{{Talbot}
  et~al.}{2024}]{Talbot+2024}
{Talbot} R.~Y.,  {Sijacki} D.,   {Bourne} M.~A.,  2024, \mn@doi [\mnras]
  {10.1093/mnras/stae392}, \href
  {https://ui.adsabs.harvard.edu/abs/2024MNRAS.528.5432T} {528, 5432}

\bibitem[\protect\citeauthoryear{{Tchekhovskoy}, {Narayan}  \&
  {McKinney}}{{Tchekhovskoy} et~al.}{2011}]{Tchekhovskoy+2011}
{Tchekhovskoy} A.,  {Narayan} R.,   {McKinney} J.~C.,  2011, \mn@doi [\mnras]
  {10.1111/j.1745-3933.2011.01147.x}, \href
  {https://ui.adsabs.harvard.edu/abs/2011MNRAS.418L..79T} {418, L79}

\bibitem[\protect\citeauthoryear{{Topping} et~al.,}{{Topping}
  et~al.}{2024}]{Topping+2024}
{Topping} M.~W.,  et~al., 2024, \mn@doi [\mnras] {10.1093/mnras/stae682}, \href
  {https://ui.adsabs.harvard.edu/abs/2024MNRAS.529.3301T} {529, 3301}

\bibitem[\protect\citeauthoryear{{Trani}, {Mapelli}  \& {Ballone}}{{Trani}
  et~al.}{2018}]{Trani+2018}
{Trani} A.~A.,  {Mapelli} M.,   {Ballone} A.,  2018, \mn@doi [\apj]
  {10.3847/1538-4357/aad414}, \href
  {https://ui.adsabs.harvard.edu/abs/2018ApJ...864...17T} {864, 17}

\bibitem[\protect\citeauthoryear{{Tremaine} \& {Ostriker}}{{Tremaine} \&
  {Ostriker}}{1999}]{Tremaine+1999}
{Tremaine} S.,  {Ostriker} J.~P.,  1999, \mn@doi [\mnras]
  {10.1046/j.1365-8711.1999.02558.x}, \href
  {https://ui.adsabs.harvard.edu/abs/1999MNRAS.306..662T} {306, 662}

\bibitem[\protect\citeauthoryear{{Tremaine}, {Ostriker}  \&
  {Spitzer}}{{Tremaine} et~al.}{1975}]{Tremaine+1975}
{Tremaine} S.~D.,  {Ostriker} J.~P.,   {Spitzer} Jr. L.,  1975, \mn@doi [\apj]
  {10.1086/153422}, \href
  {https://ui.adsabs.harvard.edu/abs/1975ApJ...196..407T} {196, 407}

\bibitem[\protect\citeauthoryear{{Walcher}, {B{\"o}ker}, {Charlot}, {Ho},
  {Rix}, {Rossa}, {Shields}  \& {van der Marel}}{{Walcher}
  et~al.}{2006}]{Walcher+2006}
{Walcher} C.~J.,  {B{\"o}ker} T.,  {Charlot} S.,  {Ho} L.~C.,  {Rix} H.~W.,
  {Rossa} J.,  {Shields} J.~C.,   {van der Marel} R.~P.,  2006, \mn@doi [\apj]
  {10.1086/505166}, \href
  {https://ui.adsabs.harvard.edu/abs/2006ApJ...649..692W} {649, 692}

\bibitem[\protect\citeauthoryear{{Weinberg}}{{Weinberg}}{1985}]{Weinberg+1985}
{Weinberg} M.~D.,  1985, \mn@doi [\mnras] {10.1093/mnras/213.3.451}, \href
  {https://ui.adsabs.harvard.edu/abs/1985MNRAS.213..451W} {213, 451}

\bibitem[\protect\citeauthoryear{{Wheeler} et~al.,}{{Wheeler}
  et~al.}{2017}]{Wheeler+2017}
{Wheeler} C.,  et~al., 2017, \mn@doi [\mnras] {10.1093/mnras/stw2583}, \href
  {https://ui.adsabs.harvard.edu/abs/2017MNRAS.465.2420W} {465, 2420}

\bibitem[\protect\citeauthoryear{{Zubovas}}{{Zubovas}}{2019}]{Zubovas+2019}
{Zubovas} K.,  2019, \mn@doi [\mnras] {10.1093/mnras/sty3211}, \href
  {https://ui.adsabs.harvard.edu/abs/2019MNRAS.483.1957Z} {483, 1957}

\bibitem[\protect\citeauthoryear{{den Brok} et~al.,}{{den Brok}
  et~al.}{2014}]{den_Brok+2014}
{den Brok} M.,  et~al., 2014, \mn@doi [\mnras] {10.1093/mnras/stu1906}, \href
  {https://ui.adsabs.harvard.edu/abs/2014MNRAS.445.2385D} {445, 2385}

\makeatother
\end{thebibliography}




\appendix

\section{Circumgalactic medium properties}
\label{sec:cgm}
\begin{figure}
\includegraphics[width=0.48\textwidth]{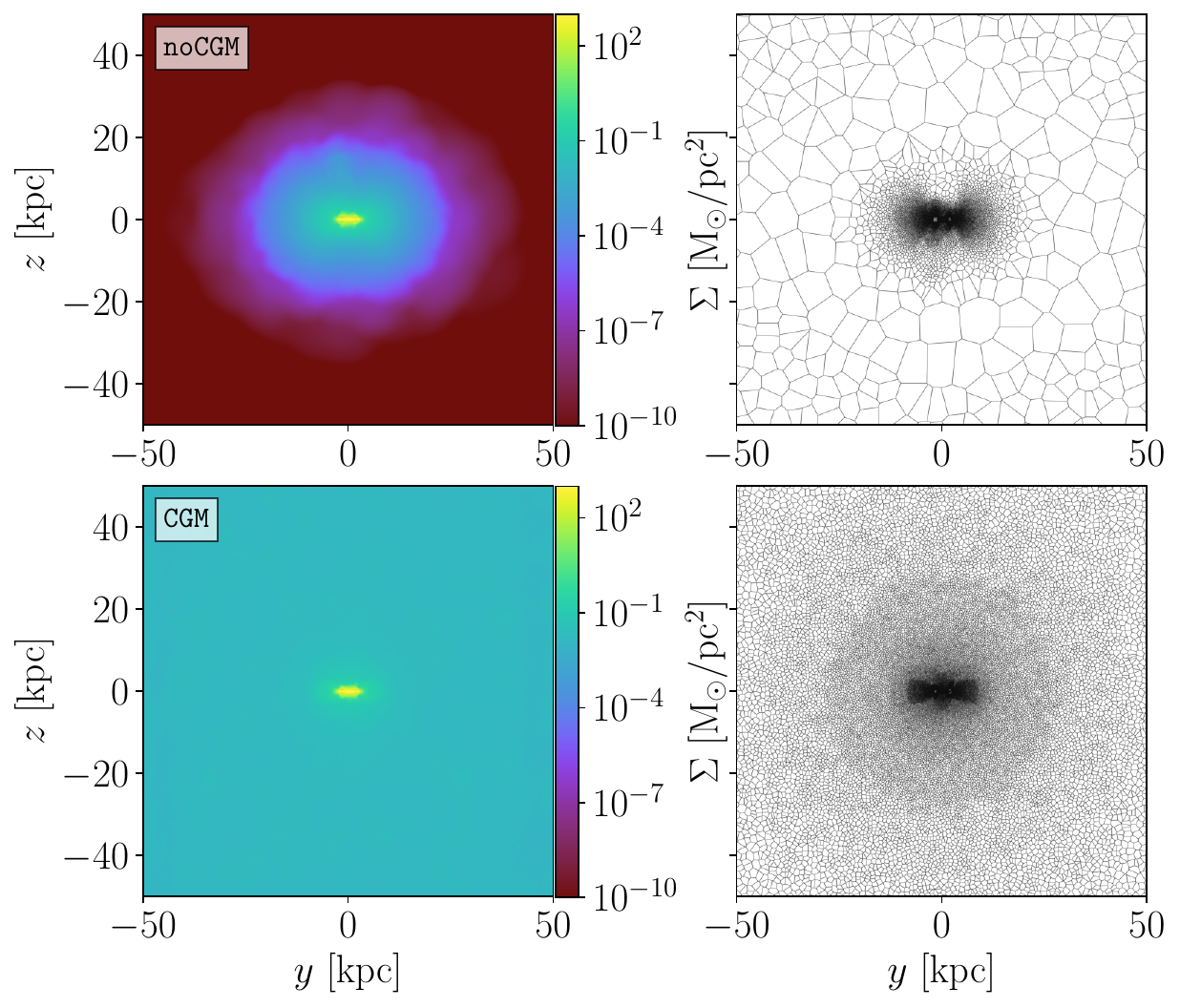}
\centering
    \caption{Visuals of {\tt noCGM} ({\it top}) and {\tt CGM} ({\it bottom}) runs at $t=200$~Myr.
    {\it Left}: edge-on gas density projection plots in the (100~kpc)$^3$ box.
    {\it Right}: Voronoi slice plots of gas cells. The gas cells size as a function of radius is shown in the Fig.~\ref{fig:SL-pdf}. 
    For more information, see Section~\ref{sec:refinement-cgm} and Appendix~\ref{sec:cgm}.}
    \label{fig:cgm-map}
\end{figure}
\begin{figure*}\includegraphics[width=\textwidth]{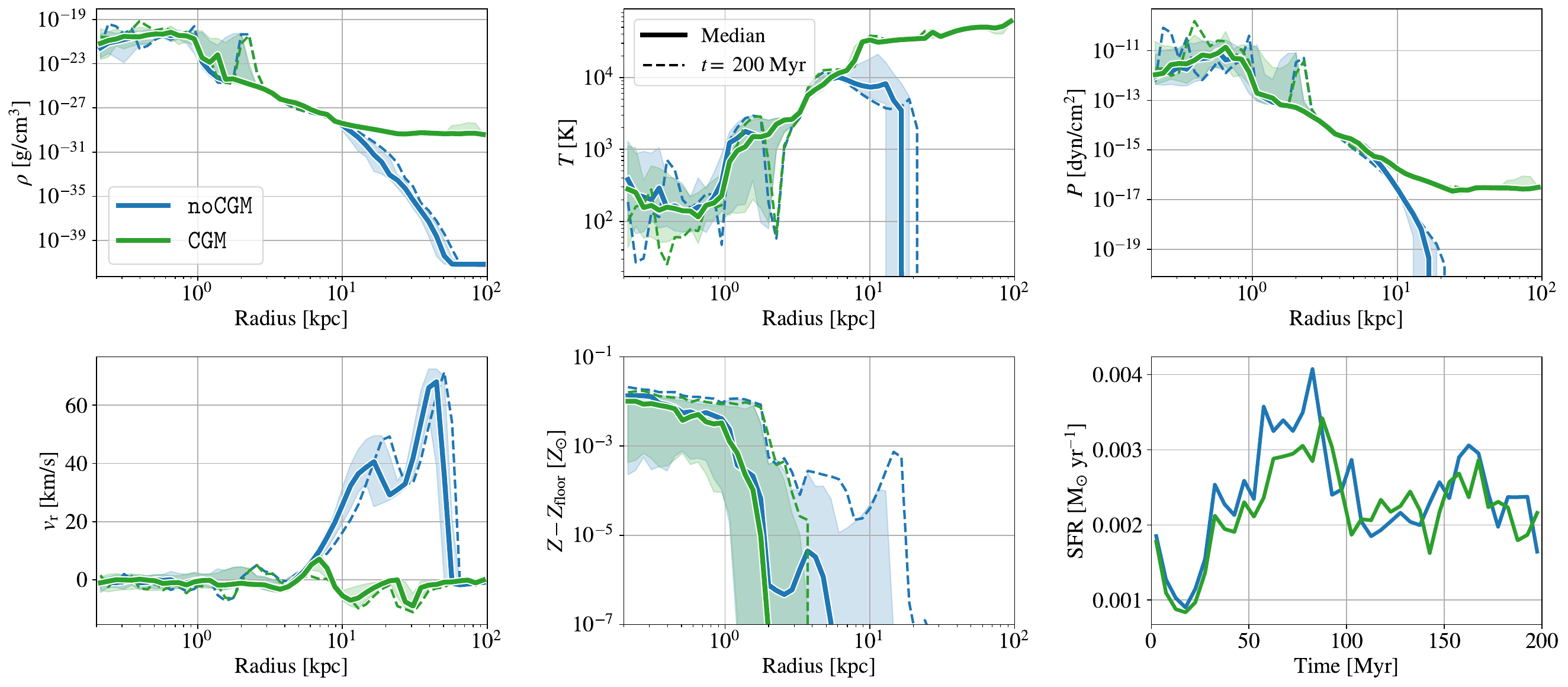}
    \caption{Radial profiles of gas density (density-weighted), temperature (density-weighted), pressure (density-weighted), radial velocity (density-weighted), metallicity (density-weighted) as well as the time evolution of SFR for {\tt noCGM} and {\tt CGM} runs. The metallicity difference, $Z-Z_{\rm floor}$, represents the deviation of metallicity from the initial value, $Z_{\rm floor}=0.1 Z_{\odot}$. The {\it dashed} lines indicate the results at 200~Myr, while the {\it solid} lines show the median values over 200~Myr. The {\it shaded} regions represent the 16th and 84th percentiles during 200~Myr. For more information, see Appendix~\ref{sec:cgm}.}
    \label{fig:CGM-profile}
\end{figure*}

Fig.~\ref{fig:cgm-map} shows gas density projections and Voronoi mesh slice plots for simulations with and without CGM. In the {\tt noCGM} run, the gas density in the central region ($r < 7$ kpc) is comparable to that in the {\tt CGM} run, but it decreases significantly with radius and has large cell sizes. 
Note that cell sizes in the {\tt noCGM} run remain uniform across the entire box, except in regions containing hot, diffuse outflows from the disc.
In contrast, the gas distribution in the {\tt CGM} run maintains an almost constant gas surface density of approximately $0.1 \,\msun\,{\rm pc}^{-2}$ throughout the entire box. The cell sizes are configured to increase with radius according to the spatially dependent resolution scheme, allowing CGM cells to be refined or derefined when the outflow mass fraction in the cell exceeds 10\% (see Section~\ref{sec:refinement-cgm}).

Fig.~\ref{fig:CGM-profile} compares the time-averaged radial profiles of gas density, temperature, metallicity, pressure, and radial velocity, as well as the time evolution of the star formation rate, between the two simulations. In both simulations, the gas distribution in the disc region, particularly within $r < 3$~kpc, exhibits significant variation over time, driven by intense star formation and stellar feedback. In the {\tt noCGM} run, the density distribution around $r = 10$~kpc fluctuates with time due to SNe-driven outflows. In contrast, the outer disc region in the {\tt CGM} run remains stable, as its gas distribution is approximate thermodynamic equilibrium. As shown in Fig.~\ref{fig:cgm-map}, the {\tt noCGM} run exhibits significant radial variations, characterized by a more diffuse, cold, and low-pressure medium compared to the {\tt CGM} run. In the outer disc of the {\tt CGM} run, the gas maintains a constant density of $10^{-29} {\rm g}~{\rm cm}^{-3}$, a temperature of $3 \times 10^4$ K, and a pressure of $10^{-17}~{\rm dyn~cm}^{-2}$.

Examining the metallicity distribution reveals that stellar feedback-driven outflows in the {\tt noCGM} run extend to approximately 20~kpc, with outflow velocities reaching $70~{\rm km}~{\rm s}^{-1}$ at a radius of around 50~kpc. In contrast, in the {\tt CGM} run, most metals are confined within a radius of 3~kpc at $t = 200$~Myr. This confinement is caused by ram pressure between the metal-rich outflows and the CGM. The ram pressure further reduces outflow velocity to approximately $10~{\rm km}~{\rm s}^{-1}$ \citep[see e.g.,][]{Shin+2021}. We note that these CGM properties are consistent with the CGM distribution observed in cosmological zoom-in simulations of dwarf galaxies by \citet{Koudmani+2022}.

Lastly, in the {\it bottom-right} panel, we present the time evolution of the star formation rate (SFR) for the {\tt CGM} and {\tt noCGM} runs. Both runs exhibit a similar SFR trend. Initially, the SFRs decrease during the first 20~Myr due to photo-ionization feedback driven by massive stars formed during the initial star formation episode \citep[see Figure 6 and discussion in][]{Smith+2021}. Subsequently, the SFR increases until $t = 100$~Myr and stabilizes at approximately $2.5 \times 10^{-3} \msun~{\rm yr}^{-1}$. This smooth evolution of the SFR is attributed to early stellar feedback processes, namely photo-ionization and photoelectric heating \citep{Smith+2021}. The total masses of newly formed stars are $4.86 \times 10^5 \msun$ and $4.48 \times 10^5 \msun$ for the {\tt CGM} and {\tt noCGM} runs, respectively. Since the CGM was initialized in an equilibrium state consistent with the halo structure, it does not exert additional pressure on the disc. However, we note that if the CGM is not in hydrostatic equilibrium with the halo potential, the SFR could be increased by CGM pressure or inflows, though this is not the case in our simulation. 
\section{Evolution of $\alpha$-accretion disc: self-gravity radius and angular momentum}
\label{sec:appendix-alpha-disc}
\begin{figure*}
\includegraphics[width=0.9\textwidth]{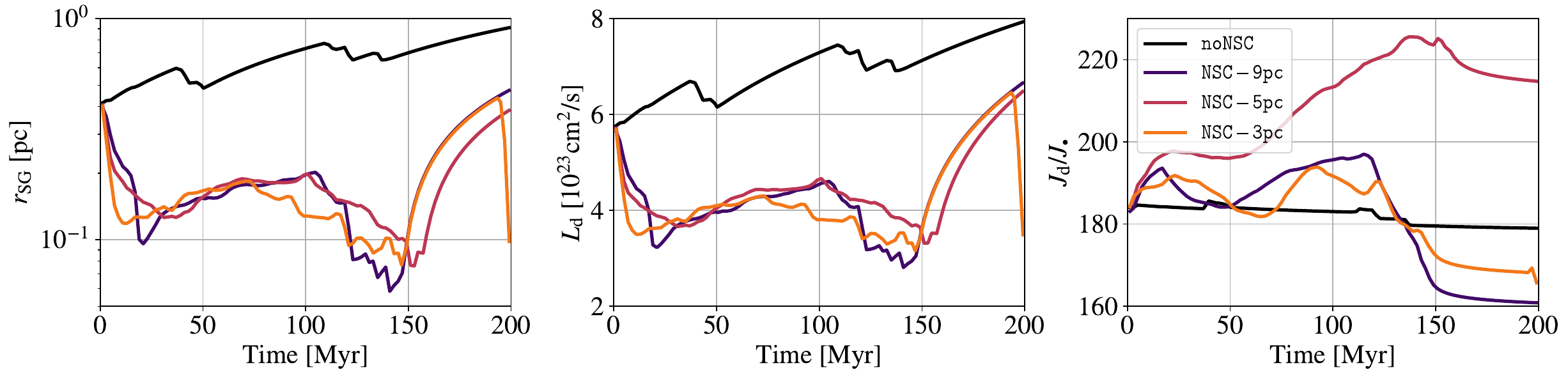}
    \caption{Time evolution of the accretion disc's self-gravity radius ({\it left}), specific angular momentum ({\it middle}), and the angular momentum ratio between the accretion disc and the BH ($J_{\rm d}/J_{\bullet}$; {\it right}), For more details, see Appendix~\ref{sec:appendix-alpha-disc}.}
\label{fig:appendix-bh-evolution}
\end{figure*}

In this section, we briefly discuss the time evolution of accretion disc properties. The self-gravity radius of accretion disc, $r_{\rm SG}$ is defined where the Toomre parameter $Q$ below the critical value, $Q(r_{\rm SG})=1$; we limit the accretion disc mass to be smaller than 
\begin{equation}
    \frac{r_{\rm SG}}{r_{\rm Sch}}\approx1.25\times10^5 \left(\frac{\alpha}{0.1}\right)^{28/45} \left(\frac{f_{\rm Edd}}{\eta_{0.1}}\right)^{-22/45} \left(\frac{M_{\bullet}}{10^6\msun}\right)^{-52/45}\,,
    \label{eq:r_SG}
\end{equation}
where $\alpha=0.1$ is the Shakura-Sunyaev viscosity parameter and $r_{\rm Sch}=2G{M_{\bullet}}/c^2$ is the Schwarzschild radius, while the specific angular momentum of the accretion disc follows:
\begin{equation}
\begin{aligned}
    {L_{\rm d}}&\approx1.23\times10^{24}{\rm cm}^2{\rm s}^{-1}\\
    &\times\left(\frac{\alpha}{0.1}\right)^{8/25}\left(\frac{M_{\rm d}}{10^4\msun}\right)^{2/5}  \left(\frac{f_{\rm Edd}}{\eta_{0.1}}\right)^{-7/25} \left(\frac{M_{\bullet}}{10^6\msun}\right)^{-3/25}\,.
\end{aligned}
\label{eq:specificAM-alpha-disc}
\end{equation}
For the detailed derivation, we refer to Appendix A in \cite{Fiacconi+2018}.

The left panel of Fig.~\ref{fig:appendix-bh-evolution} shows the evolution of the self-gravity radius in the simulations with and without an NSC. As can be seen from Eq.~\ref{eq:r_SG}, the self-gravity radius exhibits an inverse correlation with the Eddington fraction, $f_{\rm Edd}$. Initially, the self-gravity radius is approximately 0.4~pc, determined by the initial accretion disc mass and the Eddington fraction. 
In runs with an NSC, the accretion disc is continuously supplied with material from the CND, which increases the Eddington fraction and consequently causes the self-gravity radius to shrink to $\gtrsim0.1$~pc. For the run without an NSC, the self-gravity radius increases as the Eddington fraction decreases (see Fig.~\ref{fig:bondi-vs-adspin}), except during episodic accretion events at $t=40-50$~Myr and $t=100-140$~Myr. By $t=200$~Myr, the self-gravity radius reaches $0.9$~pc.

The specific angular momentum of the accretion disc, shown in the middle panel of  Fig.~\ref{fig:appendix-bh-evolution}, follows similar trends, as expected from Eq.~\ref{eq:specificAM-alpha-disc}. In the presence of a NSC, the specific angular momentum fluctuates between $3\times10^{23}-6\times10^{23}~{\rm cm}^2~{\rm s}^{-1}$, whereas in the absence of a NSC, it increases and reaches to $8\times10^{23}~{\rm cm}^2~{\rm s}^{-1}$ at $t=200$~Myr.

The right panel of Fig.~\ref{fig:appendix-bh-evolution} shows the evolution of the angular momentum ratio between the accretion disc and the BH ($J_{\rm d}/J_{\bullet}$). In all models, the angular momentum of the accretion disc is dominant, and the BH has only $\lesssim0.5\%$ of the total angular momentum over the entire simulated timespan. In the absence of an NSC, the ratio remains relatively constant, as BH accretion exceeds external mass inflow onto the $\alpha$-accretion disc, gradually reducing the accretion disc's angular momentum budget. In contrast, as shown in Section~\ref{sec:bh-evolution}, in runs with an NSC, the inflow from the CND increases the angular momentum of the accretion disc by 36-63\%, which results in a notable change in $J_{\rm d}/J_{\bullet}$. As shown in Fig.~\ref{fig:bh-ad-evol}, in the {\tt NSC-5pc} simulation, inflowing gas remains aligned with the rotation of the accretion disc throughout the $140$~Myr period, driving a faster increase in the angular momentum of the accretion disc. As discussed in Section~\ref{sec:bh-evolution}, when the BH spin aligns with the accretion disc, accretion from the disc onto the BH drives an increase in the angular momentum of the BH \citep[also see][]{Fiacconi+2018}. Thus, without sufficiently high angular momentum replenishment, the disc will gradually lose its angular momentum to the BH.

\section{ {Torque from CND-accretion disc misalignment}}

\label{sec:appendix-torque-by-warped-disc}

 {In this section, we estimate the viscous torque exchanged between the CND and the $\alpha$-disc due to their misalignment. According to \cite{Papaloizou+Pringle1983}, assuming the presence of effective viscosity, the specific torque exerted between the inner and outer disc can be approximated as follows:}

 {\begin{equation}
\begin{split}
    \tau_{\rm warp} 
    &\approx 2\pi\nu\Sigma r^3\omega \left|\frac{d\hat{l}}{dr} \right|m^{-1} \\
    &\approx 4.1\times10^6 {\rm cm}^2/{\rm s}^2\left(\frac{\nu}{2.2\times10^{18}{\rm cm^2/s}} \right)\left(\frac{\Sigma}{10^3 \msun{\rm pc}^{-2}} \right)\left(\frac{r}{0.2 {\rm pc}}\right)^{3}\\ 
    &\left(\frac{\omega}{6\times10^{-14}/{\rm s}} \right)\left(\frac{|d\hat{l}/dr|}{90^{\circ}/0.1{\rm pc}} \right)\left(\frac{m}{100\msun} \right)^{-1},
\end{split}
\end{equation}}

 {where $\nu$ is the effective viscosity, $m_{\rm d}$, $r$, $\Sigma$, $\omega$ is the mass, radius, surface density and angular velocity of the inner disc (i.e., $\alpha$-disc), respectively, and $|d\hat{l}/dr|$ is the magnitude of the radial derivative of the unit tilt vector between the outer disc and inner disc. For our simulation setup the resulting torque timescale due to the warp remains on the order of $\sim 1$~Gyr. Therefore, in Eq.~\ref{eq:6}, we can ignore the torque term arising from the misalignment between the CND and the accretion disc.}

\section{ {Gas rotation curve comparison}}
\label{sec:appendix-wlm-rotation}

\begin{figure}
\includegraphics[width=0.45\textwidth]{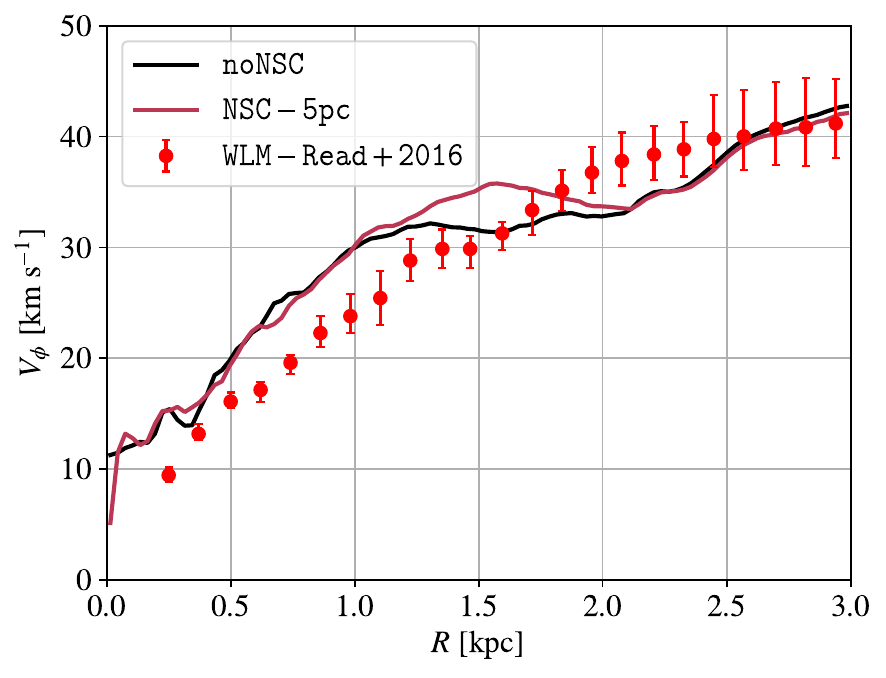}
    \caption{ {Comparison of rotation curves between our simulated dwarf galaxies ({\tt noNSC} and {\tt NSC-5pc} runs) and the WLM dwarf galaxy based on H\I\,  data analysed by \citet{Read+2016}.}}
    \label{fig:appendix-WLM-rot}
\end{figure}

 {Fig.~\ref{fig:appendix-WLM-rot} shows the rotational velocity curves of our simulated dwarf galaxy with and without NSC ({\it crimson} and {\it black} lines, respectively), along with the observed WLM galaxy ({\it red} dots). 
The WLM data are taken from \cite{Read+2016}, with dark matter halo mass of $M_{200}=8.3^{+2.1}_{-2.2}\times10^9 \msun$ and a concentration parameter of $c=17^{+3.9}_{-2.2}$, both quoted at 68\% confidence. Our halo parameters fall within the quoted error ranges. 
Overall, the rotation curves of both {\tt noNSC} and {\tt NSC-5pc} runs agree well with the observed data within $\pm5$~km~s$^{-1}$, showing that our initial conditions are consistent with the rotation curve of WLM. 
We further verified that the rotation curve remains stable throughout the simulation, exhibiting only minor variations. Moreover, \citet{Read+2016} compared concentration parameters for dwarf galaxies of similar mass and found values in the range $c=15.1-28.9$. This indicates that our adopted concentration of $c=15$ lies at the conservative end of the observed distribution. Importantly, despite this conservative choice, the CND still forms rapidly in our simulations, appearing within only $\sim$3 Myr from the simulation start, which demonstrates the robustness of this process.
}

\section{ {Torque interaction between the CND and NSC}}
\label{sec:appendix-torque-CND}
\begin{figure*}
\includegraphics[width=0.7\textwidth]{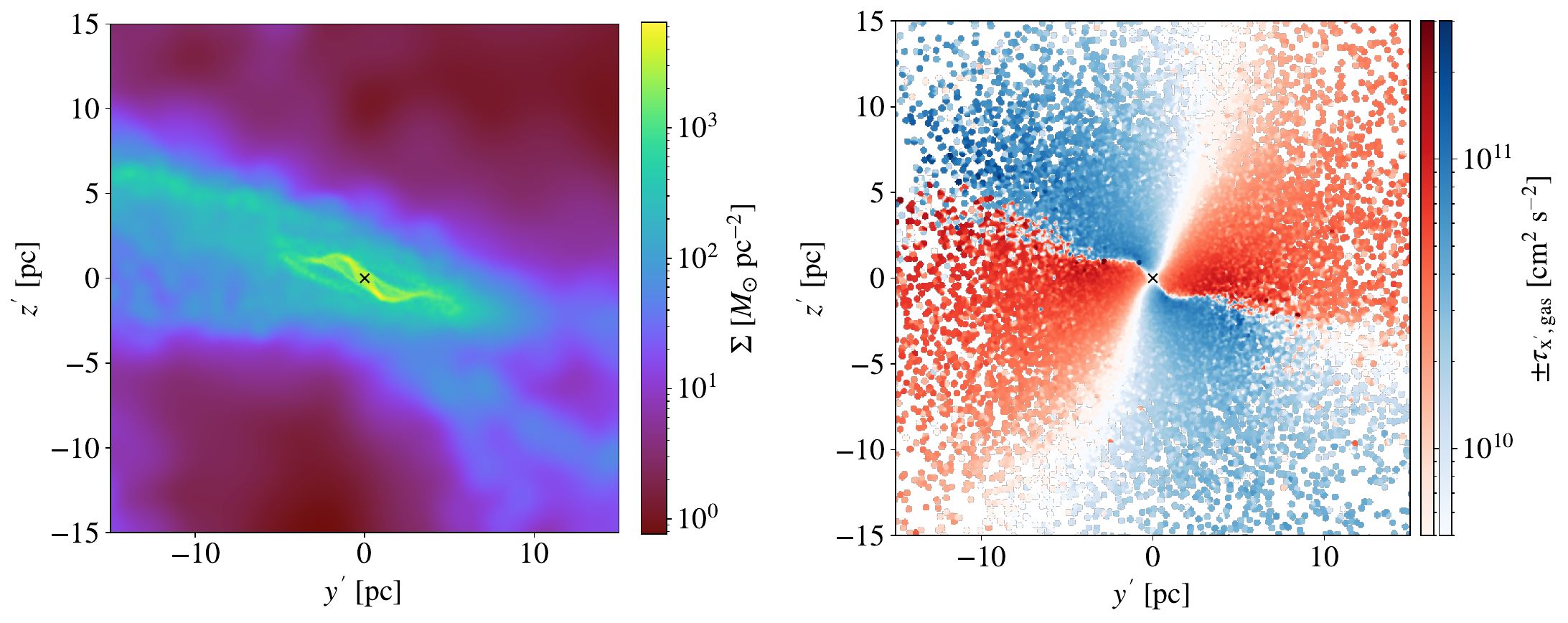}
\centering
    \caption{
    Projected gas surface density in the region where CND resides ({\it left}) and the corresponding torque map ({\it right}) showing the edge-on view of the {\tt NSC-5pc} run at $t=135$ Myr. The torque map shows the mass-weighted gravitational torque exerted by the CND on the NSC. The \textit{cross} symbol marks the location of the BH.
    }
    \label{fig:torque-cnd}
\end{figure*}
 {In this section, we provide further details on the torque interaction between the CND and the NSC. Fig.~\ref{fig:torque-cnd} shows the edge-on view of the CND surface density ({\it left}) and the corresponding map of the  gravitational torque exerted on the NSC by the CND ({\it right}) for the {\tt NSC-5pc} run at $t=135$ Myr.
The quadrupolar pattern induced by the (warped) CND, characterized by alternating signs at the quadrant boundaries where the torque changes direction, is clearly evident. 
This pattern reflects the gravitational coupling between the axisymmetric CND and the NSC, whereby CND induces a wake in the NSC which exerts a torque on the CND slowing it down. As a consequence the NSC gradually gains angular momentum and becomes aligned with the CND.
The torque strength is on the order of $\sim10^{11}\ {\rm cm^{2}\ s^{-2}}$, comparable in magnitude to the NSC torque shown in Fig.~\ref{fig:torque-prof-nsc}. 
This behaviour is analogous to the resonant torque mechanism \citep[see, e.g.,][]{Sellwood+1980, Weinberg+1985, Tremaine+1999}, in which the axisymmetric structure excites a coherent response in the surrounding spherical stellar or dark matter system, which in turn absorbs and dissipates the angular momentum of the bar or disc.}

\section{Energy distribution in the circumnuclear region}
\label{sec:appendix-energy-profile}
\begin{figure*}
\includegraphics[width=\textwidth]{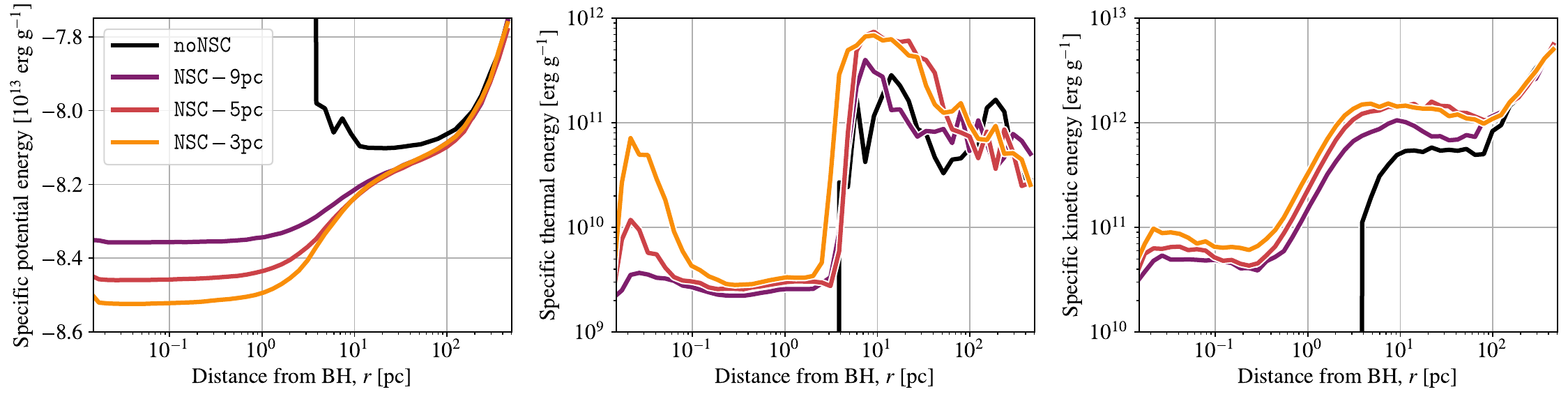}
    \caption{Radial profiles of specific potential energy (density-weighted; {\it left}), specific thermal energy (density-weighted; {\it middle}) and specific kinetic energy (density-weighted; {\it right}). The lines represent the median values during the first $135$~Myr.}
    \label{fig:energy}
\end{figure*}

Fig.~\ref{fig:energy} shows the radial distribution of specific gravitational potential, thermal and kinetic energy in the circumnuclear region for simulations with and without an NSC. For $r>100$ pc, the energy profiles remain similar regardless of the presence of an NSC. As the system approaches the NSC, a denser NSC produces a deeper gravitational potential. The potential energy difference between the most compact configuration ({\tt NSC-3pc}) and the most diffuse one ({\tt NSC-9pc}) is approximately $2\times10^{12} {~{\rm erg}~{\rm g}^{-1}}$. The denser NSC has a steeper potential gradient at $r=1-10$ pc, inducing a stronger tidal force. This, in turn, makes the denser NSC more gravitationally stable, increasing the mass threshold for the star formation as shown in Fig.~\ref{fig:sfr-evol}. As expected, the deeper gravitational potential allows the system to rotate faster, leading to a higher kinetic energy for $r<7$~pc (see also Fig.~\ref{fig:nsc-profile}). In such a potential well, the circumnuclear region can sustain a hotter and more turbulent gas. In the absence of the NSC, the potential energy does not exceed $-8.1\times10^{-13} {~{\rm erg}~{\rm g}^{-1}}$, resulting in inefficient accretion onto the CND.


\bsp	
\label{lastpage}
\end{document}